\documentclass[prd,singlecolumn,amsmath,nobibnotes,nofootinbib,superscriptaddress,preprintnumbers,floatfix]{revtex4}

\usepackage{bm} 
\usepackage{graphicx}
\usepackage{color} 
\usepackage{amssymb}
\usepackage{rotating}
\usepackage{url}

\def\ba{\begin{eqnarray}}
\def\ea{\end{eqnarray}}
\def\be{\begin{equation}}
\def\ee{\end{equation}}
\def\({\left(}
\def\){\right)}
\def\[{\left[}
\def\]{\right]}
\def\<{\left<}
\def\>{\right>}

\begin{document}

\title{Determining the outcome of cosmic bubble collisions in full General Relativity}
\date{\today}

\author{Matthew C. Johnson}
\email{mjohnson@perimeterinstitute.ca}
\affiliation{Perimeter Institute for Theoretical Physics, Waterloo, Ontario N2L 2Y5, Canada} 
\author{Hiranya V. Peiris}
\email{h.peiris@ucl.ac.uk}
\affiliation{Department of Physics and Astronomy, University College London, London WC1E 6BT, U.K.}
\author{Luis Lehner}
\email{llehner@perimeterinstitute.ca }
\affiliation{Perimeter Institute for Theoretical Physics, Waterloo, Ontario N2L 2Y5, Canada} 
\affiliation{Department of Physics, University of Guelph, Guelph, Ontario N1G 2W1, Canada}

\begin{abstract}
Cosmic bubble collisions provide an important possible observational window on the dynamics of eternal inflation. In eternal inflation, our observable universe is contained in one of many bubbles formed from an inflating metastable vacuum. The collision between bubbles can leave a detectable imprint on the cosmic microwave background radiation. Although phenomenological models of the observational signature have been proposed, to make the theory fully predictive one must determine the bubble collision spacetime, and thus the cosmological observables, from a scalar field theory giving rise to eternal inflation. Because of the intrinsically non-linear nature of the bubbles and their collision, this requires a numerical treatment incorporating General Relativity. In this paper, we present results from numerical simulations of bubble collisions in full General Relativity. These simulations allow us to accurately determine the outcome of bubble collisions, and examine their effect on the cosmology inside a bubble universe. We confirm the validity of a number of approximations used in previous analytic work, and identify qualitatively new features of bubble collision spacetimes. Both vacuum bubbles and bubbles containing a realistic inflationary cosmology are studied. We identify the constraints on the scalar field potential that must be satisfied in order to obtain collisions that are consistent with our observed cosmology, yet leave detectable signatures. 
\end{abstract}

\preprint{}

\maketitle

\section{Introduction}

Accelerated expansion of the universe and spontaneous symmetry breaking, two powerful ideas in modern physics, conspire in many theories to give rise to a phenomenon known as eternal inflation. This phenomenon occurs when a region of the universe is in a metastable vacuum with positive energy density. Classically, these vacua are stable, and drive inflation (exponential expansion). Quantum mechanically, they can decay via the nucleation of expanding bubbles \cite{Coleman:1977py,Coleman:1980aw} containing a new phase. Unless the rate of bubble formation outpaces the expansion of the inflating background, the original phase is never completely consumed, and inflation becomes eternal. In this picture, many different phases can be seeded from an eternally inflating ``parent" vacuum, leading to a patchwork-universe with diverse spacetime-dependent physical properties. For a review of eternal inflation, see e.g. Ref.~\cite{Aguirre:2007gy}. 

Surprisingly, there may be directly  observable consequences of living in an eternally-inflating multiverse. In eternal inflation, our observable universe resides inside one member of an ensemble of bubbles. These bubbles necessarily collide, perturbing the homogeneity and isotropy of our own bubble interior, and providing a direct observational test of eternal inflation~\cite{Aguirre:2007an}. Assessing the likelihood of seeing bubble collisions, determining their effects on cosmology, and predicting their observational signatures has been the subject of a large body of work~\cite{Gott:1984ps,Garriga:2006hw,Hawking:1982ga,Wu:1984eda,Moss:1994pi,Freivogel:2007fx,Aguirre:2007wm,Aguirre:2008wy,Chang:2007eq,Chang:2008gj,Czech:2010rg,Freivogel:2009it,Easther:2009ft,Giblin:2010bd,Kleban:2011yc,Lim:2011kd,Dahlen:2008rd,Johnson:2010bn}. This body of literature has established the plausibility of models where collisions are likely, compatible with our observed cosmology, and leave observable signatures. However, an existence proof does not yet exist. For a review of much of this work, see Ref.~\cite{Aguirre:2009ug}. 

The primordial inhomogeneities left by bubble collisions are imprinted in the cosmic microwave background (CMB) radiation. The signature of an individual collision is an azimuthally symmetric modulation of the CMB temperature in a localized region. In general, the theory predicts a set of observable collisions, with the properties of each drawn from an in-principle calculable probability distribution. A search for the signatures of bubble collisions in the Wilkinson Microwave Anisotropy Probe~\cite{Bennett:2003ba} (WMAP) 7-year CMB data~\cite{Komatsu:2010fb} was performed in Refs.~\cite{Feeney:2010dd,Feeney:2010jj}, which yielded an upper bound on the average number of observable collision signatures on the CMB sky. 

A key element in the analysis of Refs. \cite{Feeney:2010dd,Feeney:2010jj} is the fact that each bubble collision is expected to leave a fairly generic set of signatures in the CMB~\cite{Feeney:2010dd,Chang:2008gj,Czech:2010rg,Kleban:2011yc}, described by a phenomenological template with only a few free parameters. However, there is currently no way to connect the parameters in this phenomenological model to the parameters in the fundamental theory. In order to go from the Lagrangian to the signatures of bubble collisions, one must determine the properties of the colliding bubbles, the immediate outcome of the collision, and the cosmological evolution of the perturbed bubble interior. Because of the non-linear nature of the field equations, a full and unambiguous determination of the cosmological signatures of bubble collisions requires numerical relativity. See  e.g. Ref.~\cite{2001CQGra..18R..25L} for a modern review of this field. 

In this paper, we present the results of fully-relativistic simulations of the collision between Coleman-de Luccia (CDL) bubbles containing a realistic inflationary cosmology. Although any given bubble undergoes many collisions, the future domains of influence of each of the collision events typically do not overlap until relatively late times inside the bubbles. It is therefore a good approximation to treat each collision as occurring in isolation, and we restrict ourselves to the study of collisions between two bubbles. 

The spacetime resulting from the collision between two CDL vacuum bubbles has SO(2,1) symmetry~\cite{Hawking:1982ga,Wu:1984eda}, allowing us to treat the problem of bubble collisions in full generality using a 1+1D simulation. This symmetry is expected to be mildly broken by fluctuations on the walls  of the colliding bubbles~\cite{Adams:1989su,Garriga:1991tb,Garriga:1991ts,Aguirre:2005sv}. However, expanding bubbles become more and more spherical (the perturbations remain much smaller than the overall size of the bubbles), and so our truncation to 1+1 dimensions should capture the most important dynamics. Einstein's equations in 1+1D are fully constrained by the scalar field configuration, allowing for a straightforward and efficient numerical implementation. 

The goal of the present paper is to study the outcome of bubble collisions and identify the properties that models must have in order to be consistent with our observed cosmology. This has been addressed to a certain extent in previous work using the Israel Junction Condition formalism~\cite{Hawking:1982ga,Wu:1984eda,Freivogel:2007fx,Aguirre:2007wm,Chang:2007eq,Johnson:2010bn}, as well as in numerical simulations that neglected gravitational effects~\cite{Aguirre:2008wy,Easther:2009ft,Giblin:2010bd}. In addition, Ref.~\cite{Blanco-Pillado:2003hq} studied collisions in full numerical relativity for the collision of vacuum bubbles (see also Ref.~\cite{Carone:1989nj} for a relativistic treatment of single bubbles). Our fully relativistic treatment allows us to go beyond these previous studies and completely determine the phenomenology of bubble collisions given a particular model of the scalar field theory driving eternal inflation. In addition, we determine how well the approximations used in previous work hold up when compared against the fully relativistic solutions. In a forthcoming paper, we will extend this numerical framework to extract the observational signatures of bubble collisions.  

The remainder of the paper is organized as follows. We begin in Sec.~\ref{sec:summary} by reviewing conclusions obtained from the Israel Junction Condition formalism and summarizing the set of questions we attempt to address with our simulations. In Sec.~\ref{sec:equationsofmotion}, we perform the truncation to the 1+1D system of equations to be solved numerically, and discuss the generation of initial data. We then briefly outline the numerical implementation and tests of the code in Sec.~\ref{sec:numerical_implementation}, and present the results of our simulations in Sec.~\ref{sec:results}. More detailed treatments of a few technical aspects of the setup are presented in a set of appendices. We work in units where $M_{\rm Pl} = 1$ unless otherwise noted; Newton's constant is given by $G_N = M_\mathrm{Pl}^{-2}$.

\section{General properties of bubble collisions}\label{sec:summary}

In this section, we outline the general setup for bubble collisions using the Israel Junction Condition formalism~\cite{Israel:1966rt}. This also serves to lay out a set of expectations for the numerics and illustrate the approximations made in previous work. After this introduction to bubble collisions, we summarize the main questions which are addressed by our numerical studies. 

\subsection{The Israel Junction Condition formalism}\label{sec:junctionconditions}

\begin{figure*}
   \includegraphics[width=10 cm]{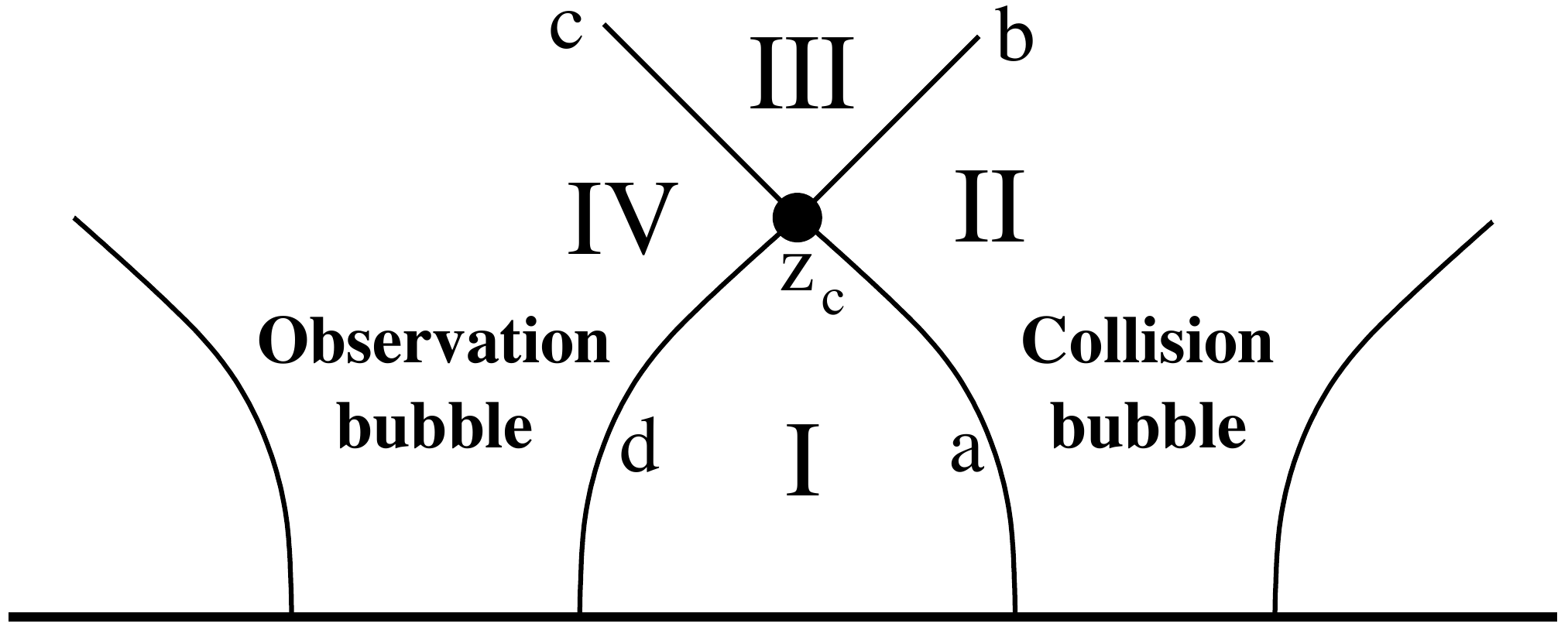}
\caption{The generic bubble collision spacetime in the thin-wall approximation is split into four regions. Region I is the false vacuum, Region II is the interior of the Collision bubble, Region III is the future domain of influence of the collision (which might be composed of further sub-regions), and Region IV is the interior of the Observation bubble. In the Israel junction condition formalism, the Observation and Collision bubbles are separated from the false vacuum by infinitesimally thin walls (a and d) characterized only by their tensions. In addition, Region III is enclosed (depending on the context) by either domain walls or shells of scalar radiation (b and c), each characterized by their tension and energy density. 
\label{fig-coll_diagram}}
\end{figure*}

The generic bubble collision spacetime is shown in Fig.~\ref{fig-coll_diagram}. We work in the ``Collision frame," the frame in which the two colliding bubbles nucleate simultaneously in some timeslicing~\cite{Aguirre:2007wm}. It is always possible to perform a Lorentz boost to go to the collision frame in a spacetime with two bubbles. As shown in the figure, the two bubbles form (Regions II and IV in the figure) from the false vacuum (Region I in the figure), expand, and then collide. The bubble interiors are perturbed in the future of the collision (Region III in the figure); the boundary of the affected region can carry energy density, and the interior of this region might be comprised of multiple subregions. We refer to the bubble on the left, which we assume to contain an inflationary cosmology consistent with our observable universe, as the ``Observation bubble." The ``Collision bubble," which we always place on the right, can have more general properties, and serves to perturb the Observation bubble interior. 

In general, there are a number of possibilities for the region to the future of the collision (Region III), and the interfaces separating this region from each bubble interior:
\begin{itemize}
\item $H_{\rm III} = H_{\rm IV}$: Region III belongs to the Observation bubble; the $c$ interface is a shell of radiation required to conserve energy and momentum; the $b$ interface is a domain wall separating the interior of the Observation and Collision bubbles if they are different vacua, or a shell of radiation if they are identical vacua.
\item $H_{\rm III} = H_{\rm II}$: Region III belongs to the Collision bubble; $c$ is a domain wall if the Observation and Collision bubbles contain different vacua (it is a shell of radiation if they are identical); $b$ is a shell of radiation.
\item $H_{\rm III} \neq H_{\rm II}, H_{\rm IV}$: Region III is in a distinct vacuum, different from that of the Observation or Collision bubbles; $b$ and $c$ are domain walls separating Region III from the Collision and Observation bubble interiors. This is known as a classical transition~\cite{Easther:2009ft,Giblin:2010bd}. 
\end{itemize}

The collision spacetime in the Israel junction condition formalism is constructed by matching vacuum solutions of Einstein's equations across interfaces characterized by some tension $\sigma_i$ (where $i=a,b,c,d$ denotes the walls separating the various regions in Fig.~\ref{fig-coll_diagram}). Each region corresponds to a different vacuum of the underlying potential, and the tensions are determined by the potential barriers between vacua. As discussed in the introduction, the collision spacetime has SO$(2,1)$ (hyperbolic) symmetry. Using a hyperbolic version of Birkhoff's theorem~\cite{Wu:1984eda,Freivogel:2007fx,Aguirre:2007wm,Chang:2007eq}, we can find the most general vacuum solutions to Einstein's equations possessing this symmetry:
\begin{equation}\label{eq:thin-wallmetric}
ds^2 = - A_\gamma (z)^{-1} dz^2 + A_\gamma (z) dx^2 +z^2 dH_2^2~,  
\end{equation}
where $dH_2^2 = d\chi^2 + \sinh^2 \chi d\phi^2$ is the metric on a unit hyperboloid, 
and the coordinates generally range from $0 < z < \infty$, 
$-\infty < x < \infty$, $0 < \chi < \infty$, and $0 < \phi < 2 \pi$. 
The metric functions $A_\gamma (z)$ are 
\begin{equation}\label{eq:vacuummetricfunction}
A_\gamma (z) = 1 - \frac{2 M_\gamma}{z} + H_\gamma^2 z^2~,
\end{equation}
where we have included the subscript $\gamma = {\rm I,II,III,IV}$ to label the different regions in the collision spacetime. There are two constants in each metric function, the mass parameter $M_\gamma$ (akin to the Schwarzschild mass parameter) and the Hubble parameter $H_\gamma$ (related to the cosmological constant by $\Lambda = 3H^2$). When $M_\gamma=0$, the metric describes a hyperbolic foliation of de Sitter space (hyperbolic de Sitter, which we denote by HdS). In our conventions, $H_\gamma^2 > 0$ $(< 0)$ for positive (negative) vacuum energy. In the following, we only consider the case where $H_\gamma^2 \geq 0$.

Returning to Fig.~\ref{fig-coll_diagram}, we can identify this sketch as a slice through the full bubble collision spacetime in the $x-z$ plane defined by $\chi =0$ and $\phi = {\rm const}$. All the plots shown in later sections are along a similar slice through the collision spacetime. Integrating Einstein's equations across each interface and requiring continuity of the metric yields an equation of motion for each interface. The continuity of the metric across each interface requires $z$ to be continuous. However, the $x$ coordinate is not continuous, and we therefore have to take into account different coordinates, $x_L$ and $x_R$, on either side of each interface. Defining $k_{i} \equiv 4 \pi \sigma_i$, the equations of motion in terms of the proper time measured by observers along the interface are given by:
\begin{eqnarray}
\dot{z}^2 &=&
\frac{1}{4}\bigg[z^2k_{i}^2+2(A_L+A_R)+
\frac{(A_L-A_R)^2}{z^2k_{i}^2}\bigg]~,  \nonumber \\
\dot{x}_L &=& \frac{\beta_L}{A_L}~, \label{eq-z} \\
\dot{x}_R &=& \frac{\beta_R}{A_R}~; \nonumber
\end{eqnarray}
where $A_L$ and $A_R$ are the metric functions on the left (L) and right (R) side of a particular interface in Fig.~\ref{fig-coll_diagram} and 
\begin{eqnarray}\label{eq:betas}
\beta_L   &=& \frac{A_R-A_L+z^2k_{i}^2}{2zk_{i}}~, \label{eq-L} \\
\beta_R   &=& \frac{A_R-A_L-z^2k_{i}^2}{2zk_{i}}~. \label{eq-R} 
\end{eqnarray}
The sign of $\dot{x}_{L,R}$ determines the direction in which the wall is moving from the perspective of observers on either side. From Eq.~\ref{eq:vacuummetricfunction} for the metric functions, there are two possible behaviors for the wall at large $z$: normal and repulsive. Normal walls require $|H_R^2 - H_L^2| > k_i^2$, in which case $\dot{x}$ has the same sign on both sides of the wall, and the wall accelerates towards the side with the larger Hubble parameter. Repulsive walls occur for $|H_R^2 - H_L^2| < k_i^2$. In this case, $\dot{x}$ takes opposite signs on either side of the interface, implying that it is accelerating away from both sides. This behavior arises due to the gravitationally repulsive nature of domain walls~\cite{Vilenkin:1984hy,Ipser:1983db}. The simplest application of these equations of motion is to calculate the trajectory of the wall (of the normal type) of a single bubble from the perspective of the false vacuum. This is given by
\begin{equation}\label{eq:thinwalltraj}
z = H_F^{-1} \left[ \left( 1 - H_F^2 R_0^2 \right) \sec^2 \left( H_F \left[ x + \Delta x \right]  \right)  \right]^{1/2} \, ,
\end{equation}
where $H_F$ is the Hubble parameter in the false vacuum from which the bubble is formed, and $R_0$ is the critical radius of the bubble in the thin wall approximation. More generally, Eq.~\ref{eq-z}  determines the motion of a post-collision domain wall, should one be produced by the collision.

The kinematics of a collision are fully determined by the location $z_c$ of the collision. This is equivalent to specifying the distance between the bubble centers in the Collision frame. The surface tensions of the interfaces and vacuum energies are determined by the underlying potential, but the mass parameters in Eq.~\ref{eq:thin-wallmetric} for each region remain undetermined. Imposing the condition that there are no conical singularities at $z_c$ (equivalent to imposing energy and momentum conservation~\cite{Langlois:2001uq}) yields one equation relating these mass parameters to the vacuum energies and tensions. In the limit where the incoming and outgoing interfaces are highly relativistic, this relation takes a particularly simple form~\cite{Moss:1994pi}, which with four distinct post-collision regions as shown in Fig.~\ref{fig-coll_diagram} is given by
\begin{equation}\label{eq:kinematics}
A_{\rm I} (z_c) A_{\rm III} (z_c) = A_{\rm II} (z_c) A_{\rm IV} (z_c).
\end{equation}
For vacuum bubbles, the mass parameters in regions ${\rm I,II,}$ and ${\rm IV}$ are zero. Assuming that region ${\rm III}$ is not subdivided, there is only one mass parameter, which is determined by Eq.~\ref{eq:kinematics}:
\begin{equation}\label{eq:massparameterprediction}
M_{\rm III} = \frac{z_c^3}{2} \frac{ H_{\rm III}^2 - H_{\rm IV}^2 + H_{\rm I}^2 \left( 1 + H_{\rm III}^2 z_c^2 \right) - H_{\rm II}^2 \left( 1 + H_{\rm IV}^2 z_c^2  \right)  }{1+H_{\rm I}^2 z_c^2} \, .
\end{equation}
Although the mass parameter increases without bound as the collision occurs at increasing $z_c$, the collision between two bubbles can never produce a black hole~\cite{Hawking:1982ga,Moss:1994pi,Freivogel:2007fx}; this would be indicated if there was a solution with $A_{\rm III}=0$ which, from Eq.~\ref{eq:kinematics}, is not possible. 

The thin-wall solutions provide an excellent way of characterizing the possible outcomes of the collision between two vacuum bubbles. However, these solutions do not take into account field dynamics, and therefore are not sufficient to compute the effects of a bubble collision on the cosmological evolution inside a bubble. Before tackling this question using numerical methods, let us first summarize some of the specific questions we address.

\subsection{Questions for the relativistic simulations}

The set of questions we are primarily  interested in answering using our numerics are as follows:
\begin{itemize}
\item How valid is the thin-wall approximation, and does the Israel junction condition formalism miss any dynamics that are important for determining the outcome of bubble collisions? How important are the internal dynamics of the walls themselves?
\item In the thin-wall approximation, the interior of the colliding bubbles is treated as having constant energy density. Since one would ultimately like to embed a cosmology into the bubble interiors (to make contact with our own observable universe), this is clearly a key  limitation. Previous work used the thin-wall solutions as a background in which to evolve a scalar field~\cite{Chang:2008gj} but to date, there has been no full solution of cosmological evolution inside colliding bubbles. Taking into account full General Relativity, how does the collision affect cosmology? 
\item Non-gravitational simulations~\cite{Aguirre:2008wy} predict that the outcome of a bubble collision might depend sensitively on the underlying potential. In particular, Ref.~\cite{Aguirre:2008wy} found that models with inflaton potentials of the ``small-field'' type (those driving inflation near a maximum or inflection point in the potential) were far more sensitive to the effects of bubble collisions than models with inflaton potentials of the ``large-field" type (such as inflation in a quadratic potential). If true, this would have important implications for the connection between the possible presence of bubble collisions and other cosmological observables, in particular the tensor-to-scalar ratio~\cite{Lyth:1996im}. Can inflation of the small-field type occur to the future of a bubble collision? What properties must a potential have in order to produce collisions that are consistent with our observed cosmology, yet produce observable signatures? 
\item Depending on the structure of the potential underlying eternal inflation, a collision between two bubbles can push the field into the basin of attraction of a new vacuum, causing a ``classical transition" to a new phase~\cite{Easther:2009ft}. In the absence of gravity, classical transitions can only give rise to lasting regions with a lower energy density than the colliding bubble interiors. However, in the presence of gravity, the thin-wall analysis of Ref.~\cite{Johnson:2010bn} showed that it is kinematically possible to create lasting regions of higher energy density in the collision between two vacuum bubbles. Such transitions can even prevent a portion of the interior of a bubble undergoing a big crunch from encountering a singularity. Do these result hold true once the detailed field dynamics are included? 
\end{itemize}

\section{Equations of motion}\label{sec:equationsofmotion}

We now consider non-vacuum solutions of Einstein's equations, and derive the equations of motion used in our implementation within the Cauchy approach (see e.g. Ref.~\cite{2001CQGra..18R..25L}). Adopting coordinates $x^a=(z,x^i)=(z,x,\chi,\phi)$, the most general line element with hyperbolic symmetry can be written as:
\begin{equation}\label{eq:generalmetric}
ds^2 = (-\alpha^2 + a^2 \beta^2) dz^2 + 2 a^2 \beta dz dx + a^2 dx^2 + z^2 b^2 \left[ d\chi^2 + \sinh^2 \chi d\phi^2 \right] \, ,
\end{equation}
with $\{a^2 ,b^2\}$ metric variables describing the intrinsic curvature 
$\gamma_{ij} = {\rm diag}[a^2,z^2 b^2,z^2 b^2 \sinh^2 \chi]$ of (spacelike) hypersurfaces defined
by $z=$ const and $\{ \alpha,\beta \}$ coordinate conditions. It is convenient to introduce the extrinsic curvature $-2 \alpha K_{ij}=(\partial_t - {\cal L}_{\beta})\gamma_{ij}$ to express the Einstein equations within the ADM formalism~\cite{1993PhRvL..70....9C}. The equation governing the behavior of the scalar field $\varphi$ is
\begin{equation}
\square \varphi =\frac{1}{\sqrt{-g}} \partial_{\mu} \left( \sqrt{-g} g^{\mu \nu} \partial_{\nu} \phi \right) = \partial_{\varphi} V \, .
\end{equation}
We reduce this equation to first order form by introducing the variables 
\begin{eqnarray}
\Phi &=& \varphi' \, , \\
\Pi &=& \frac{a}{\alpha} \left( \dot{\varphi} - \beta \varphi' \right) \, .
\end{eqnarray}
The general equations of motion for hyperbolic symmetry and arbitrary $\{ \alpha,\beta \}$ are derived 
in Appendix~\ref{sec:eomappendix}. For our current purposes we exploit the coordinate freedom to
set
\begin{equation}
b = 1, \ \ \beta = 0 ,
\end{equation}
motivated by the vacuum solutions to Einstein's equations with metric Eq.~\ref{eq:thin-wallmetric}.
With this choice the Einstein equations (of motion and constraints) result in:
\begin{eqnarray}
\dot{a} &=& - \alpha a {K^{x}}_x \label{basicadot}\, , \\
\dot{b} & = & -\alpha {K^\chi}_\chi - \frac{1}{z} = 0 \label{basicbdot} \, , \\
\dot{{K^\chi}_\chi} &=& \alpha \left[ -\frac{1}{z^2} + K K^\chi_\chi  - 8 \pi V \right] \label{basickchidot} \, , \\
\dot{{K^x}_x} &=& - \frac{1}{a} \left( \frac{\alpha'}{a} \right)' + \alpha \left( K {K^x}_x - 8 \pi \frac{\Phi^2}{a^2} - 8 \pi V(\varphi)  \right) \, , \label{basickxdot} \\
16 \pi \left[ \frac{\Phi^2 + \Pi^2}{2 a^2 } + V(\varphi) \right] &=&  -\frac{2}{z^2} + 4 {{K^x}_x} {K^\chi}_\chi + 2 {{K^\chi}_\chi}^2 \label{basicham} \, , \\
\partial_x {K^\chi}_\chi &=& 4 \pi \frac{\Pi \Phi}{a} \, . \label{basicmom} 
\end{eqnarray}
The remaining freedom can be tied to the choice of the lapse function. Again,
a particularly simple option is to define
\begin{equation}
\alpha = - \frac{1}{{K^\chi}_\chi z} 
\end{equation}
which, when substituted into the momentum constraint equation (Eq.~\ref{basicmom}), gives 
\begin{equation}\label{eq:spatialconstraint}
\alpha' = \frac{4 \pi z \alpha^2}{a M_{\rm Pl}^2} \Pi \Phi \, ,
\end{equation}
where we have restored factors of $M_{\rm Pl}$.

Next, we can solve for ${K^x}_x$ in the Hamiltonian constraint (Eq.~\ref{basicham}),
\begin{equation}\label{eq:kxx}
{K^x}_x = \frac{1}{2z\alpha} -\frac{\alpha}{2z} - 4 \pi z \alpha \left(  \frac{\Phi^2 + \Pi^2}{2 a^2 } + V(\varphi) \right) \, ,
\end{equation}
and replace it in the time-evolution equation for $a$ to obtain (restoring factors of $M_{\rm Pl}$)
\begin{equation}
\dot{a} = - \frac{a}{2z} \left[ 1 - \alpha^2 \left( 1 + \frac{8 \pi}{M_{\rm Pl}^2} z^2 \left[ \frac{\Phi^2 + \Pi^2}{2 a^2 } + V(\varphi) \right] \right) \right] \, .
\end{equation}
Now, an evolution equation for $\alpha$ can be defined by inserting ${K^x}_x$ and  ${K^\chi}_\chi$ into the time-evolution equation for ${K^\chi}_\chi$. Restoring factors of $M_{\rm Pl}$, this is given by
\begin{equation}
\dot{\alpha} = \frac{1}{2 z} \left[ \alpha - \left( 1+ \frac{8 \pi}{M_{\rm Pl}^2} z^2 V \right) \alpha^3  \right]  + \frac{2 \pi z \alpha^3}{M_{\rm Pl}^2 a^2} \left[\Phi^2 + \Pi^2 \right] \, .
\end{equation}
In addition, we have the equations of motion for $\Phi$,
\begin{equation}
\dot{\Phi} = \partial_x \left( \frac{\alpha}{a} \Pi \right) \,  , 
\end{equation}
 and $\Pi$, 
\begin{equation}
\dot{\Pi} = - \frac{2}{z} \Pi + \left( \frac{\alpha'}{a} - \frac{a' \alpha}{a^2} \right) \Phi + \frac{\alpha}{a} \Phi' - \alpha a \partial_\varphi V \, .\\
\end{equation}

For the simulation, it is convenient to work with a set of units where time, distance, and energy are measured in terms of the false-vacuum Hubble constant $H_F$, defined by 
\begin{equation}
H_F^2  = \frac{8 \pi}{3 M_\mathrm{Pl}^2} V(\varphi_F) \, . 
\end{equation}
We now define the following dimensionless variables:
\begin{equation}
\tilde{z} = H_F z, \ \ \tilde{x} = H_F x, \ \ \tilde{\varphi} = \frac{\varphi}{M_\mathrm{Pl}}, \ \ \tilde{\Phi} = \frac{\Phi}{M_\mathrm{Pl} H_F}, \ \ \tilde{\Pi} = \frac{\Pi}{M_\mathrm{Pl} H_F}, \ \ \tilde{V} = \frac{V}{M_\mathrm{Pl}^2 H_F^2} \, .
\end{equation}
The evolution equations become:
\begin{equation}\label{eq:adot}
\frac{da}{d\tilde{z}} = - \frac{a}{2 \tilde{z}} \left[ 1 - \alpha^2 \left( 1+ 8 \pi \tilde{z}^2 \left[ \frac{\tilde{\Pi}^2 + \tilde{\Phi}^2 }{2 a^2} + \tilde{V} (\tilde{\varphi}) \right]   \right)  \right] \, , 
\end{equation}
\begin{equation}\label{eq:alphadot}
\frac{d\alpha}{d\tilde{z}} = \frac{\alpha}{2 \tilde{z}} \left[ 1 - \left( 1+ 8 \pi \tilde{z}^2 \tilde{V}(\tilde{\varphi}) \right) \alpha^2  \right] + \frac{2 \pi \tilde{z} \alpha^3}{a^2} \left[ \tilde{\Phi}^2 + \tilde{\Pi}^2 \right] \, , 
\end{equation}
\begin{equation} \label{eq:pi}
\frac{d\tilde{\Pi}}{d \tilde{z}} = - \frac{2}{\tilde{z}} \tilde{\Pi} + \left( \frac{1}{a} \frac{d\alpha}{d\tilde{z}} - \frac{\alpha}{a^2} \frac{da}{d\tilde{z}}   \right) \tilde{\Phi} + \frac{\alpha}{a} \frac{d \tilde{\Phi}}{d \tilde{z}} - \alpha a \frac{d \tilde{V}}{d \tilde{\varphi}} \, , 
\end{equation}
\begin{equation}\label{eq:phi}
\frac{d \tilde{\Phi}}{d \tilde{z}} = \frac{d}{d \tilde{x}} \left( \frac{\alpha}{a} \tilde{\Pi}  \right) \, .
\end{equation}
In the following, we suppress the tildes for notational clarity. 

The system of equations~(\ref{eq:adot}--\ref{eq:pi}) fully determine the future evolution of the field and metric variables given consistent initial data (e.g. satisfying the constraint equation Eq.~\ref{eq:spatialconstraint}). The simplest test of the validity of these equations is to reproduce the vacuum solutions (Eq.~\ref{eq:thin-wallmetric}) in the limit where all derivatives of the field are zero and $V = {\rm const}$. We demonstrate this in Appendix~\ref{sec:HdSsection}. It is also illustrative to consider the evolution of fluctuations in the field in the false vacuum background. This is outlined in 
Appendix~\ref{sec:fieldeqnsHdS}, where we show that the characteristic speed of fluctuations is given by:
\begin{equation}\label{eq:characteristics}
c = \pm (1+H_F^2 z^2)^{-1} \, ,
\end{equation}
which ranges from unity at small $z$ to zero as $z$ becomes large compared to $H_F$. 
 
\section{Determining the Initial data}
Our strategy for defining the initial data for bubble collisions proceeds in two steps. First, given a potential 
for the scalar field, we construct the CDL instanton for each possible transition from the false vacuum. A slice through the CDL instanton fully determines the initial data for a single-bubble spacetime. When the bubbles are small  compared with the horizon size defined by the false vacuum, and have tensions small compared with the energy scale 
of the false vacuum, they introduce small perturbations to the false vacuum background. In this limit, we 
can construct the two-bubble spacetimes by linear superposition. Next, we need to express the field and metric 
configuration defined by the CDL instanton in terms of the variables used in the simulation. To do so, we adopt an
 initial hypersurface at $z=0$, where the $x$-coordinate in the hyperbolic slicing can be identified with a
 proper distance. It is then straightforward to map the field configuration from the CDL instanton onto the
 $z=0$ hypersurface, and perturbatively determine the field and metric functions at $z=dz$. 
This provides consistent initial data to evolve the solution forward from $z=dz$: we confirm the validity of our initial data  
by evaluating the Hamiltonian and momentum constraints and checking for the convergence of the full solutions. In what follows, we briefly describe the procedure adopted and defer to Appendix~\ref{sec:numericalCDL} for further details.

\subsection{Solving for the CDL instanton} \label{instanton_solve}

The CDL instanton~\cite{Coleman:1977py,Coleman:1980aw} is a compact solution to the Euclidean equations of motion with Euclidean metric:
\begin{equation}\label{eq:CDLmetric}
ds^2 = dt^2 + \rho(t)^2 d\Omega_3^2 \, .
\end{equation}
The metric function $\rho(t)$ has two zeros at $t=0$ and $t=t_\mathrm{max}$, and evolves according to the Euclidean Friedmann equation
\begin{equation}
\frac{d^2 \rho}{dt^2} = - \frac{8 \pi}{3 M_{\rm Pl}^2} \rho \ \left( \left( \frac{d \varphi}{dt} \right)^2 + V(\varphi) \right) \, . \label{eq:instanton}
\end{equation}
As described in Appendix~\ref{sec:numericalCDL} (and further in Refs.~\cite{Aguirre:2006ap,Banks:2005ru}), the importance of gravity in the computation of the instanton is characterized by the parameter
\begin{equation}
\epsilon^2 \equiv \frac{8 \pi M^2}{3 M_{P}^{2}}
\end{equation}
where $M$ is a scale parameterizing the width of the potential barrier. In the numerical solutions that follow, a typical value is $\epsilon = 0.01$. With the metric Eq.~\ref{eq:CDLmetric}, the field equation is:
\begin{equation}\label{eq:eucfield}
\frac{d^2 \varphi}{dt^2} + \frac{3}{\rho} \frac{d \rho }{dt} \frac{d \varphi}{dt} = \frac{dV}{d \varphi}. 
\end{equation}
The CDL instanton is a non-singular solution to the equations of motion that interpolates between the basins of attraction of the true and false vacua. Non-singular solutions satisfy the boundary conditions: 
\begin{equation}\label{eq:cdl_initialc}
\varphi(t=0) \simeq \varphi_T, \ \ \frac{d\varphi}{dt}(t=0) = 0, \ \ \rho(t=0) = 0, \ \ \varphi(t=t_\mathrm{max}) \simeq \varphi_F, \ \ \frac{d\varphi}{dt}(t=t_\mathrm{max}) = 0, \ \ \rho(t=t_\mathrm{max}) = 0.
\end{equation}
A review of CDL instantons and their construction can be found in e.g. Ref.~\cite{Banks:2002nm}. 

\begin{figure*}
   \includegraphics[width=8 cm]{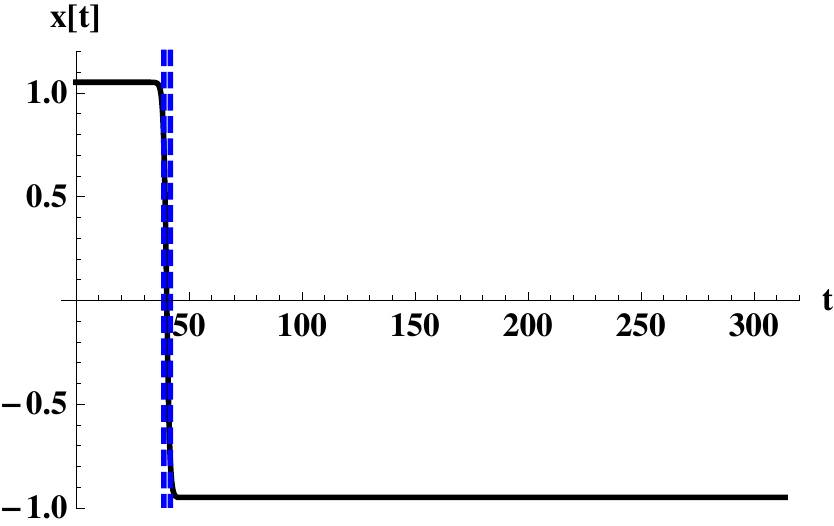} \hfill
   \includegraphics[width=8 cm]{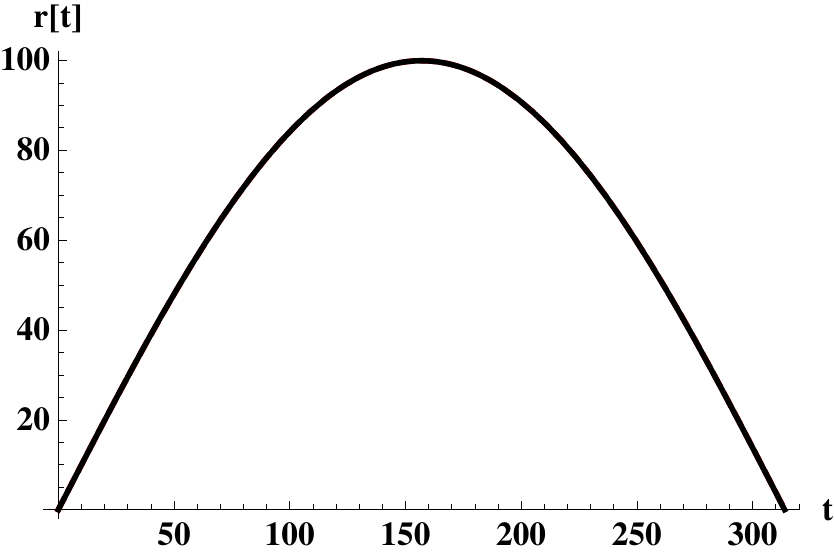}
\caption{The numerically generated instantons for a potential of the form Eq.~\ref{eq:V} with $a=0.1$ for $\epsilon = 0.0075$. The dimensionless variable $x$ is related to the field $\varphi$, $r$ to the metric function $\rho$, and $s$ to $t$ as defined in Eq.~\ref{eq:CDLvariables}. The prediction for the critical radius (based on Eq.~\ref{eq:thinwallcriticalradius} in the thin-wall approximation) is shown as the blue dashed line.}
\label{fig-inst_examples}
\end{figure*}

In general, it is not possible to solve analytically for the CDL instanton (although see Ref.~\cite{Dong:2011gx} for some special cases where an analytic solution can be found). We therefore adopt a numerical scheme to solve the double-boundary condition problem defined by Eq.~\ref{eq:cdl_initialc}. We describe our numerical methods in Appendix~\ref{sec:numericalCDL}. An example output is shown in Fig.~\ref{fig-inst_examples}.

\subsection{Obtaining initial data in the hyperbolic slicing}
We are now in a position to determine the initial data for the field and metric functions in the hyperbolic slicing used in the simulation. We first discuss how initial data for a single bubble is constructed, and then add a colliding bubble to the picture. The CDL instanton can be analytically continued to obtain the post-tunnelling field configuration $\varphi (t)$. The metric is given by:
\begin{equation}
ds^2 = dt^2 + \rho(t)^2 \left[ -d \Psi^2 + \cosh^2 \Psi d\Omega_2^2   \right] \, .
\end{equation}
In the absence of a bubble, the slice at $\Psi=0$ corresponds to the throat of de Sitter. When a bubble is present, for small $\epsilon$, this slice is a small perturbation to the throat of de Sitter space (see Ref.~\cite{Aguirre:2009ug} for more detailed discussion of the coordinates relevant to the CDL bubble spacetime). Here, $t$ measures a proper distance. In the hyperbolic slicing, $z=0$ can also be identified with a portion of the perturbed throat of de Sitter. In the limit where $z \rightarrow 0$, the metric is given by
\begin{equation}
ds^2 = -dz^2 + dx^2 + z^2 dH_2^2 \, .
\end{equation}
Therefore, since both $x$ and $t$ measure a proper distance on the throat of de Sitter, 
we can simply substitute $x$ for $t$ in the instanton solution. This is the field configuration 
on the $z=0$ slice. Further, since the curvature of the transverse components of the slices dominates
 the energy density at $z=0$, the metric functions do not concern us here. Examining the equations
 of motion, there are potentially various singularities at $z=0$. We must therefore find the initial conditions
 after one time step $dz$. To do so, we first consider the necessary asymptotics for the various functions.
 The field $\varphi$ is expanded as 
\begin{equation}\label{eq:phii}
\varphi = \varphi_0 (x) + \varphi_2 (x) z^2 \, , 
\end{equation}
where the first order term is excluded since we require $d\varphi / dz = 0$ for a non-singular solution. We also have
\begin{equation}\label{eq:aali}
a = 1+ a_2(x) z^2, \ \ \ \alpha = 1 - \alpha_2(x) z^2 \, .
\end{equation}
Using these asymptotic relations, and working to lowest order in $z$, we have:
\begin{equation}\label{eq:piphii}
\Phi = \frac{d \varphi_0}{dx} + \frac{d \varphi_2}{dx} z^2, \ \ \ \Pi = 2 \varphi_2 z \, .
\end{equation}

Now, substituting into the equations of motion (Eqs.~\ref{eq:adot}--\ref{eq:phi}), we can solve for the unknowns $\varphi_2(x)$, $a_2(x)$, $\alpha_2 (x)$, in terms of $\varphi_0(x)$. From the constraint equation for $\alpha$,
\begin{equation}
\frac{d\alpha_2}{dx} z^2 = - 8 \pi \frac{d \varphi_0}{dx} \varphi_2 z^2 \, ,
\end{equation}
while from the evolution equation for $\alpha$,
\begin{equation}
-2 z \alpha_2 = \left[ \alpha_2 - 4 \pi V + 2 \pi \left(\frac{d \varphi_0}{dx}\right)^2 \right] z, \ \ \ \rightarrow \ \ \alpha_2 = \frac{4 \pi}{3} V(\varphi_0) - \frac{2 \pi}{3} \left(\frac{d \varphi_0}{dx}\right)^2 \, .
 \end{equation}
 This matches with the expectation for a constant potential $V$ in the small-z limit:
 \begin{equation}
\alpha_{\rm HdS} = 1 - \frac{4 \pi V}{3} z^2 + \ldots  \, .
\end{equation} 
From the evolution equation for $a$, we have
\begin{equation}
2 a_2 z = \left[ 4 \pi V - \alpha_2 + 2 \pi \left( \frac{d \varphi_0}{dx}\right)^2 \right]  z, \ \ \ \rightarrow \ \ a_2 = \frac{4 \pi}{3} V(\varphi_0) + \frac{4 \pi}{3} \left( \frac{d \varphi_0}{dx} \right)^2 \, .
\end{equation}
We also have the field equations. From the evolution equation for $\Pi$, we find:
\begin{equation}
2 \varphi_2 = \frac{d^2 \varphi_0}{dx^2} - 4 \varphi_2 - dV, \ \ \ \ \rightarrow \ \ \varphi_2 = \frac{1}{6} \left( \frac{d^2 \varphi_0}{dx^2} - \partial_{\varphi} V (\varphi_0) \right) \, .
\end{equation}
The evolution equation for $\Phi$ yields an identity. Note that substituting the definition of $\varphi_2$ into the constraint equation for $\alpha$ (Eq.~\ref{eq:spatialconstraint}), we find consistency. Substituting the expressions for $\varphi_2(x)$, $a_2(x)$, $\alpha_2 (x)$, and $\varphi_0(x)$ into Eqs.~\ref{eq:aali} and~\ref{eq:piphii}, we set initial data at $z = dz \ll 1$.

The above procedure is perfectly rigorous for initial data involving a single bubble. In order to discuss configurations with two bubbles, we must make a few approximations. Assuming that the bubbles represent minor perturbations to the false vacuum background, we determine the initial data for two different bubbles independently, add the two field configurations, and calculate the quantities in Eqs.~\ref{eq:piphii} and~\ref{eq:aali} as above. To the extent that each of the bubbles evolves before the collision just as it would in isolation, this is a good approximation. In practice, this amounts to choosing potentials where one instanton endpoint is very close to the false vacuum, the initial radii of the bubbles are small compared to $H_F^{-1}$, and ensuring that the bubbles are sufficiently far apart on the $z=0$ hypersurface for the field to be very close to the false vacuum between the bubbles. For bubbles satisfying the thin-wall approximation described in Appendix~\ref{sec:numericalCDL}, these conditions are almost always satisfied. Our choice of $\epsilon \ll 1$ ensures that we are well within this regime.

\subsection{Potentials}

\begin{figure*}
   \includegraphics[width=10 cm]{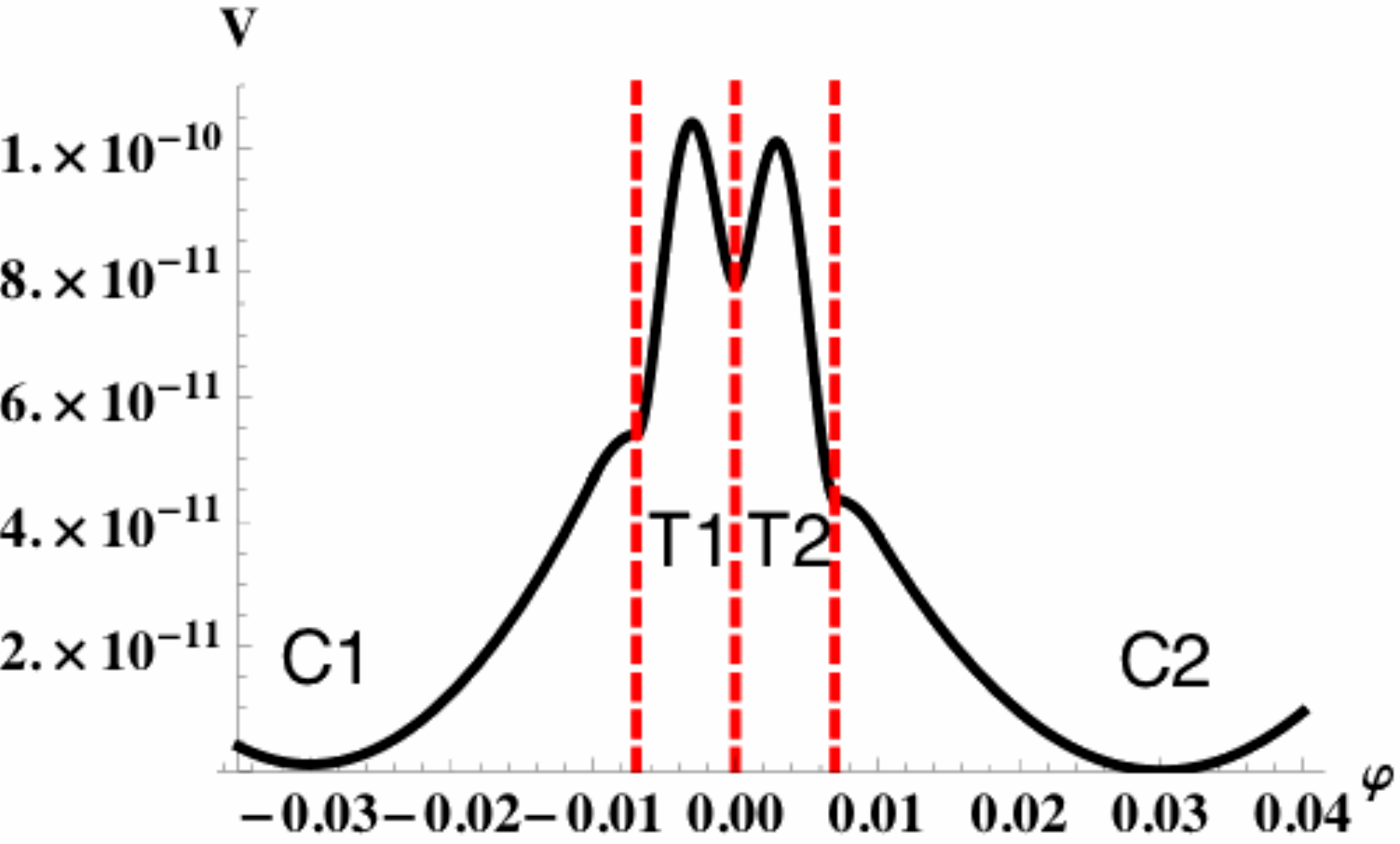}
\caption{The generic form of the potentials used in our simulations consists of four components. $T1$ and $T2$ describe the potential barriers between the three vacua of the potential. $C1$ and $C2$ describe regions of the potential leading to the vacua and can drive a period of inflation.}
\label{fig-potentials}
\end{figure*}

For our simulations, we must supply a set of scalar potentials. We define a class of piecewise potentials constructed in a manner similar to the procedure outlined in Ref.~\cite{Aguirre:2008wy}. There are 4 possible segments for each potential, as shown in Fig.~\ref{fig-potentials}. These segments correspond to two potential barriers (labeled $T_1$ and $T_2$ in the figure) and two slopes leading away from each potential barrier towards a set of minima (labeled $C_1$ and $C_2$ in the figure). 

By convention, $T2$ and $C2$ correspond to the portions of the potential relevant for the Observation bubble. For studies of the collision between two vacuum bubbles, we take $C_1$ and $C_2$ simply to be minima of the potential. For bubbles with a cosmology, the $C2$ portion of the potential is chosen so that an undisturbed Observation bubble contains an epoch of inflation which lasts at least $60$ $e$-folds. We consider two broad classes of inflationary potentials $C2$: those for which inflation occurs over a field range larger than or comparable to $M_{\rm Pl}$ (large-field inflation) and those for which inflation occurs over a field range smaller than $M_{\rm Pl}$ (small-field inflation). 

The functional form for the various segments is given by 
\begin{equation}\label{eq:pot1}
V =
\left\{
	\begin{array}{ll}
		V_{C1}  & \mbox{if } \varphi \leq \varphi_{T1} \\
		V_{T1}  & \mbox{if } \varphi_{T1} < \varphi \leq 0 \\
		V_{T2}  & \mbox{if } 0 < \varphi \leq\varphi_{T2} \\
		V_{C1} & \mbox{if }  \varphi_{T2} < \varphi
	\end{array}
\right.
\end{equation}
where, when there is a cosmology inside the Collision bubble, we have
\begin{equation}
V_{C1}  = 
\left\{
	\begin{array}{ll}
	      	\frac{m_{11}^2}{2} (\varphi - \varphi_{m1})^2 + V_{01}  & \mbox{if } \varphi \leq \varphi_{j1} \\
               - \frac{ m_{12}^2 }{2} ( \varphi - \varphi_{T1} )^2 + V_{T1} & \mbox{if } \varphi_{j1} < \varphi < \varphi_{T1}	\, .
        \end{array}
\right.
\end{equation}
In all cases, we have
\begin{equation}
V_{T_1} = - \frac{M^2}{2} (\varphi - \varphi_{1})^2 + \frac{a_1 M}{3} (\varphi - \varphi_{1})^3 + \frac{(\varphi - \varphi_{1})^4}{4} + M^4 C_1\, , 
\end{equation}
\begin{equation}
V_{T_2} = - \frac{M^2}{2} (\varphi - \varphi_{2})^2 - \frac{a_2 M}{3} (\varphi - \varphi_{2})^3 + \frac{(\varphi - \varphi_{2})^4}{4} + M^4 C_2\, .
\end{equation}
For large-field inflation, we assume a form of the potential given by: 
\begin{equation}\label{eq:largefieldc2}
V_{C2}  = 
\left\{
	\begin{array}{ll}
	      - \frac{ m_{21}^2 }{2} ( \varphi - \varphi_{T2} )^2 + V_{T2} & \mbox{if } \varphi_{T2} < \varphi < \varphi_{j2} \\
               \frac{m_{22}^2}{2} (\varphi - \varphi_{m2})^2 & \mbox{if } \varphi > \varphi_{j2}	
        \end{array}
\right.
\end{equation}
while for small-field inflation, we use:
\begin{equation}\label{eq:smallfieldc2}
V_{C2}  = V_{T2} - \frac{\nu M}{3} \left(\varphi - \varphi_{T2} \right)^3 + \frac{\lambda}{4} \left( \varphi - \varphi_{T2} \right)^4 \, .
\end{equation}

\begin{figure}
   \includegraphics[width=18cm]{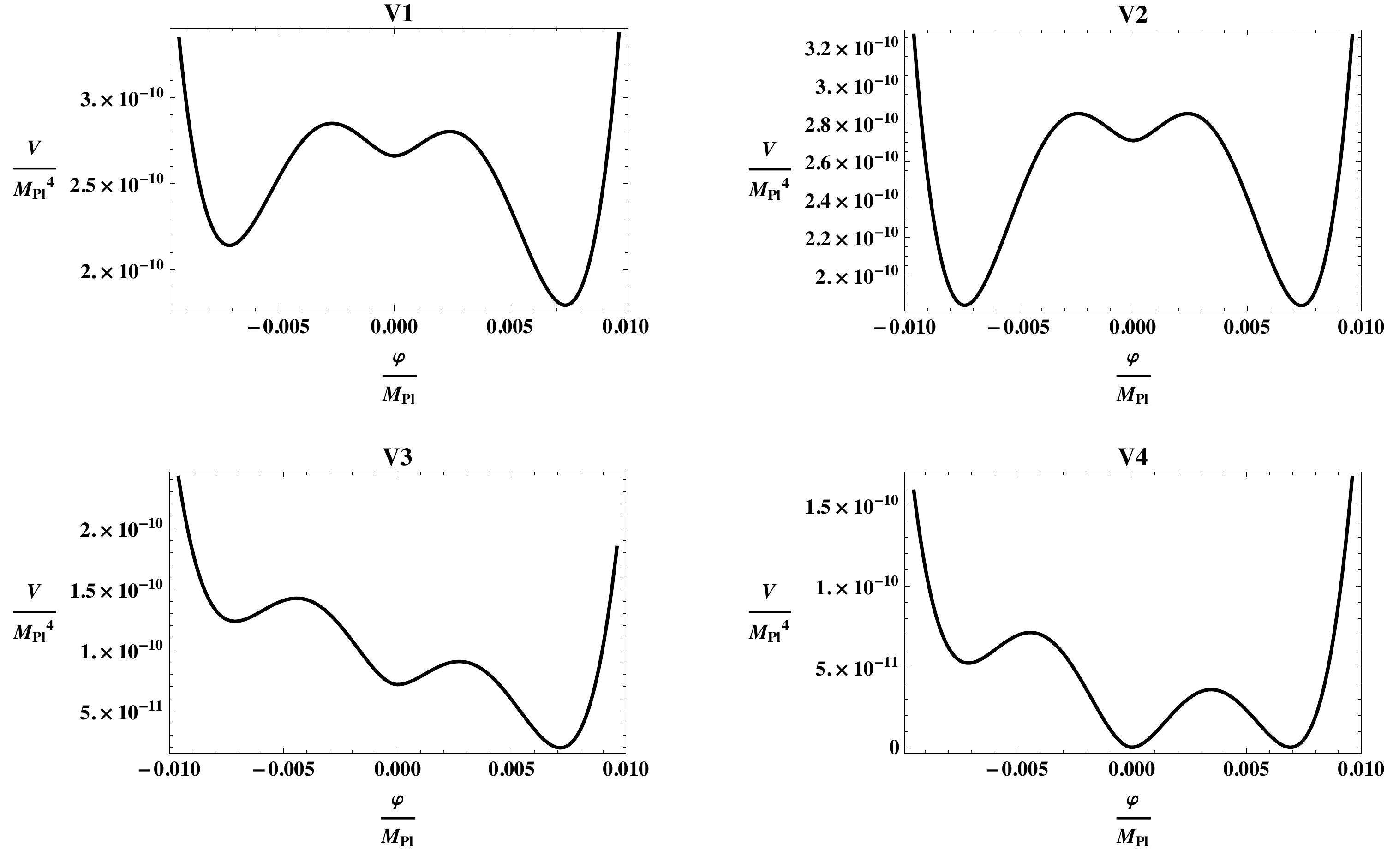}
 \caption{Potentials giving rise to vacuum bubbles. For the potentials on the top row, the false vacuum is located at $\varphi=0$, and there are two possible CDL solutions, connecting the false vacuum to positive or negative $\varphi$. In both cases, the lowest energy vacuum is the one at positive $\varphi$. The potentials on the bottom row have a false vacuum at negative $\varphi$. There is only one CDL solution, connecting the false vacuum to the vacuum at $\varphi=0$. The lowest energy vacuum for the potential V3 is at positive $\varphi$ while the lowest energy vacuum for the potential V4 is at $\varphi = 0$.}
\label{fig:vacpots}
\end{figure}

\begin{figure}
   \includegraphics[width=17cm]{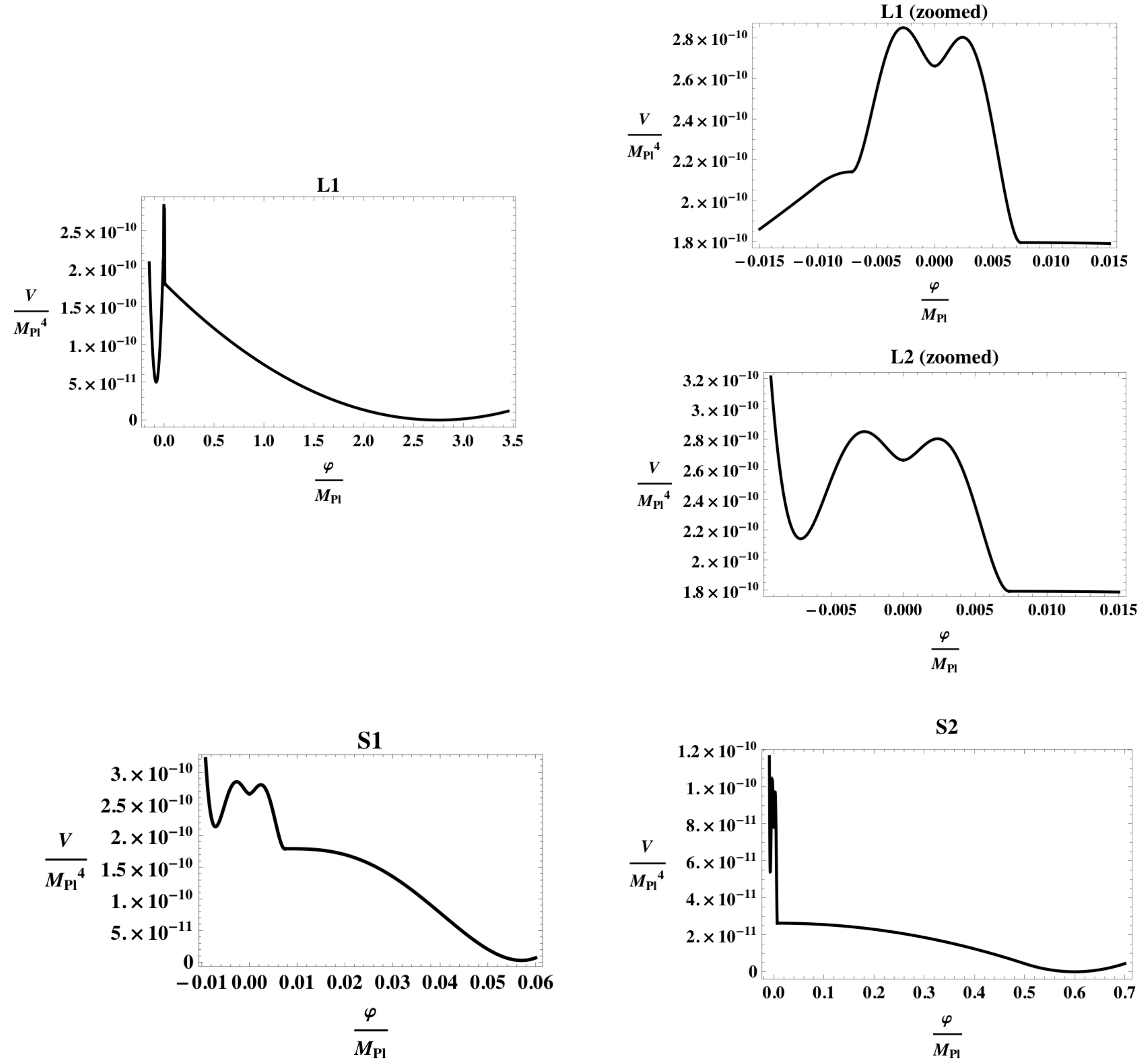}
 \caption{Potentials giving rise to bubbles with an interior cosmology. The potentials on the top row are of the large-field type, both have the false vacuum at $\varphi=0$, and both allow for two possible CDL solutions connecting the false vacuum to regions of positive or negative $\varphi$. The potential L1 has cosmological evolution inside both bubbles, with a much steeper slope at negative $\varphi$ than positive $\varphi$. We show both the full potential, and a plot zoomed in around the position of the false vacuum. The potential L2 omits the cosmological evolution at negative $\varphi$, but has the same potential at positive $\varphi$ as L1. We show only the portion of the potential L2 in the vicinity of the false vacuum. The potentials on the bottom row are of the small-field type, and again, have the false vacuum at $\varphi = 0$, with two possible CDL solutions. The potential S1 is defined using Eq.~\ref{eq:smallfieldc2} while S2 is defined using Eq.~\ref{eq:largefieldc2}; inflation occurs in the vicinity of an inflection point for S1 while it occurs in the vicinity of a hilltop for S2. }
\label{fig:SRpots}
\end{figure}

Here, the parameters we must specify are $a_1$, $a_2$, $M$, $V_{01}$,  $\varphi_{m1}$, $\varphi_{m2}$, $\varphi_{j1}$, $\varphi_{j2}$, $C_1$ (and for the small-field model, $\nu$ and $\lambda$). This corresponds to specifying the false vacuum energy density, barrier widths, and the length of the inflationary regions of the potential on either side. All other parameters are fixed by requiring continuity of the potential and its derivatives. We have simulated a wide variety of potentials, but consider only a representative sample here. The parameters defining each of these examples are recorded in Appendix~\ref{sec:potentialpars}. The potentials giving rise to vacuum bubbles are shown in Fig.~\ref{fig:vacpots}; the potentials giving rise to bubbles with an interior cosmology are shown in Fig.~\ref{fig:SRpots}. These particular choices are motivated in the following sections. 

\section{Numerical Implementation}\label{sec:numerical_implementation}

\subsection{Testing for convergence}
In this section, we discuss the details of our numerical solution of the equations of motion (Eqs.~\ref{eq:adot}--\ref{eq:phi}). Since the bubbles we consider are small perturbations to the false vacuum, we can use the behavior of null geodesics in
 HdS to determine the spatial size of the simulation. This is discussed in Appendix~\ref{sec:box_size},
where we show that a box of size $x\in[-\pi/2,x_c + \pi/2]$, with $x_c$ denoting the center of the Collision bubble,
 contains the two colliding bubbles in their entirety. For larger bubble separations, we also 
 consider a box extending from the origin of the Observation bubble at $x=0$ to the origin of the
 Collision bubble at $x_c$.

To integrate the equations of motion forward in time, we use the method of lines on a fixed uniform grid with 
a 4th order Runge-Kutta scheme (see, e.g. Ref.~\cite{Calabrese:2003vx}). To discretize the equations in space, we use Finite Difference approximations for the spatial derivatives (satisfying summation by parts) which are 4th order accurate at all interior points, with the exception of two points at the boundaries where the convergence is 2nd order. The resulting implementation, as discussed in Refs.~\cite{Calabrese:2003vx,Lehner:2004cf}, ensures a consistent treatment of the problem and guarantees stability in the linearized case. The obtained solutions should converge to 3rd order accuracy in the discretization length. We confirm this through several tests, which we now describe. 

\begin{figure}
   \includegraphics[width=12 cm]{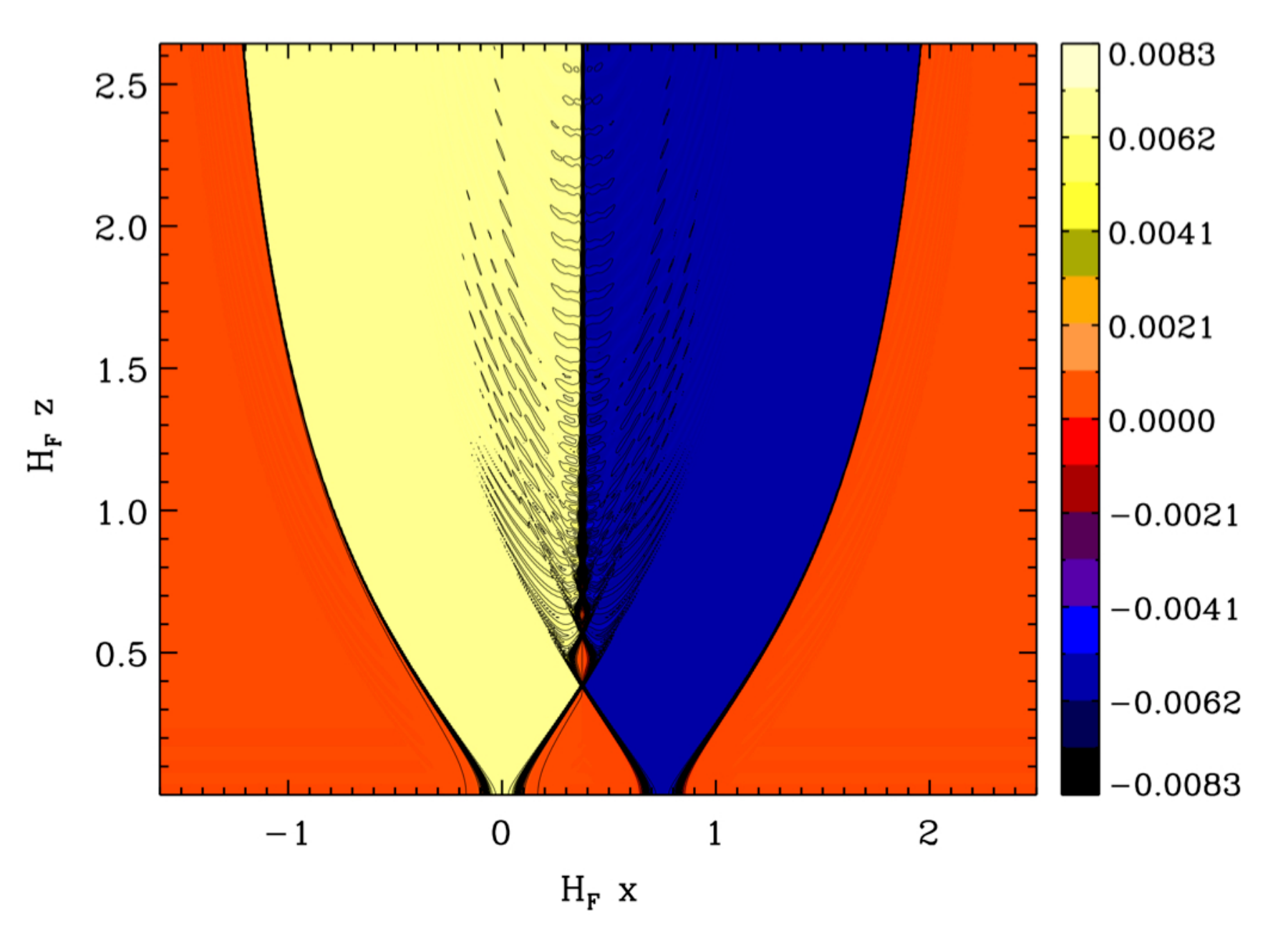} 
\caption{A contour plot of the field $\varphi (x,z)$ for the collision of the two different types of bubbles allowed by the potential V2 shown in Fig.~\ref{fig:vacpots}. The color bar indicates the value of the field at each point in spacetime. In this case, the post-collision domain wall is repulsive, and accelerates away from the interior of both bubbles. }
\label{fig-coll_stationary}
\end{figure}

\begin{figure}
   \includegraphics[width=8cm]{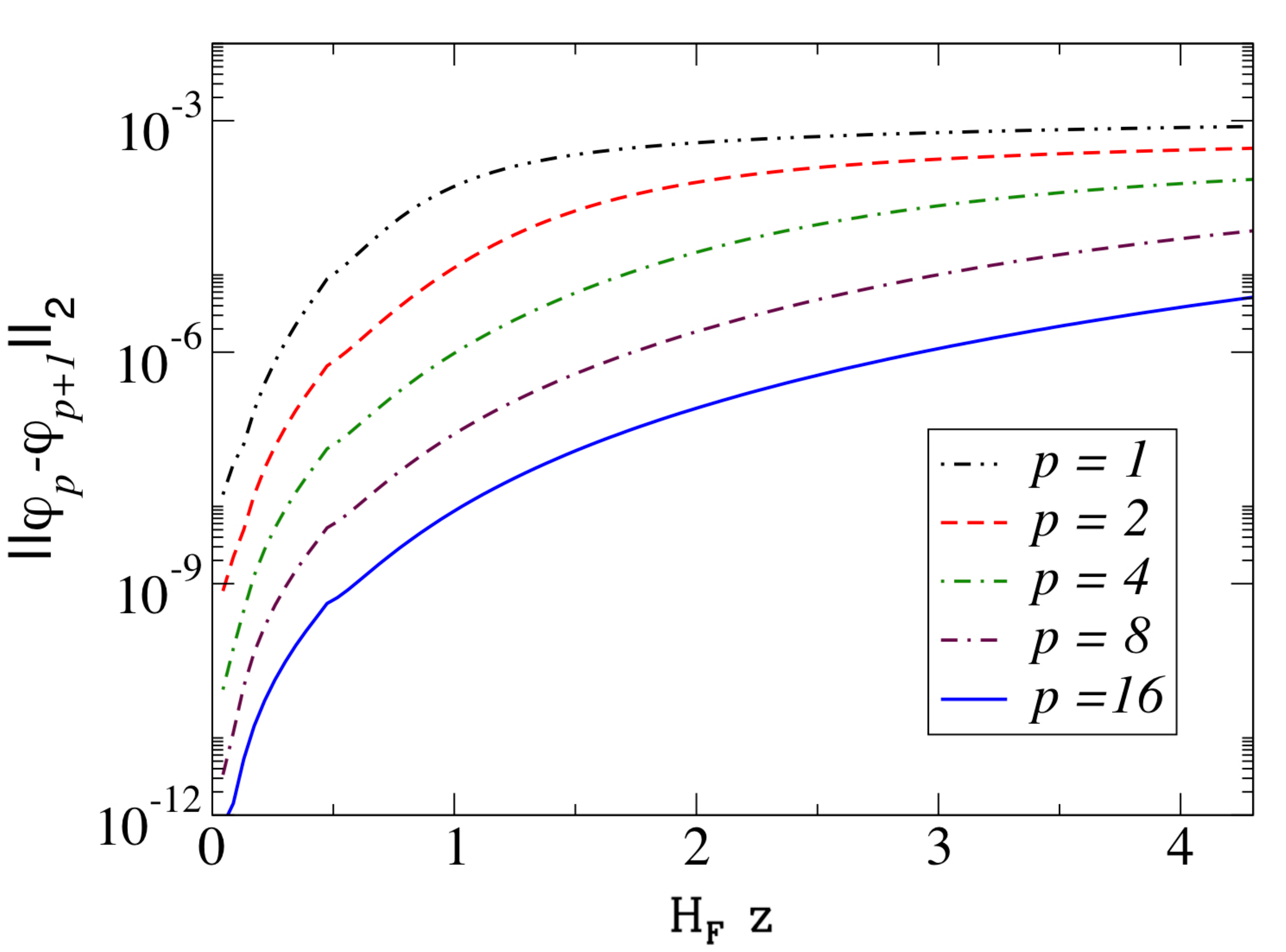} \hfil
      \includegraphics[width=8cm]{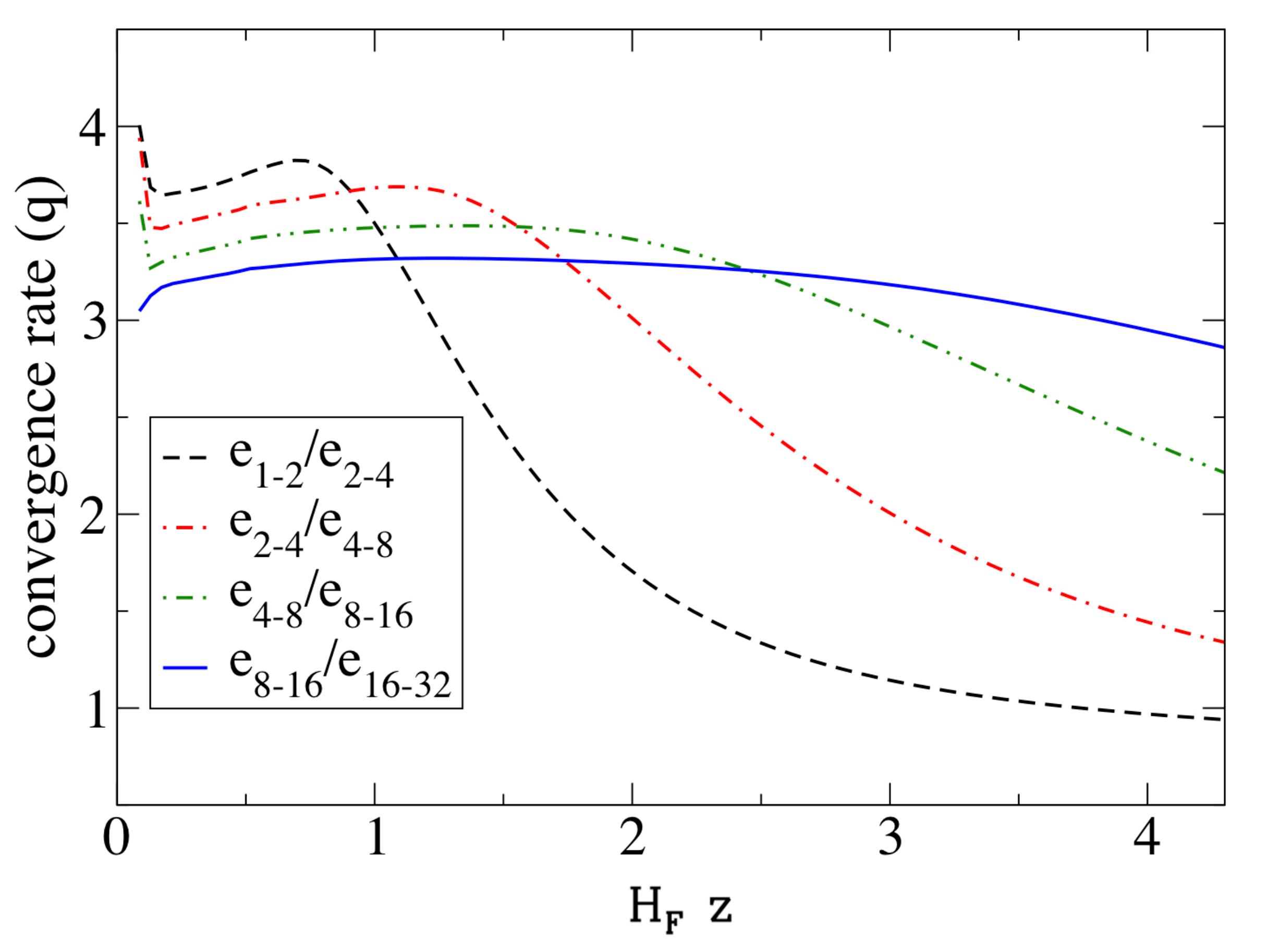}
\caption{Demonstrations of the convergence of our code for the simulation shown in Fig.~\ref{fig-coll_stationary}. On the left, we show the absolute value of the difference between solutions
with discretizations obtained for $p$ and $p+1$, illustrating how the difference decreases with increasing resolution as expected for a convergent implementation. On the right, we plot the estimated convergence rate as calculated from Eq.~\ref{convvalue}: as the resolution increases, the value obtained approaches $3$, which is consistent with the expected value for our implementation.}
\label{fig:errors_test}
\end{figure}

We choose the potential V2 shown in Fig.~\ref{fig:vacpots}, and set up initial conditions consisting of two different bubbles with the Collision bubble located at $H_F x_c = 0.75$. A contour plot of the field $\varphi$ is shown in Fig.~\ref{fig-coll_stationary}. The details of the solution are not immediately relevant, but this is a fairly representative example of a collision spacetime. We monitor the solution up to $z=4.3$, which includes the collision and the post-collision domain wall. We evolve this scenario on a uniform grid with number of points $N_p = 2^p \times 10^3$ ($p=0\,..\,5$) and compare the obtained solutions. In particular we monitor the ($l_2$ norm of the) difference of solutions with successive values of $p$ (i.e. for a given field, we define $e_{p \mbox{-} p+1} \equiv ||\varphi_p - \varphi_{p+1}||_2$). These are shown in Fig. \ref{fig:errors_test} (left panel) highlighting how, as resolution is increased, the difference decreases as expected for a convergent solution. We can further test the convergence rate $q$ as estimated by
\begin{equation}
2^q \equiv \frac{||\varphi_p - \varphi_{p+1}||_2}{|| \varphi_{p+1} - \varphi_{p+2}||_2}=\frac{e_{p \mbox{-} p+1}}{e_{p+1 \mbox{-} p+2}} \label{convvalue}\, .
\end{equation}
The results are shown in Fig. \ref{fig:errors_test} for the field $\varphi$, indicating that the solution converges; further, as the resolution is increased, the overall convergence rate tends to the expected value of $3$~\cite{Lehner:2004cf}. In addition, we have confirmed that the constraint Eq.~\ref{eq:spatialconstraint} is well preserved throughout the simulations. The resolutions we use here are representative of what is computationally feasible with our current implementation (the run-time for the highest resolution shown here was approximately 48 hours on a single core).

Unless otherwise noted, all of the simulations in the following sections are run with a total of $N_x = 32 \times 10^3$ spatial grid points. Having demonstrated a convergent code, we now perform a number of cross-checks of the expectation from the thin-wall single bubble-solutions outlined in Sec.~\ref{sec:junctionconditions}. 

\subsection{Simulation of a single bubble}

In this section, we study the evolution of single bubbles. Our purpose is threefold: we check the numerical solutions against the predictions of the Israel junction condition formalism outlined in Sec.~\ref{sec:junctionconditions}; discuss the general features of the coordinates used in our simulation; and illustrate the limitations of our numerical implementation.

\begin{figure}
    \includegraphics[width=5.5 cm]{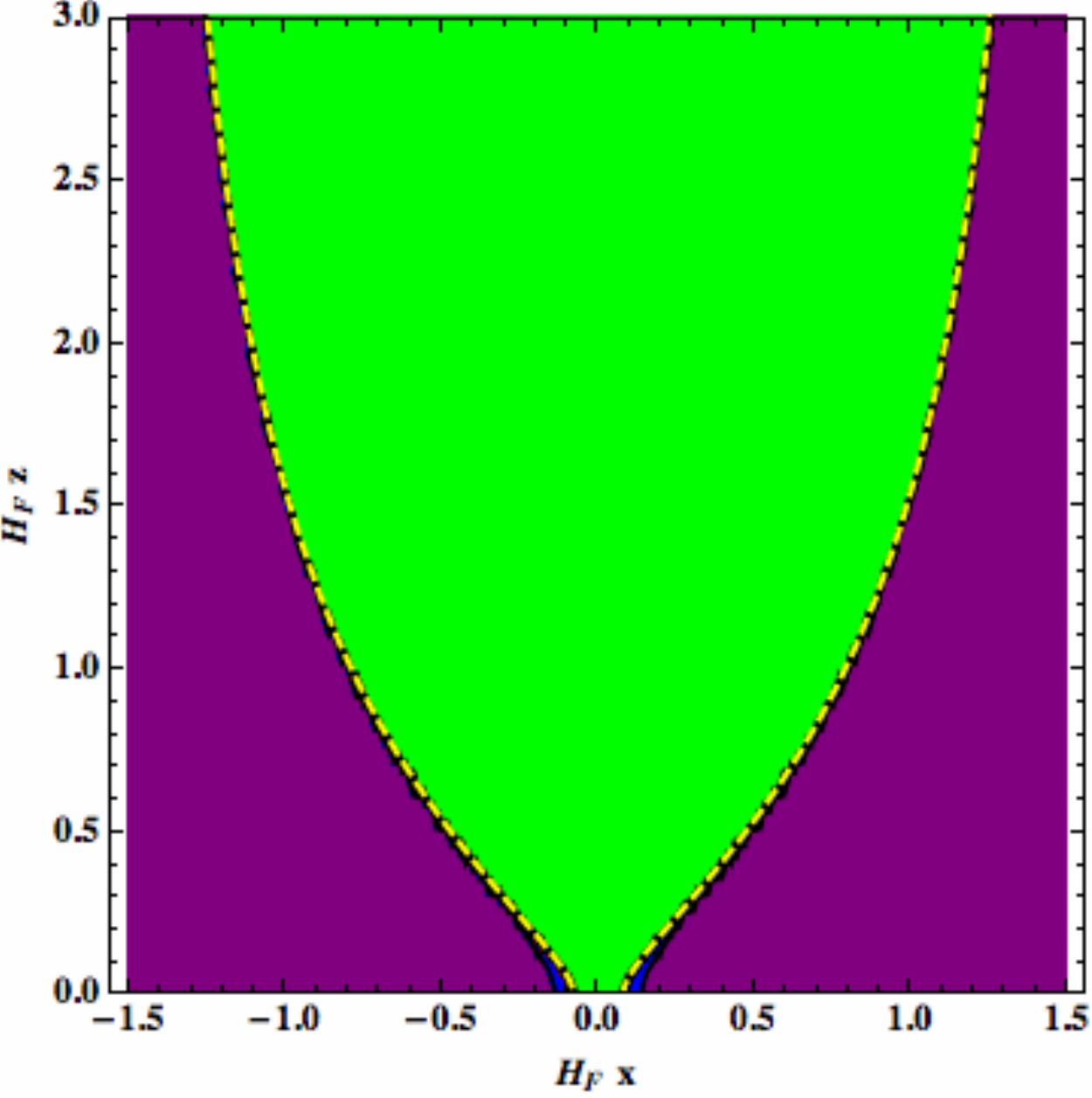}
    \includegraphics[width=5.5 cm]{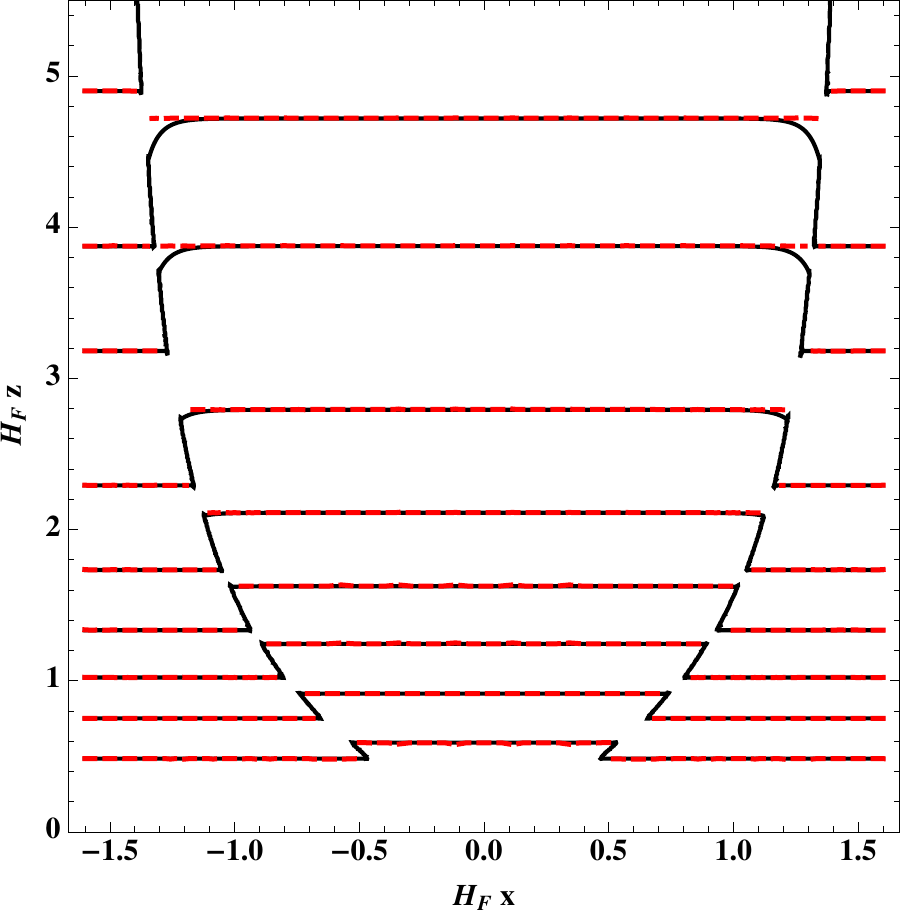}
    \includegraphics[width=5.5 cm]{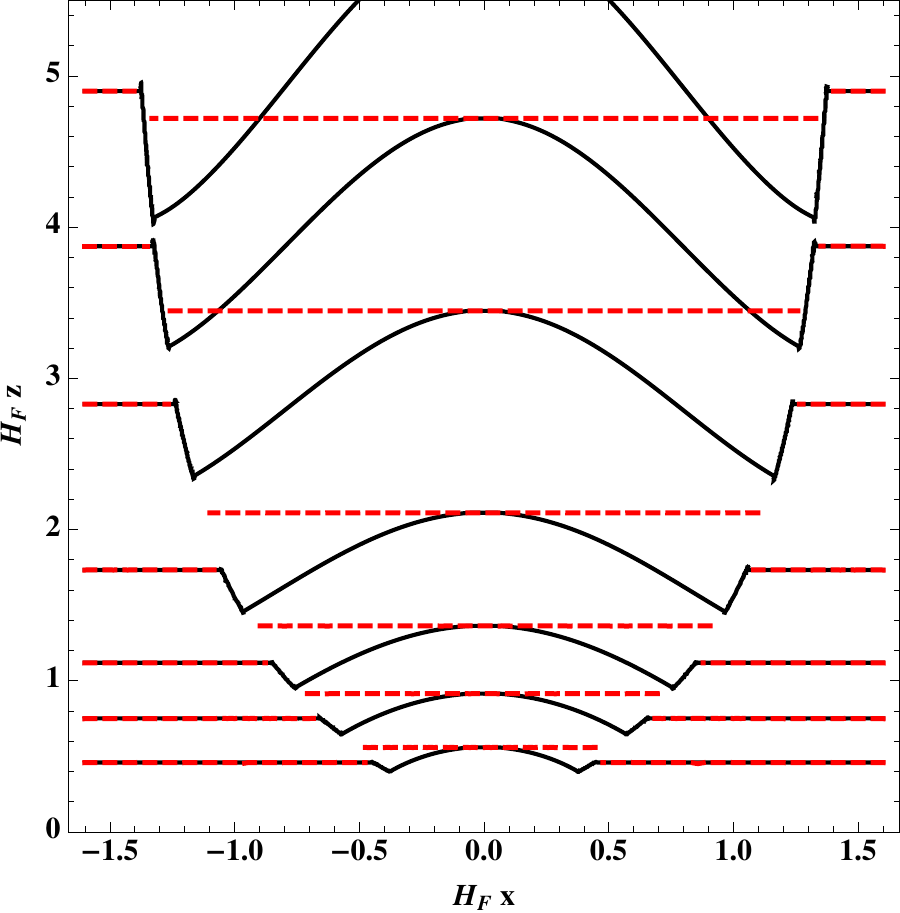}
 \caption{The numerical simulation of a single Observation bubble formed from the potential V1. On the left, we show a contour plot of $\varphi(x,z)$. As this is a vacuum bubble, the field is constant on either side of the bubble wall. The yellow dashed line is the wall trajectory predicted by the thin-wall approximation, illustrating the excellent agreement between the numerical and junction condition solutions. In the middle, we show a contour plot of the numerical solution for $\alpha(x,z)$ (black solid lines) as well as the junction condition solution (red-dashed lines). The agreement is excellent, except in the vicinity of the bubble walls at large $z$. On the right, we show a contour plot of the numerical solution for $a(x,z)$ as well as the junction condition solution (red-dashed lines). The disagreement between these two solutions is due to the fact that in the junction condition formalism, the $x$-coordinate is discontinuous across the wall, while in the simulation it is treated as a continuous coordinate. The solutions agree in the bubble center and the bubble exterior, which is expected. In the numerical solution, $a$ must continuously interpolate between the solution in the bubble center and the bubble exterior.}
 \label{fig-vacuumbubble_tw_comp}
 \end{figure}

The expectation from the Israel junction condition formalism is that the walls of vacuum bubbles follow the trajectory given by Eq.~\ref{eq:thinwalltraj}. The left panel of Fig.~\ref{fig-vacuumbubble_tw_comp} shows a contour plot of the field for a single Observation bubble generated from the potential V1. The field is in the vacuum at $\varphi = 0$ everywhere outside the bubble (dark, purple area), and in the vacuum at positive $\varphi$ everywhere inside the bubble (light, green area). Even though the bubble wall is expanding at an increasing velocity as time goes on, it reaches a constant comoving size. Since $x$ is a comoving coordinate, the bubble walls asymptote to a constant position as $z \rightarrow \infty$ (as can be seen from Eq.~\ref{eq:thinwalltraj}). This can also be understood from the fact that the characteristic speed of fluctuations given in Eq.~\ref{eq:characteristics} goes to zero as $z \rightarrow \infty$. In Fig.~\ref{fig-vacuumbubble_tw_comp}, the thin-wall solution is over-plotted as the dashed yellow line, showing that the agreement between the numerical and thin-wall solutions at this resolution is excellent. 
 
We can also determine how the metric functions $a$ and $\alpha$ compare with the expectation from the junction condition solutions. Identifying $\alpha = A_{\gamma}^{-1/2}$ from Eq.~\ref{eq:thin-wallmetric}, the thin-wall solution has $\alpha = (1+ H_{\rm I}^2 z^2)^{1/2}$ outside the bubble and $\alpha =  (1+ H_{\rm II}^2 z^2)^{1/2}$ inside it (using the conventions of Fig.~\ref{fig-coll_diagram}). In the middle panel of Fig.~\ref{fig-vacuumbubble_tw_comp} we show a contour plot of $\alpha$ (solid black lines) along with the junction condition solution (red-dashed lines). The agreement is good, except at large $z$ in the vicinity of the wall. We return to this below. In the case of the metric function $a$, the comparison is not as straightforward. In the junction condition solution, the $x$ coordinate is discontinuous across the bubble wall, while for the numerics we have (appropriately!) chosen a coordinate system in which $x$ is continuous. A contour plot of $a$ is shown in the right panel of Fig.~\ref{fig-vacuumbubble_tw_comp} (solid black lines), and compared to the expectation from the junction condition solution (red-dashed lines), where $a =A_{\gamma}^{1/2}$. Outside the bubble, and at its center, the thin-wall and numerical solutions match, as they should. Between the bubble center and the wall, the behavior of $a$ differs from the junction condition solution in order to enforce the continuity of $x$ across the wall. 
 
\begin{figure}
 \begin{center}
   \includegraphics[height=7 cm]{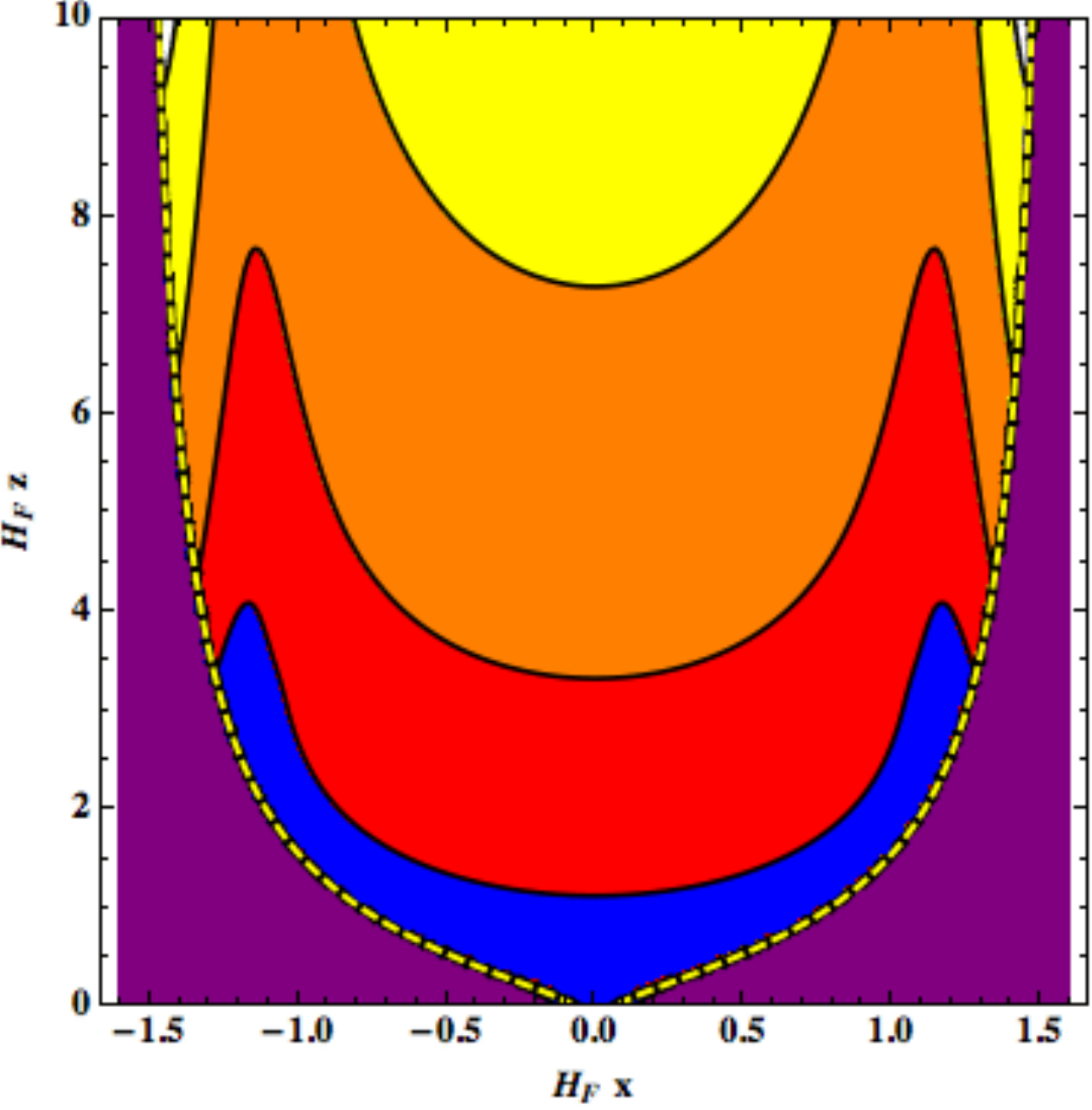} \hfill
   \includegraphics[height=5 cm]{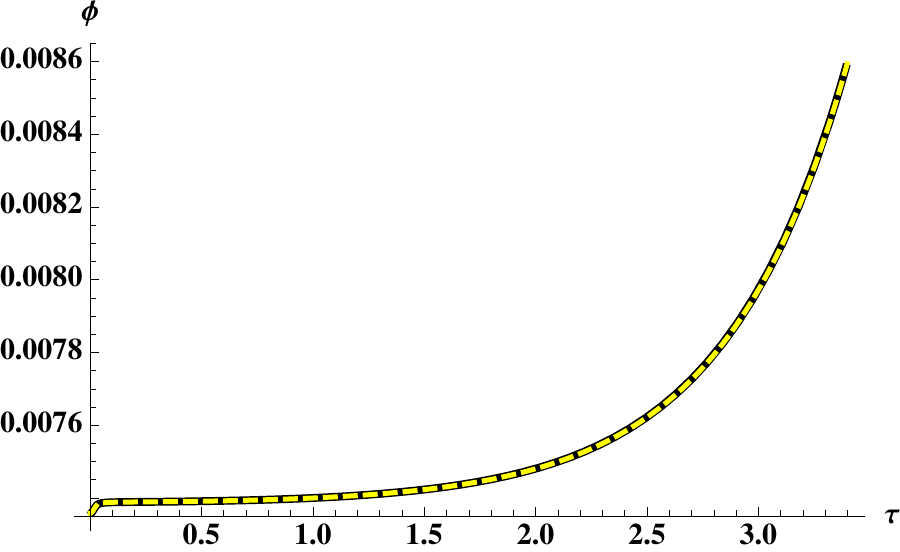}
\caption{Numerical simulation of a single Observation bubble formed from the potential L1. On the left, we show a contour plot of $\varphi(x,z)$. The trajectory for the wall in the junction-condition solution (yellow-dashed line) can be found under the assumption that the bubble interior has a constant vacuum energy; the agreement with the numerical solution is excellent. The field is in the false vacuum outside the bubble, and evolves on the bubble interior down the potential towards increasingly positive values of $\varphi$. The numerical solution has the correct qualitative behavior except in the vicinity of the bubble walls. As the resolution is increased, we expect that the contours in the bubble interior to go up to $z \rightarrow \infty$. In the right panel, we plot the field in the center of the simulated bubble $\varphi(x=0,z)$ as a function of the elapsed proper time (black-solid line). The slow-roll solution obtained by integrating the field and Friedmann equations is overplotted (dashed-yellow line); the agreement is excellent.}
\label{fig-cosmocompare}
\end{center}
\end{figure}

We now turn to bubbles containing a cosmology. As a representative example, we simulate a single Observation bubble in the potential L1. A contour plot of the field is shown in the left panel of Fig.~\ref{fig-cosmocompare}. The metric functions $a$ and $\alpha$ are qualitatively similar to those shown in Fig.~\ref{fig-vacuumbubble_tw_comp}, since in this example the field is undergoing slow-roll inflation inside the bubble, and the solution (over the timescale of the simulation) is therefore quite close to pure HdS. The theoretical expectation for the evolution of the field inside the bubble is found by integrating the field and Friedmann equations for a scalar field in an open FRW universe, with an initial condition set by the CDL instanton. In our numerical solution, the field at the center of the bubble as a function of the elapsed proper time $\tau = \int dz \ \alpha(z,x=0)$ should match this theoretical expectation for the time evolution of the field. This is indeed the case, as shown in the right panel of Fig.~\ref{fig-cosmocompare}. Moving away from the bubble center, we should find that the surfaces of constant field extend all the way up to $z \rightarrow \infty$. This would be true if the bubble interior truly contained an open FRW universe. The numerical solution shown in the left panel of Fig.~\ref{fig-cosmocompare} does not have this behavior at large $z$, although it converges to it with increasing resolution.

In our simulations of both vacuum bubbles and bubbles containing a cosmology, the numerical solutions are worst near the bubble walls. Physically, this is to be expected. As the bubble expands, the wall becomes thinner and thinner due to Lorentz contraction (the velocity of the wall is continually increasing) and the fact that $x$ is a comoving coordinate. In Appendix~\ref{sec:box_size}, we show that if the metric is approximately HdS, the width of the wall in terms of the simulation coordinate $x$ is inversely proportional to $z$. For our standard resolution of $32 \times 10^3$ spatial grid points, and for the potentials we study (which fixes the initial width of the bubble walls), this imposes an upper limit on the run-time of roughly $z \sim 10$ for our simulations. In fact, our ability to accurately track the solution in the vicinity of the walls breaks down somewhere in the range between $z=2-4$ at the standard resolution, depending on the potential (see the contour plots in Figs.~\ref{fig-vacuumbubble_tw_comp} and~\ref{fig-cosmocompare}). Thus, in order to accurately model bubble collisions, we must require that the collision occurs before our description of the walls breaks down, and we always impose $z_c \alt 4$. 

The numerical solutions we obtain still exhibit an overall convergent behavior, and so Richardson extrapolation~\cite{richardson} could be employed for better accuracy. Nevertheless, it is clear that for even later times the solution drifts away from the convergent regime, which limits the application of this approach. In addition, computation time restricts the number of grid points to perhaps two to four times our standard resolution, resulting in an increase by only a factor of two in run time. Thus, the main limitation of our current implementation is the restriction on run time, and the associated restriction on the kinematics of the collision. Because significantly longer run times are necessary to accurately extract the cosmological signatures of bubble collisions, we restrict ourselves in the current paper to exploring the immediate outcome of bubble collisions. To move beyond this, we are extending our implementation to exploit adaptive mesh refinement and alternative coordinate conditions (for lapse and shift). Our results on the cosmological signatures of bubble collisions will appear in a forthcoming publication.

\section{Results}\label{sec:results}

We now discuss the results of our simulations, beginning with the collision between vacuum 
bubbles (i.e. those not containing the inflationary plateaus $C1$ and $C2$ in Fig.~\ref{fig-potentials}), and then
 considering the outcome of collisions between bubbles that contain an interior cosmology. 

\subsection{Colliding identical vacuum bubbles}

In this section, we discuss the collision of identical vacuum bubbles generated from the potentials shown in Fig.~\ref{fig:vacpots}. Since the vacuum in each of the colliding bubbles is identical, the two bubble interiors must eventually merge. However, to satisfy energy-momentum conservation, the energy of the collision must be dissipated in some way. The form of energy loss is an input to the junction condition formalism described in Sec.~\ref{sec:junctionconditions}, and it is typically assumed~\cite{Hawking:1982ga,Wu:1984eda,Freivogel:2007fx,Aguirre:2007wm,Chang:2007eq} that the energy of the collision is released in the form of scalar radiation. However, there are many other possibilities once the full dynamics of the field and metric are considered. 

\begin{figure}
   \includegraphics[width=12 cm]{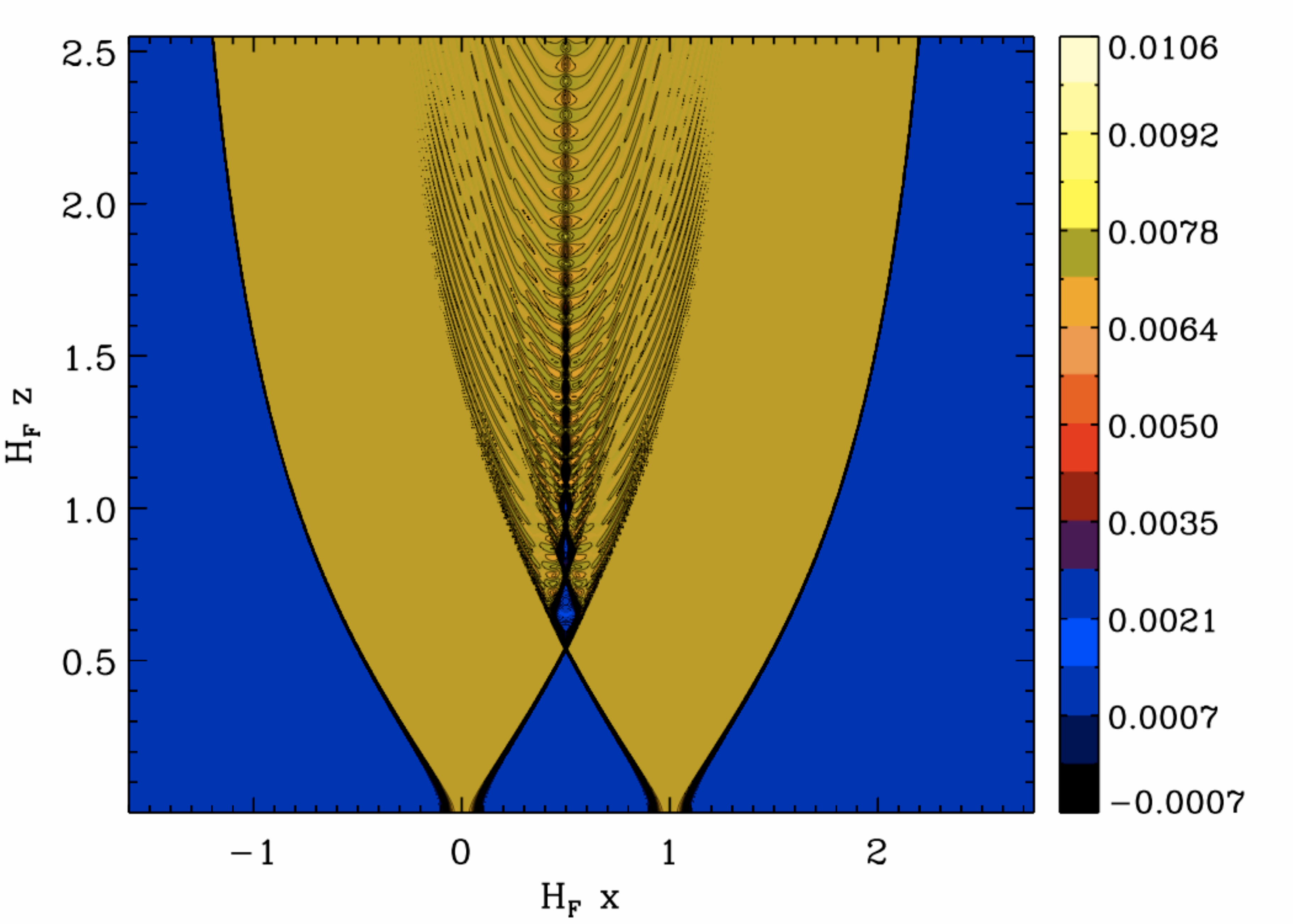} 
\caption{A contour plot of the field for a collision between identical bubbles formed in the potential V1. The solid lines and colorbar indicate the value of the field. In the immediate aftermath of the collision, a pocket of the false vacuum is formed. This pocket expands, re-collapses, and then re-forms. After a few cycles of this, a coherent, oscillating field configuration known as an oscillon is formed. }
\label{fig-oscillon_contours_f}
\end{figure}

\begin{figure}
   \includegraphics[width=8.9 cm]{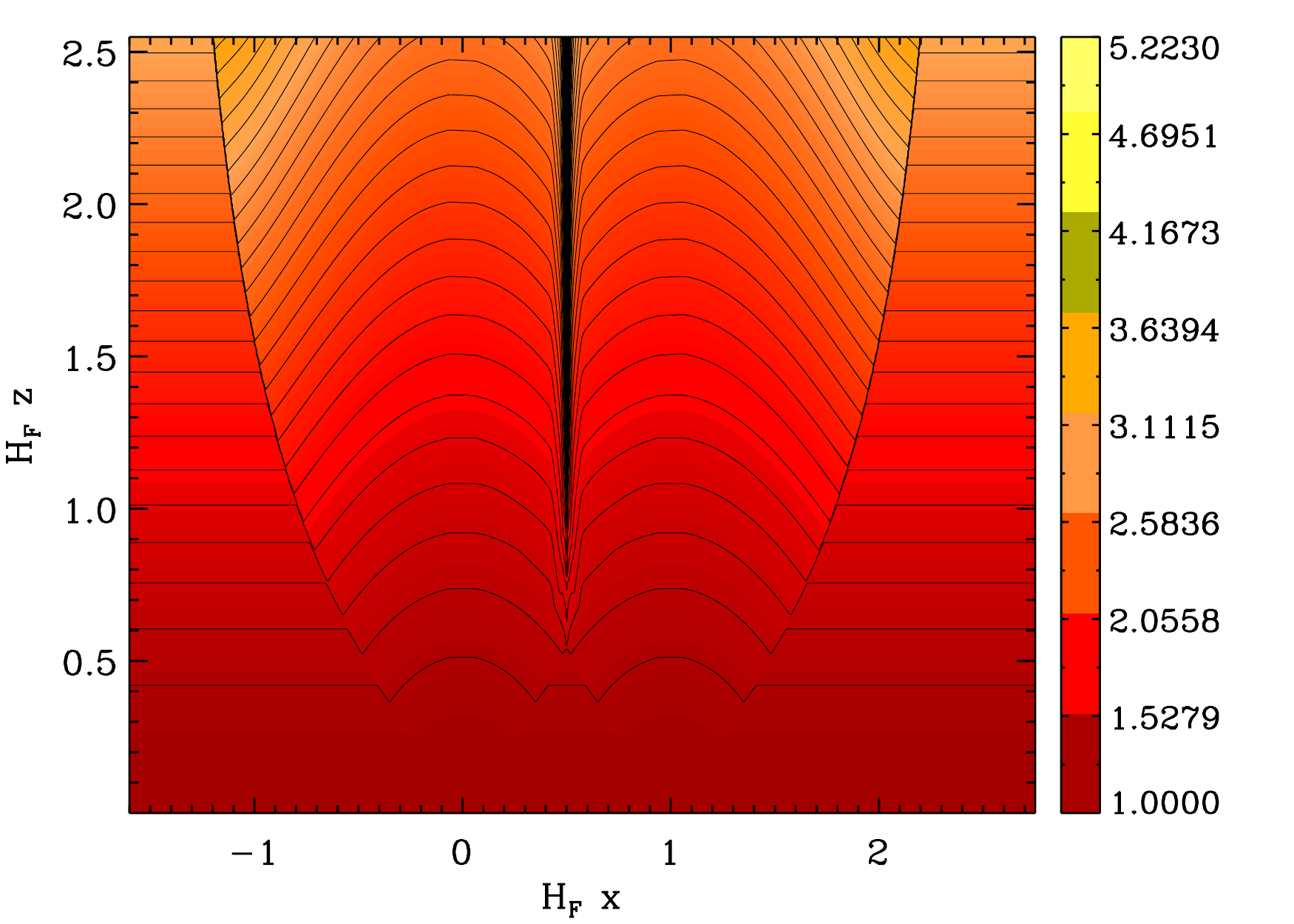} 
    \includegraphics[width=8.9 cm]{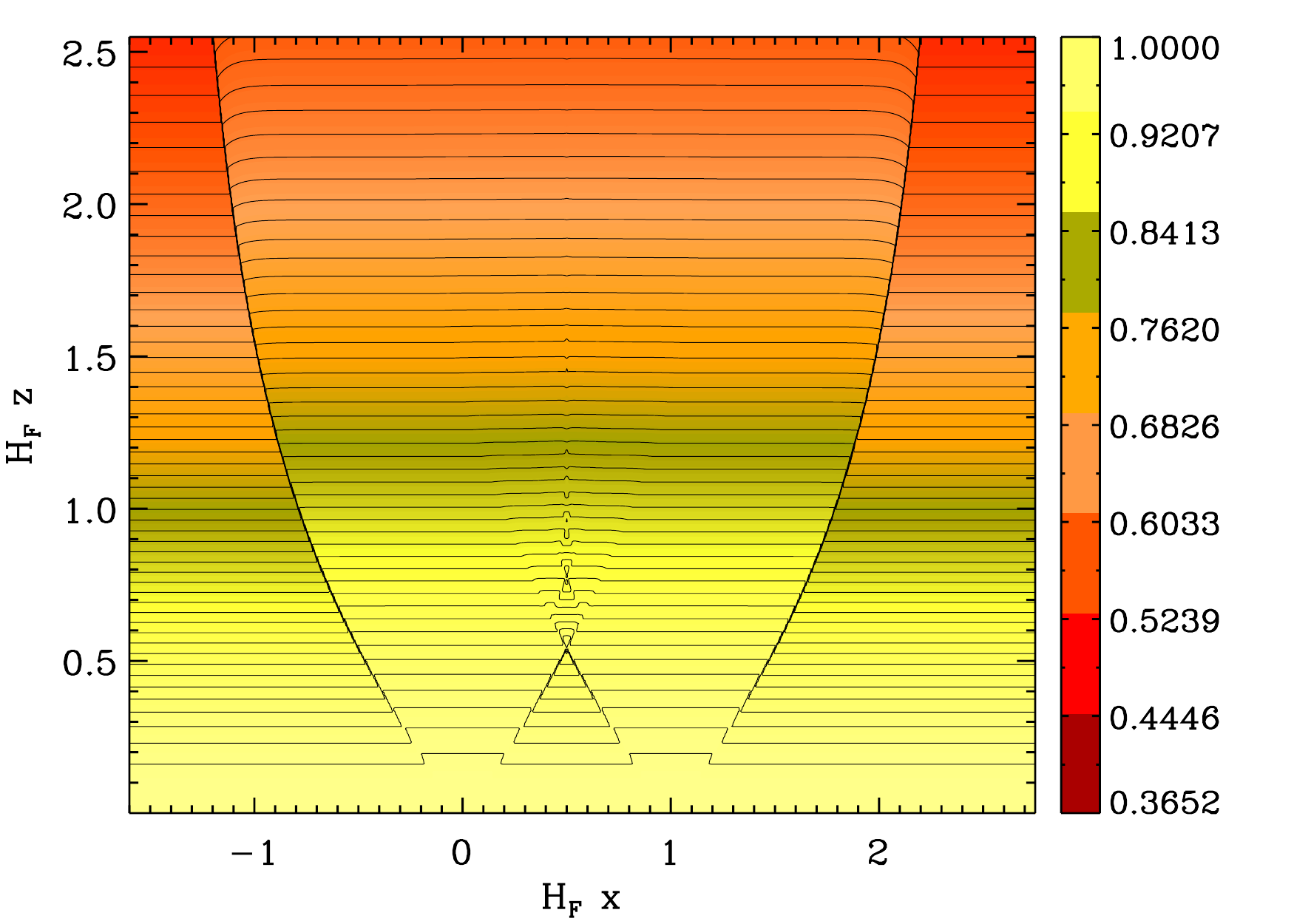} 
\caption{A contour plot of the metric functions $a$ (left) and $\alpha$ (right) for the collision shown in Fig.~\ref{fig-oscillon_contours_f}.} 
\label{fig-oscillon_contours_aal}
\end{figure}

In Fig.~\ref{fig-oscillon_contours_f}, we show a contour plot of the field for the collision between two identical vacuum bubbles formed from the potential V1 with an initial separation of $H_F x_c = 1$. Contour plots of the metric functions are shown in Fig.~\ref{fig-oscillon_contours_aal}. In the immediate aftermath of the collision, a small pocket containing the false vacuum is formed. This region expands, re-collapses, and forms again when the walls re-collide~\footnote{This behavior was first observed in Ref.~\cite{Hawking:1982ga}, and is referred to in the literature on thin-wall solutions as an Oscillatory solution (see Ref.~\cite{Johnson:2010bn}).}. In addition, some (relatively small) amount of scalar radiation is emitted at each collision. After a few cycles of expansion and contraction, there is no longer enough energy for the field to return back up to the false vacuum. The energy does not dissipate, but instead goes into the formation of a coherent, oscillating structure centered about the location of the collision. Such structures are known as oscillons~ \cite{Copeland:1995fq,Fodor:2009kg,Amin:2011hj,Amin:2010dc,Amin:2010xe,Amin:2010jq}, and arise due to non-linear interaction terms in the scalar potential. In this case, if we extend the solution along the symmetry direction, the oscillating field configuration extends indefinitely in $\chi$ (the coordinate labelling the spatial position along the hyperbolae of constant $z$). The formation of oscillons in the collision between bubbles was first discussed in Ref.~\cite{Dymnikova:2000dy} (see also Ref.~\cite{Hindmarsh:2007jb,Khlopov:1998nm}). For all sets of kinematics that we have studied, an oscillon ultimately results. This is quite different from the standard assumption, based on the thin-wall solutions, that the colliding domain walls eventually dissipate into scalar radiation.

\begin{figure}
   \includegraphics[width=8 cm]{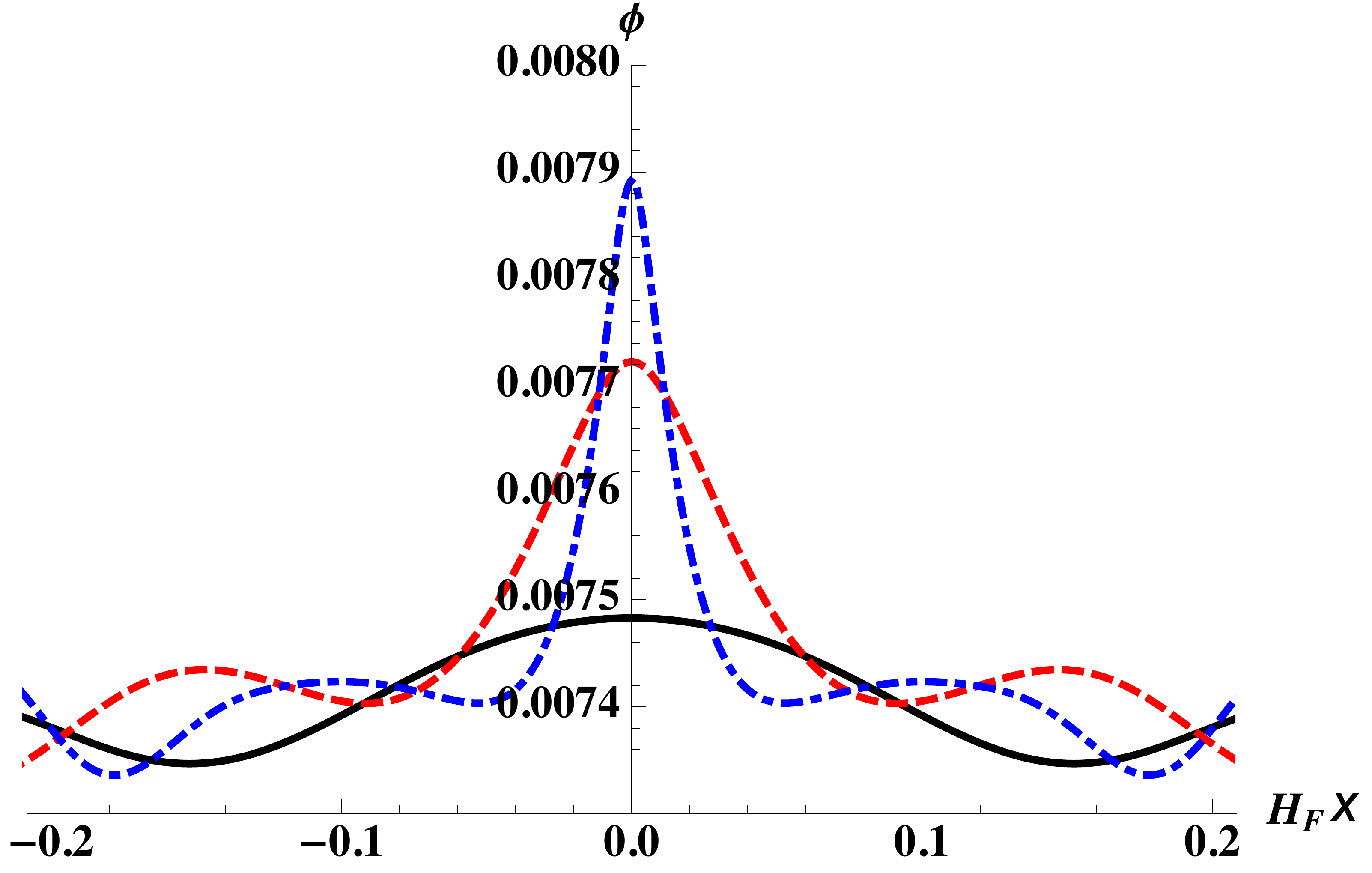} 
\caption{The field on constant $z \simeq 3$ slices for simulated collisions between identical bubbles formed in the potential V1 for three initial separations $H_F x_c = 0.25$ (black-solid line), $H_F x_c = 0.50$ (red-dashed line), and $H_F x_c = 1.00$ (blue dot-dashed line). The solutions have been translated so that the collisions each occur at $H_F x=0$. On subsequent time-slices, the profile retains its shape, but oscillates in amplitude. Such structures are known as oscillons.}
\label{fig-oscillon_profiles}
\end{figure}

Let us now analyze each stage of evolution in this collision. In the immediate future of the collision, the expectation is that the ``free-passage" approximation is satisfied~\cite{Easther:2009ft,Giblin:2010bd}. In this approximation, the bubble walls do not initially interact (as long as they are  sufficiently relativistic), but instead superpose linearly. Immediately after the collision, the field in a small region is at a position $\varphi \sim 2 \phi_{T2}$ (twice the width of the potential barrier). This is high enough on the potential that the field overshoots the minimum, bounces off the rising potential at positive $\varphi$, and forms an expanding pocket of the false vacuum. Because this pocket has higher energy than the bubble interior, it eventually collapses. If there is enough energy left, a new pocket of false vacuum is formed after the collapse of the first. This continues until there is no longer enough energy to create a pocket of false vacuum, at which point an oscillon forms. In Fig.~\ref{fig-oscillon_profiles}, we show the oscillon profiles on constant-$z$ slices through simulated collisions with three different initial separations ($H_F x_c=0.50$, $0.75$, $1.00$). The main effect of increasing the initial separation is a larger-amplitude, smaller-width profile. In each case, the profile retains its shape as $z$ increases, but oscillates with decreasing amplitude.

We can understand some quantitative aspects of the oscillon solution analytically. Neglecting the gravitational backreaction of the field on the metric, Eq.~\ref{eq:Hbgrndeom} gives the field equation for $\varphi$. We assume that the solution can be written as a stably oscillating packet plus a correction:
\begin{equation}
\varphi(z,x) = f_0 (z) g_0 (x) + \delta \varphi (z,x)\, .
\end{equation}
Substituting this ansatz into the equation of motion, Taylor-expanding the potential about the true vacuum minimum, and keeping only the lowest-order terms, we have:
\begin{equation}
\partial_{zz} f_0 =- \frac{2 (1+ 2 H^2 z^2)}{z (1+H^2 z^2)}  \partial_z f_0 - \left( \partial_{\varphi \varphi} V(\phi_T) - \frac{\partial_{xx} g_0}{(1+H^2 z^2 )g_0} \right) \frac{f_0}{1+H^2 z^2} \, .
\end{equation}
In the limit where $z \gg H^{-1}$, we can neglect the contribution of the gradients in the field, and the general solution to the equation of motion is:
\begin{equation}\label{eq:predicted_oscillon_time}
f_0 = \frac{A}{z^{3/2}} \cos \left[ \sqrt{ \frac{\partial_{\phi \phi} V(\phi_{T2})}{H^2} - \frac{9}{4} } \ \log \left( z \right) \right] + \frac{B}{z^{3/2}} \sin \left[ \sqrt{\frac{\partial_{\phi \phi} V(\phi_{T2})}{H^2}  - \frac{9}{4}} \ \log \left( z \right) \right] \, .
\end{equation}
Therefore, the expectation is that we have fluctuations of decreasing amplitude and increasing period as a function of $z$. The $\delta \varphi (z,x)$ term in our ansatz captures any additional effects, such as the dissipation of the oscillon into scalar radiation.

\begin{figure}
      \includegraphics[width=10 cm]{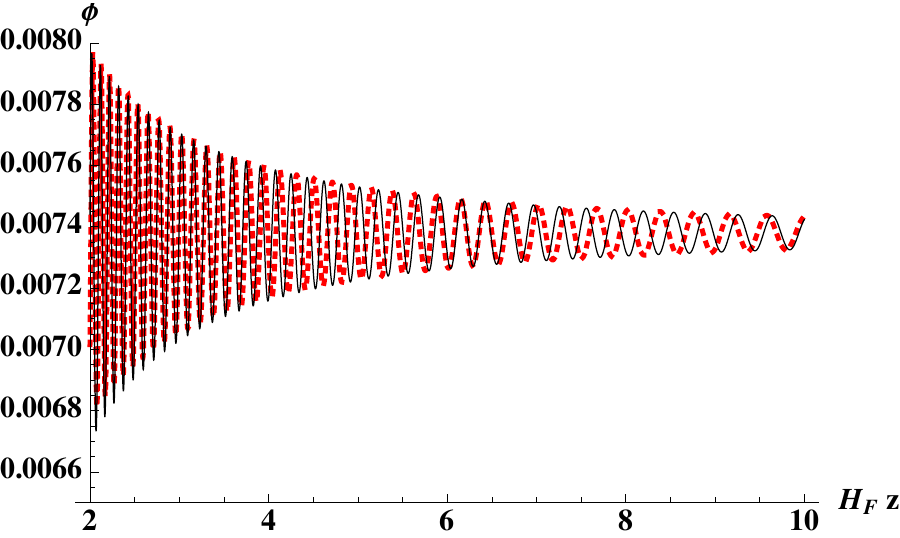} 
\caption{Cutting through the center of the intermediate oscillon profile shown in Fig.~\ref{fig-oscillon_profiles}, we plot the field as a function of $z$. Overplotted (red-dashed lines) is the analytic solution Eq.~\ref{eq:predicted_oscillon_time}. The qualitative agreement is quite good, although the analytic solution does not get the drift of the frequency of oscillation exactly correct. As the initial bubble separation is increased, the packet amplitude becomes larger. }
\label{fig-oscillon_time}
\end{figure}

 Cutting through the oscillon profile at constant $x$ in our simulations, we show in Fig.~\ref{fig-oscillon_time} how the amplitude decreases with time for the intermediate case plotted in Fig.~\ref{fig-oscillon_profiles}. We also plot the solution Eq.~\ref{eq:predicted_oscillon_time} (red dotted line) obtained in the limit where the gravitational back-reaction of the field configuration is negligible, matching the analytic and numerical solutions at $H_F z = 2$. First, note that the analytic solution approximates the $z$-dependence of the amplitude of the oscillations very well. This indicates that the dominant form of energy-loss for the oscillon is due to the background expansion, and not scalar radiation being shed from the configuration. The frequency of the oscillations is initially matched quite well by the analytic solutions, but there is a mismatch at larger values of $z$. This could be due to gravitational effects not captured by our analytic approximation, or non-linear corrections to the period of oscillations about the potential minimum. Nevertheless, the qualitative agreement between the analytic and numerical solutions is quite good. We also study the field as a function of proper distance on slices at $z=5$ and $z=10$, observing that the oscillon profile roughly doubles in width over this timescale. Such behavior is consistent with the profile undergoing expansion with the background, since the physical distance between two points of constant $x$ has also doubled. The oscillons formed in these collisions are therefore not gravitationally bound.

\begin{figure}
   \includegraphics[width=8 cm]{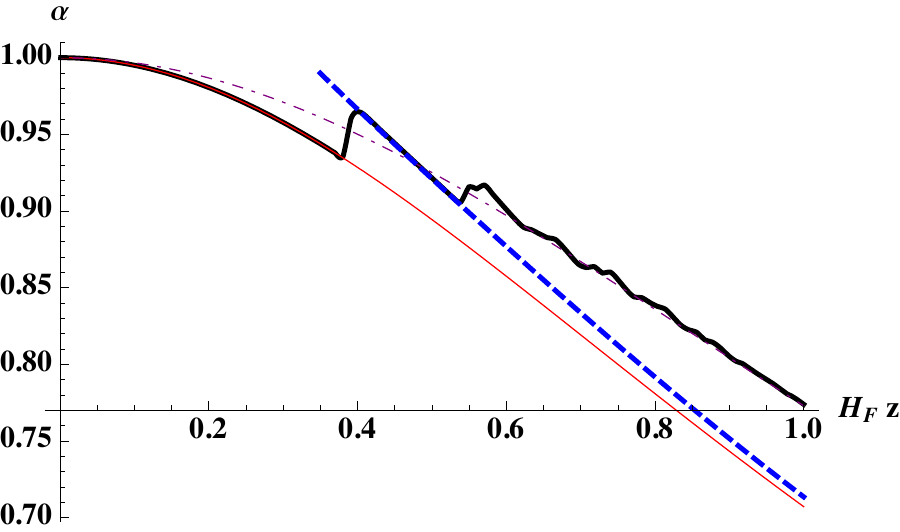}    \hfill
   \includegraphics[width=8 cm]{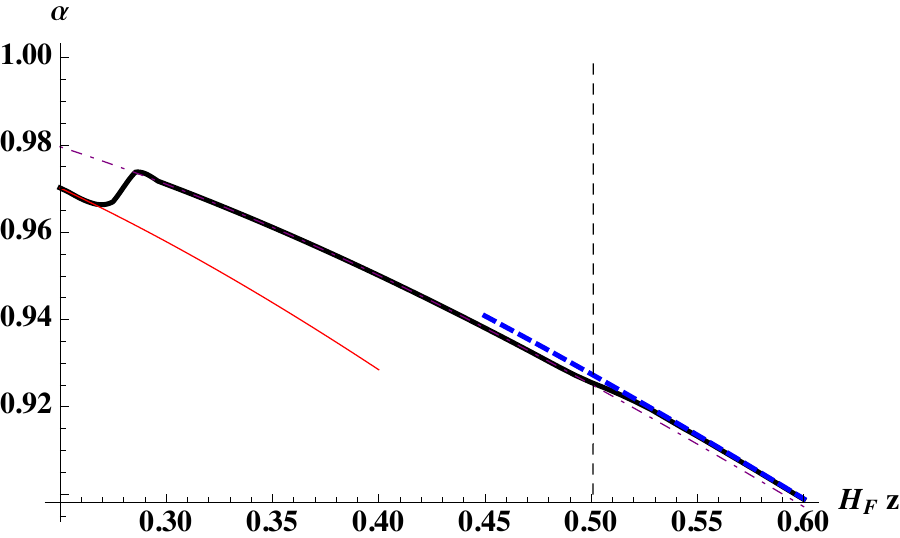} 
\caption{The metric function $\alpha$ evaluated at $x=0.375$ (left) and $x=0.280$ (right) for the collision between two identical bubbles formed in the potential V1 and initially separated by $x_c = 0.75$. The plot on the left cuts through the region that underwent a transition back to the false vacuum. We overplot the expectation for $\alpha$ from the thin-wall approximation (blue dashed line) as well as the metric in empty hyperbolic de Sitter outside (solid red line) and inside (dot-dashed purple line) the bubble. The agreement between the thin-wall solution to the immediate future of the collision (at $z \simeq 0.4$) and the numerics is excellent. The plot on the right cuts through the region to the future of the outgoing scalar radiation, but outside the pocket of false vacuum. We find the best-fit for the mass parameter, assuming that $\alpha = A_{I}^{-1/2}$ (see Eq.~\ref{eq:vacuummetricfunction}) in this region, obtaining $M_r \simeq 3.0 \times 10^{-3}$. This function is over-plotted on the simulation as the blue dashed line, and in addition we show the metric function for empty HdS inside (dot-dashed purple line) and outside (solid red line) the bubble.}
\label{fig-mass_k3}
\end{figure}

In the thin-wall approximation, the mass parameter in the metric function Eq.~\ref{eq:vacuummetricfunction} characterizes the effect of the collision on the metric describing the bubble interiors. Referring back to Fig.~\ref{fig-coll_diagram}, we can identify Region III with the false vacuum, use Eq.~\ref{eq:massparameterprediction} to predict the mass parameter, and directly compare the metric function $\alpha$ between the thin-wall and numerical solutions. In the numerical solution, at the location of the collision, the metric function should jump, and then be dominated by the $1/z$ behavior dictated by the mass parameter. In Fig.~\ref{fig-mass_k3}, we show $\alpha$ at the location of a collision between two identical vacuum bubbles with an initial separation $H_F x_c = 0.75$ (left panel), and $\alpha$ at a location which intersects the outgoing shell of scalar radiation (right panel) in the same simulation. Since the incoming and outgoing walls are relativistic and there is little scalar radiation expelled in the collision, Eq.~\ref{eq:massparameterprediction} should be a good estimate of the mass parameter in the region immediately to the future of the collision. Using the values of $H$ in the true and false vacuum, we obtain $M_{\rm III} = 0.02$ as an estimate. In the left panel of Fig.~\ref{fig-mass_k3}, we overplot (red-solid and blue dashed lines) the predicted behavior of $\alpha$ as a function of $z$; the agreement between the numerical solution and the thin-wall solution is excellent. A more detailed thin-wall construction takes into account the scalar radiation emitted by the collision. In this case, there are two mass parameters: one describing the metric in the false vacuum pocket and one describing the metric in the future of an outgoing null shell of scalar radiation. Assuming the metric in the future of the radiation shell is of the form Eq.~\ref{eq:vacuummetricfunction}, we can find the best-fit to the mass parameter for the simulated collision. Doing so, we find $M_r \simeq 0.003$; the best-fit is overplotted (blue dashed line) on the curve obtained from the simulation in the right panel of Fig.~\ref{fig-mass_k3}. The energy density in the outgoing shell of radiation is related to the mass parameter by
\begin{equation}
\sigma_r = \frac{M_r}{4 \pi z^2} \, .
\end{equation}
We can therefore estimate the initial energy density in scalar radiation released during the collision in this example to be $\sigma_r (z_c) \simeq 10^{-3} H_F^{3}$. Repeating this exercise for bubbles with several different initial separations, we find that Eq.~\ref{eq:massparameterprediction} predicts the mass parameter in the post-collision false vacuum pocket very well. The mass parameter associated with the region to the future of the outgoing radiation is roughly $1/10$ of the mass parameter associated with the false vacuum pocket. Therefore, we conclude that most of the energy released in the collision goes into the formation of the false vacuum pocket as opposed to the outgoing scalar radiation.

\begin{figure}
   \includegraphics[width=8 cm]{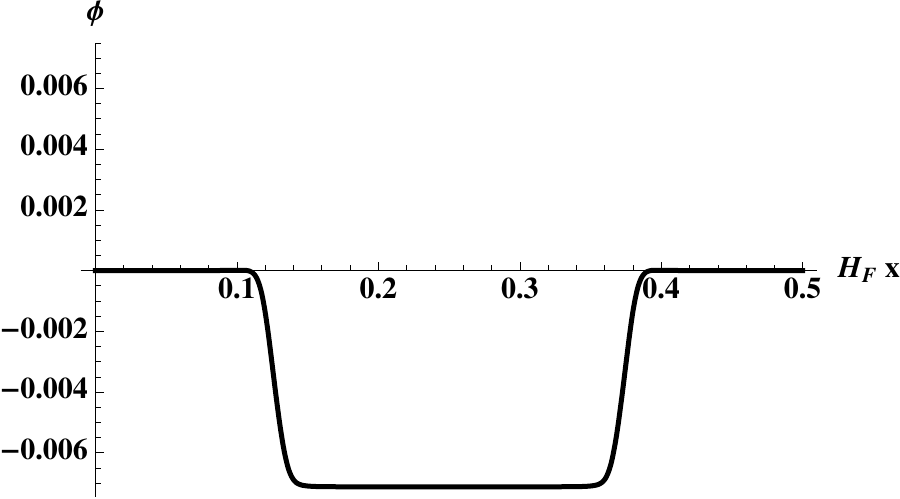} \hfill
   \includegraphics[width=8 cm]{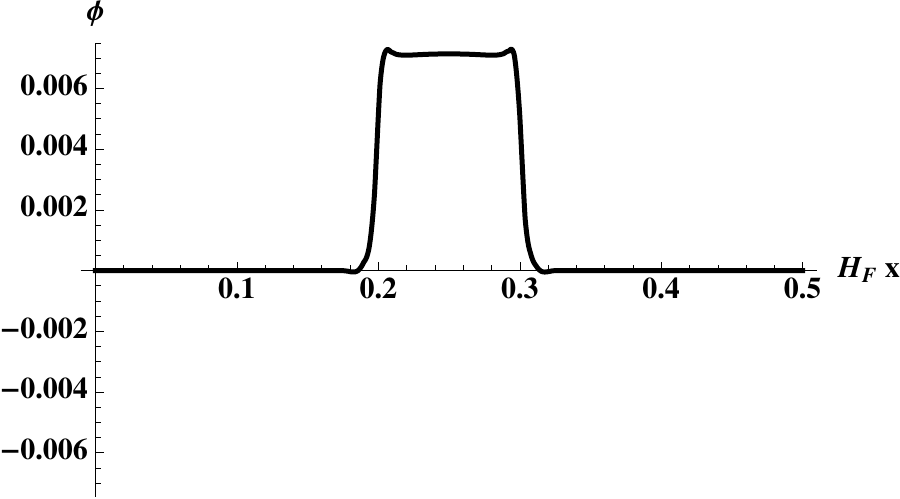} 
\caption{The field on two constant-time slices (left and right) just before and just after the collision of two identical bubbles formed in the potential V3. The field linearly superposes in the vicinity of the collision. This value of the field is in the vicinity of a third vacuum, inducing a classical transition.}
\label{fig-classtrans_const_z}
\end{figure}

\begin{figure}
   \includegraphics[width=8.9 cm]{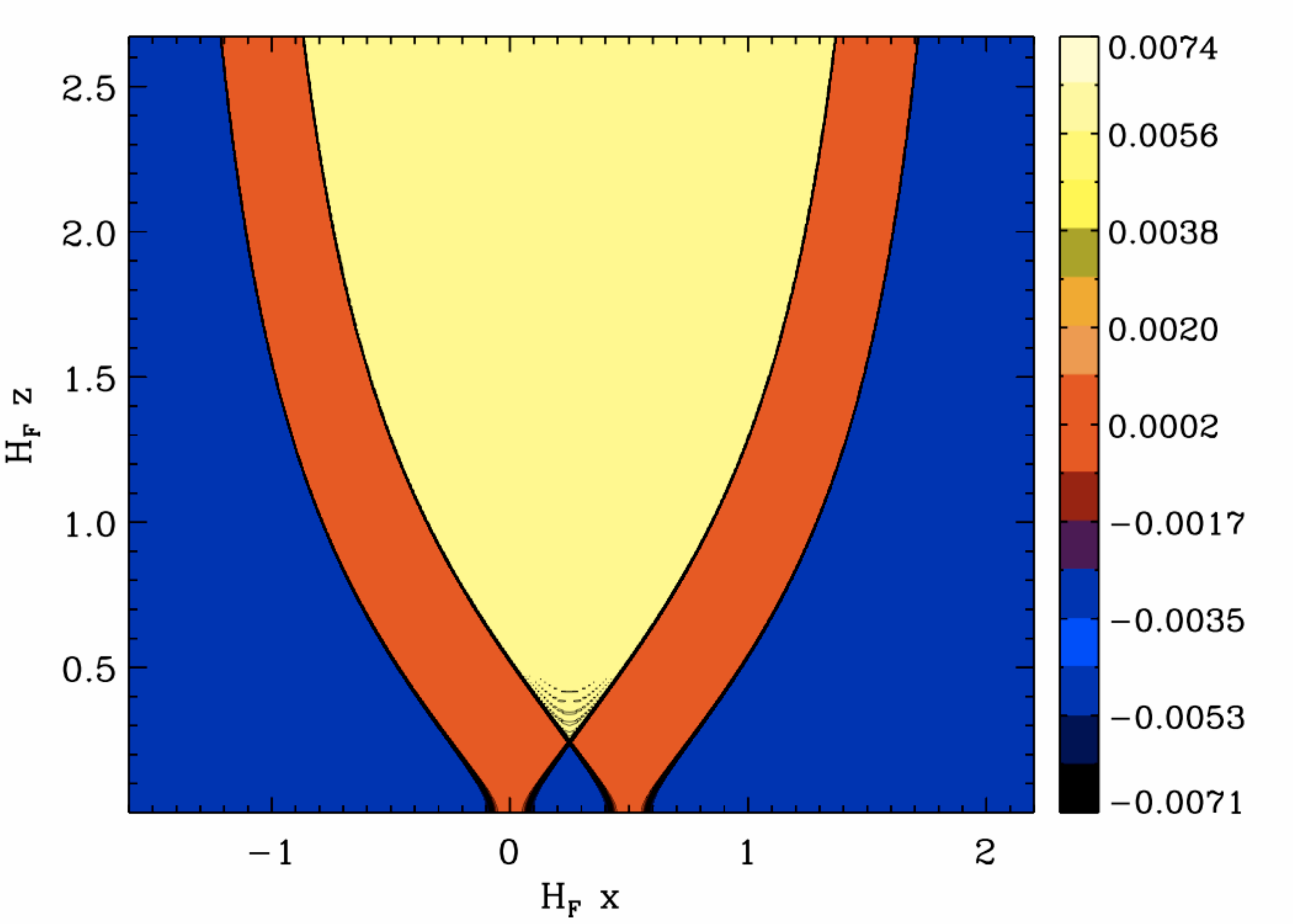} 
   \includegraphics[width=8.9 cm]{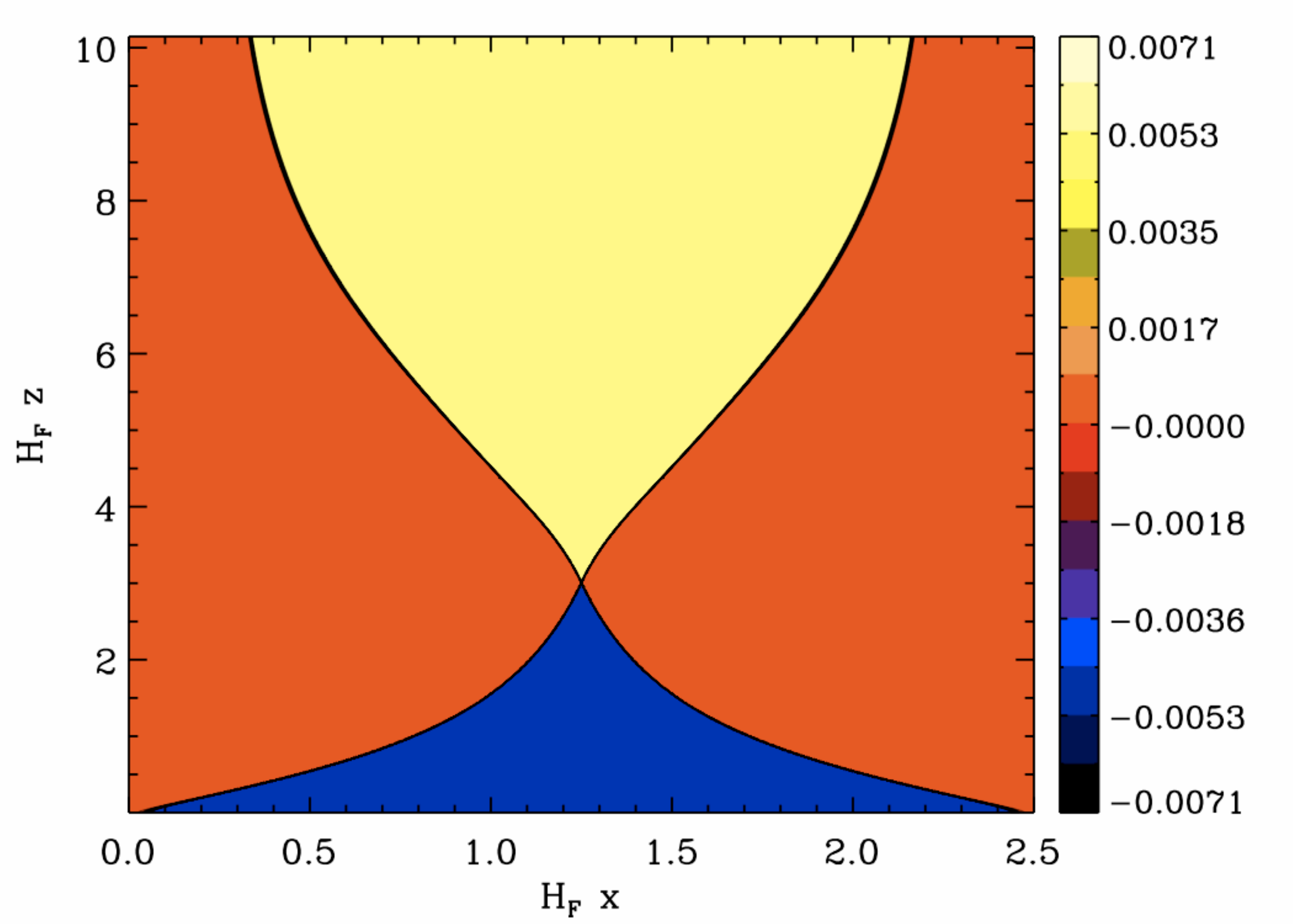}  \hfill
\caption{Contour plots of the field for the collision of two identical vacuum bubbles formed in the potential V3 (left) and V4 (right). In the right panel, the boundaries of the simulation lie at the center of each bubble. In both cases, the collision causes a classical transition to a region containing another vacuum. On the left, the new vacuum has lower energy than either bubble interior, and therefore the region that underwent the classical transition expands. On the right, the new vacuum has higher energy than either bubble interior, but because the domain walls surrounding the new vacuum are repulsive, the region that underwent the classical transition survives.}
\label{fig-classtrans_down}
\end{figure}

Finally, we discuss bubble collisions for the potentials V3 and V4. As described above, when two highly relativistic bubbles collide, the field configurations in the vicinity of the collision simply superpose. If two identical vacuum bubbles collide, and there happens to be a third vacuum located at a distance in field space equal to twice the separation between the true and false vacua, then the collision can result in a lasting region of the third vacuum. This phenomenon is known as a classical transition, and was introduced in Refs.~\cite{Easther:2009ft} and~\cite{Giblin:2010bd}. In Fig.~\ref{fig-classtrans_const_z} we plot two constant-$z$ slices through the collision of two bubbles formed from the potential V3. Just after the collision, the incoming field profiles superpose. In this potential, there is another vacuum of lower energy density at this position in field space. Therefore, a classical transition results, as shown in the contour plot in the left panel of Fig.~\ref{fig-classtrans_down}. These solutions are in good qualitative agreement with previous simulations which neglected gravitational effects~\cite{Easther:2009ft,Giblin:2010bd,Aguirre:2008wy}.

In Ref.~\cite{Johnson:2010bn}, the effect of gravity on classical transitions was discussed using the junction condition formalism. In the absence of gravitational effects, classical transitions can only lead to lasting regions of new vacua that have lower energy density than the interior of the colliding bubbles. However, the gravitationally repulsive nature of domain walls, and the background expansion of space in hyperbolic de Sitter, can allow for the formation of lasting regions that have {\em higher} energy density than the colliding bubble interiors, but no higher than the false vacuum from which the bubbles were formed. The easiest way to ensure a lasting region of new vacuum is to enclose it with domain walls that have enough energy density to be gravitationally repulsive (see the discussion surrounding Eq.~\ref{eq:betas}): $|H_R^2 - H_L^2| < k_i^2$. In the right panel of Fig.~\ref{fig-classtrans_down}, we show a simulated collision in the potential V4 which produces a lasting region of higher vacuum through the production of repulsive walls. It is also possible to create a lasting region by creating marginally repulsive domain walls (which have a tension close to the bound for repulsive walls) with sufficient kinetic energy to grow a region of new vacuum larger than the size of the interior causal horizon. In practice, for the examples we have studied, the required initial separation of the bubbles is larger than what we can access reliably with our current numerical implementation.

In summary, we find that most of the energy released in the collision of identical vacuum bubbles goes into localized field configurations at the position of the collision. Very little energy produced in the collision is radiated away, at least initially. In addition, the thin-wall solutions are in good agreement, both qualitative and quantitative, with the simulated bubble collisions. We have confirmed the existence of classical transitions in the presence of gravity, and found that lasting regions of higher energy density can indeed be produced through gravitational effects as predicted by Ref.~\cite{Johnson:2010bn}.

\subsection{Non-identical vacuum bubbles}\label{sec:vacdiffbubs}

We now move on to the case where the colliding vacuum bubbles are not identical. Since the interiors of the colliding bubbles are not identical, a domain wall must form in the future of the collision to separate the two bubbles. In the junction condition formalism, the post-collision domain wall is characterized only by its tension. To conserve energy and momentum at the collision, a shell of outgoing scalar radiation is assumed to be produced as well. The late-time behavior of the post-collision domain wall in the junction condition formalism is fully determined by its tension, and the vacuum energies on either side by Eq.~\ref{eq:betas}. The wall can either be normal (accelerating towards the side with a lower vacuum energy) or repulsive (accelerating away from both sides). In this section, we confirm these late-time properties of the solution with our numerics, and show that taking into account the full dynamics of the collision can lead to far more varied behavior in the vicinity of the collision.  

\begin{figure}
   \includegraphics[width=8.9 cm]{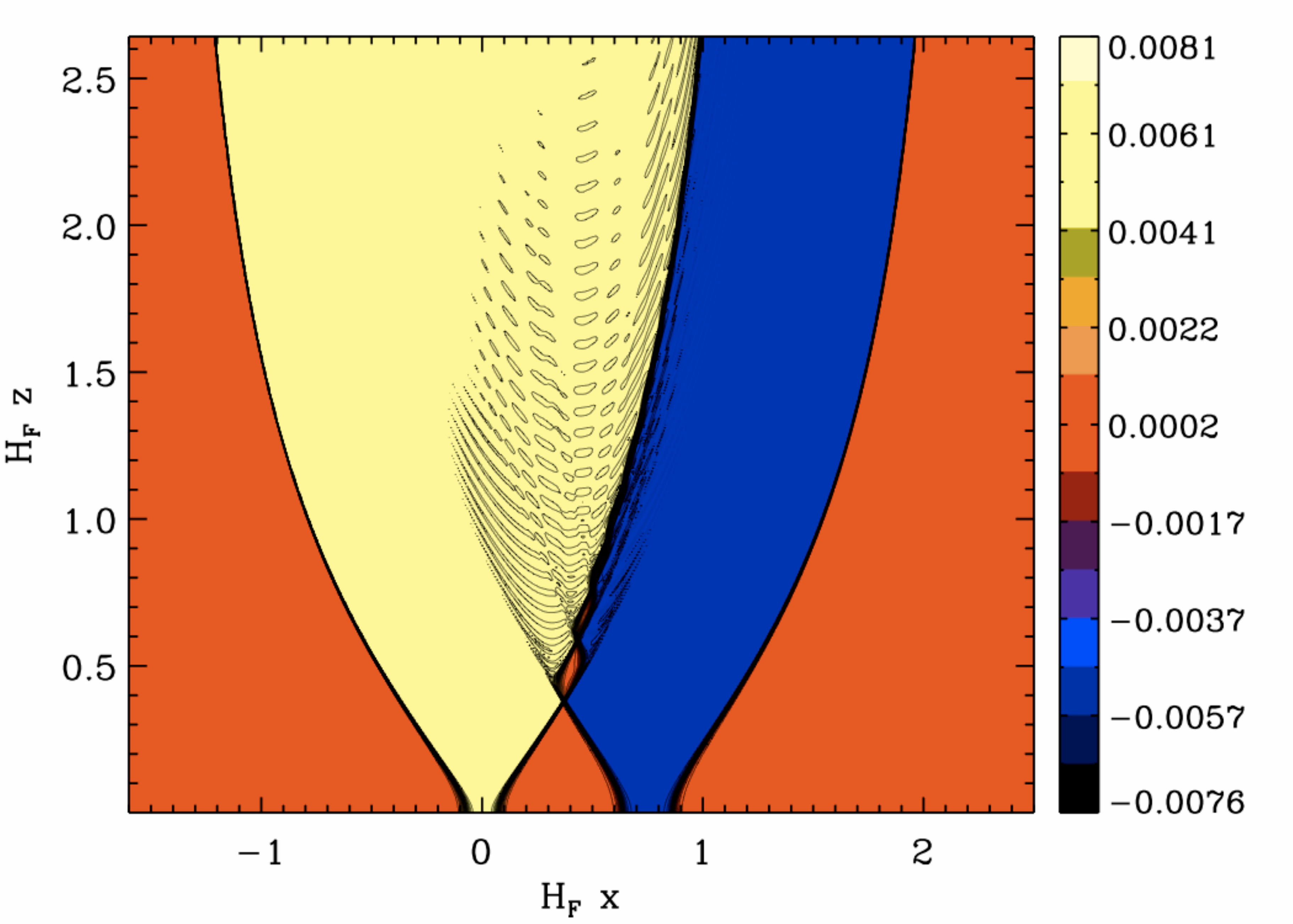}
       \includegraphics[width=8.9 cm]{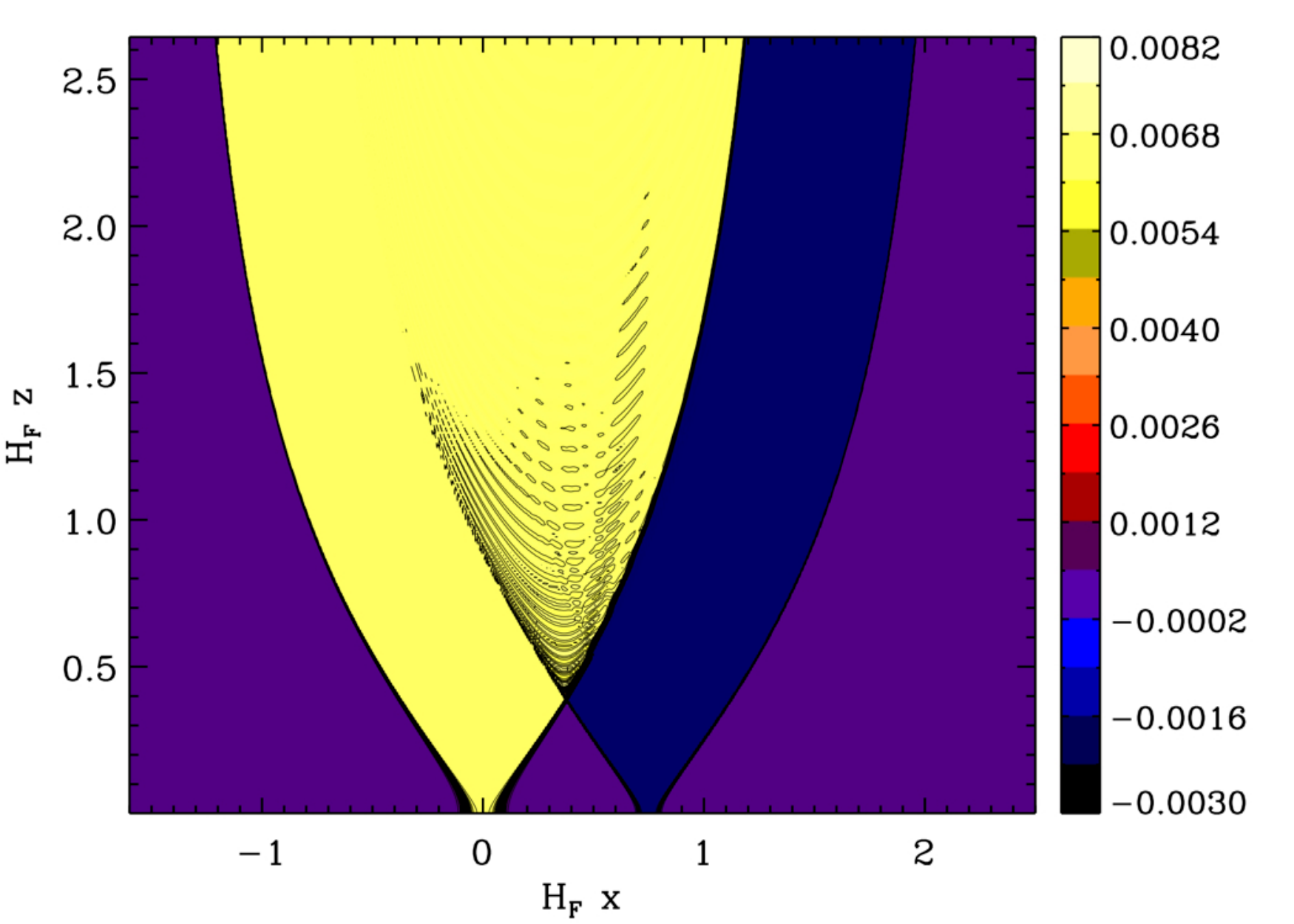} \hfill
\caption{On the left, we show a contour plot of $\varphi(x,z)$ for the collision between the two different bubbles allowed by the potential V1. On the right, the potential barrier in the region $T1$ has been compressed by a factor of two relative to the potential V1. In the free-passage approximation, the structure of the potential determines the position of the field in the immediate aftermath of the collision. For the collision on the left, the potential barriers are nearly symmetric, and the sum of the wall profiles is close to $\varphi = 0$. This leads to the production of an initially expanding  pocket of false vacuum inside the post-collision domain wall. The collision on the right was generated from a potential where the width of $T1$ is roughly half that of $T2$. The sum of the wall profiles is at positive $\varphi$ near the vacuum of the Observation bubble. Therefore, the post-collision domain wall is formed moving away from the Observation bubble interior, and no internal modes of the post-collision domain wall are excited. In both cases at late times the post-collision domain wall accelerates from the Observation bubble interior, as predicted by the junction condition formalism in Eq.~\ref{eq:betas}.}
\label{fig-coll_acc_away}
\end{figure}

As a representative example, we show a contour plot of $\varphi$ in the left panel of Fig.~\ref{fig-coll_acc_away} for the collision of the two different types of bubbles allowed by the potential V1. Focusing for the moment on the behavior of the field in the immediate aftermath of the collision, note that the post-collision domain wall has a great deal of structure. A portion of the energy released in the collision goes into a breathing mode of the wall: a pocket of the false vacuum is created during the collision, which then expands and contracts. In fact, the excitation of this mode can easily be predicted using the free-passage approximation discussed earlier. The potential barriers $T1$ and $T2$ for the potential V1 are nearly symmetric. This means that when the incoming field profiles making up the bubble walls superpose, the field is placed very near the false vacuum at $\varphi = 0$ immediately after the collision. Thus, we excite the breathing mode of the post-collision wall. Such modes increase the energy density and tension carried by the wall, and change the dynamics, causing the wall to accelerate more slowly out of the Observation bubble.

The free-passage approximation predicts that fewer internal modes of the wall are produced when the potential barriers making up the colliding bubbles are asymmetric. To test this, we generate a potential identical to V1, but where the width of the barrier $T1$ is halved. Superposing the bubble wall profiles, the field should be near the vacuum inside the Observation bubble immediately after the collision. A contour plot of $\varphi$ for the collision in this case is shown in the right panel of Fig.~\ref{fig-coll_acc_away}. Here, the wall forms already moving away from the Observation bubble interior with no breathing modes, as expected. If the width of the barrier $T1$ is {\em increased}, then superposing the bubble wall profiles, the wall is predicted to form moving {\em into} the Observation bubble with no breathing modes. This is again observed in our simulations.

After a short period of time, the breathing modes of the wall damp, and for the examples shown in Fig.~\ref{fig-coll_acc_away}, the wall accelerates out of the Observation bubble. How does this compare to our expectation from the junction condition solutions? From the potential, the difference in Hubble parameters is $H_{R}^2 - H_{L}^2 \simeq 0.13$, and we can estimate the tension of the post-collision domain wall for the potential V1 as the sum of the pre-collision domain walls, yielding $k^2 \simeq 0.001$. Since the difference in Hubble parameters is much greater than the tension (making this domain wall of the normal type), we expect that at late times the post-collision domain wall accelerates in the direction of the bubble with lower vacuum energy: in this case, the Collision bubble. This is indeed what we observe. Again from the junction condition solutions, we expect that the potential V2 gives rise to a repulsive post-collision domain wall. This is because, while the tension of the post-collision domain wall is approximately the same as for V1, the difference in Hubble parameters is much smaller: $H_{R}^2 - H_{L}^2 \simeq 0.0006$. We plot this example in Fig.~\ref{fig-coll_stationary}. Calculating the proper distance from the center of each bubble to the wall, we find that it is indeed accelerating away from both bubble interiors. 

In summary, we have confirmed that the late-time behavior of the domain wall produced in the collision between two different vacuum bubbles is well described by the junction condition formalism, but also discovered a rich dynamics in the immediate aftermath of the collision. For the potentials we have studied, the relative widths of the barriers $T1$ and $T2$ determine the outcome of the collision, as predicted by the free passage approximation. For symmetric barriers, the collision excites a breathing mode of the post-collision domain wall. For asymmetric barriers, the post-collision domain wall forms already moving in the direction of the thicker barrier, with no breathing modes.

\subsection{Colliding bubbles with an interior cosmology}

We now turn to the case where there is an inflationary potential inside the Observation and Collision bubbles. In this case, it is not possible to apply the Israel junction condition formalism to predict the outcome of bubble collisions: the resulting spacetime depends on the full non-linear dynamics. This section is also most relevant for determining the possible observational signatures of bubble collisions. As we discussed previously, our current implementation is not well-suited to run for the timescales necessary to extract the signal accurately. However, we can gain much insight into the dynamics of the collision, and the expected amplitude of any possible signals, by studying the outcome of the collision over a limited number of $e$-folds of the inner-bubble cosmology following the collision.   

\begin{figure}
   \includegraphics[width=10 cm]{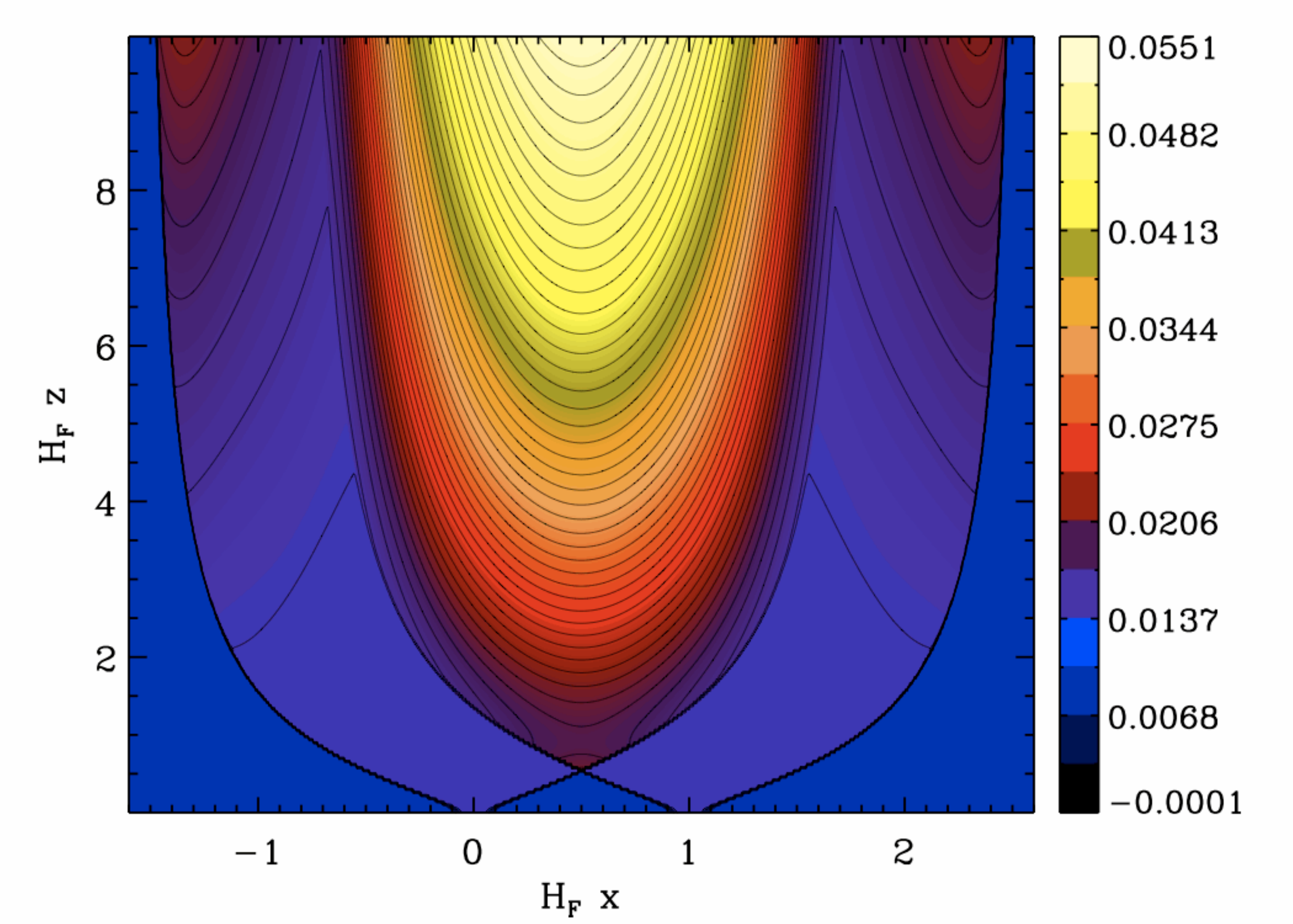} 
\caption{A contour plot of the field $\varphi$ for the collision of two identical bubbles in the large-field potential L1. The field in the future of the collision is moved down the inflationary potential, as predicted by the free passage approximation.}
\label{fig-L2A_oo}
\end{figure}

\begin{figure}
   \includegraphics[width=8 cm]{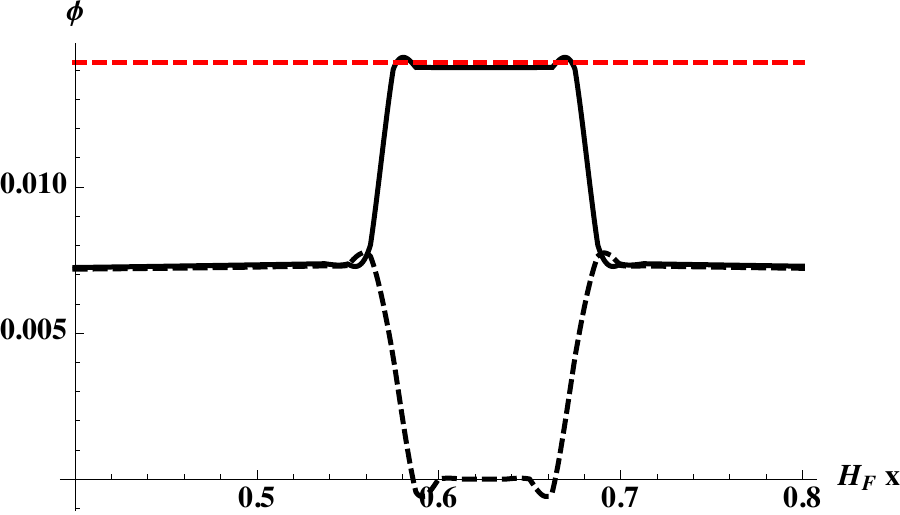} 
\caption{The field on two constant-$z$ slices just before and just after the collision in the simulation shown in Fig.~\ref{fig-L2A_oo}. The red-dashed line is the sum of the distance in field space between the false vacuum and the position the endpoint of the CDL instanton. In agreement with the free passage approximation, the field profiles in the immediate future of the collision simply superpose.}
\label{fig-superposition_L2A_oo}
\end{figure}

As a first application, we collide two identical bubbles formed from the ``large-field" potential L1. A contour plot of the field is shown in Fig.~\ref{fig-L2A_oo}. In the case of large-field models, it is necessary to design potentials with a hierarchy of scales between the width of the barrier $T2$ and the width of the inflationary segment $C2$. This is evident in Fig.~\ref{fig:SRpots}, and is necessary because potentials with a barrier width comparable to $M_\mathrm{Pl}$ typically do not support CDL bubbles, while inflation occurs as the field rolls over a distance greater than $M_\mathrm{Pl}$ (see Ref.~\cite{Aguirre:2008wy} for further discussion of this point). This is quite relevant for determining the outcome of the collision, because in the free passage approximation, the field is moved a distance comparable to the barrier width down the inflationary potential. This distance is predicted to be relatively small compared to the total field excursion in large-field models. In Fig.~\ref{fig-superposition_L2A_oo} we plot two constant-$z$ slices through the simulation just before and just after the collision. The red dashed line represents the sum of the incoming wall profiles, and it can be seen that initially the field profiles do indeed simply superpose. After the collision, inflation proceeds, although from a different initial value of the field as the rest of the bubble interior. We confirm that inflation is still occurring in this region by checking that the metric functions are still approximately HdS in the future of the collision. Examining the colorbar in Fig.~\ref{fig-L2A_oo} and Fig.~\ref{fig:SRpots}, the field has not progressed very far down the potential. The conclusion is that there are fewer $e$-folds of inflation in the future of the collision than in the undisturbed portions of the bubble. We expect that this is a generic prediction for the outcome of the collision between identical bubbles.

We have also examined collisions between bubbles with different interior cosmologies generated from the potentials L1 and L2. These potentials differ only in the choice of $T1$ and $C1$: for L1, the potential slopes down from the barrier $T1$ to a vacuum, while for L2, there is a potential minimum immediately on the other side of the barrier $T1$. Using the potential L1, we can simulate the collision between two bubbles with an interior cosmology, while using the potential L2, we can simulate the collision between a vacuum bubble and a bubble containing an interior cosmology.  

\begin{figure}
   \includegraphics[width=10 cm]{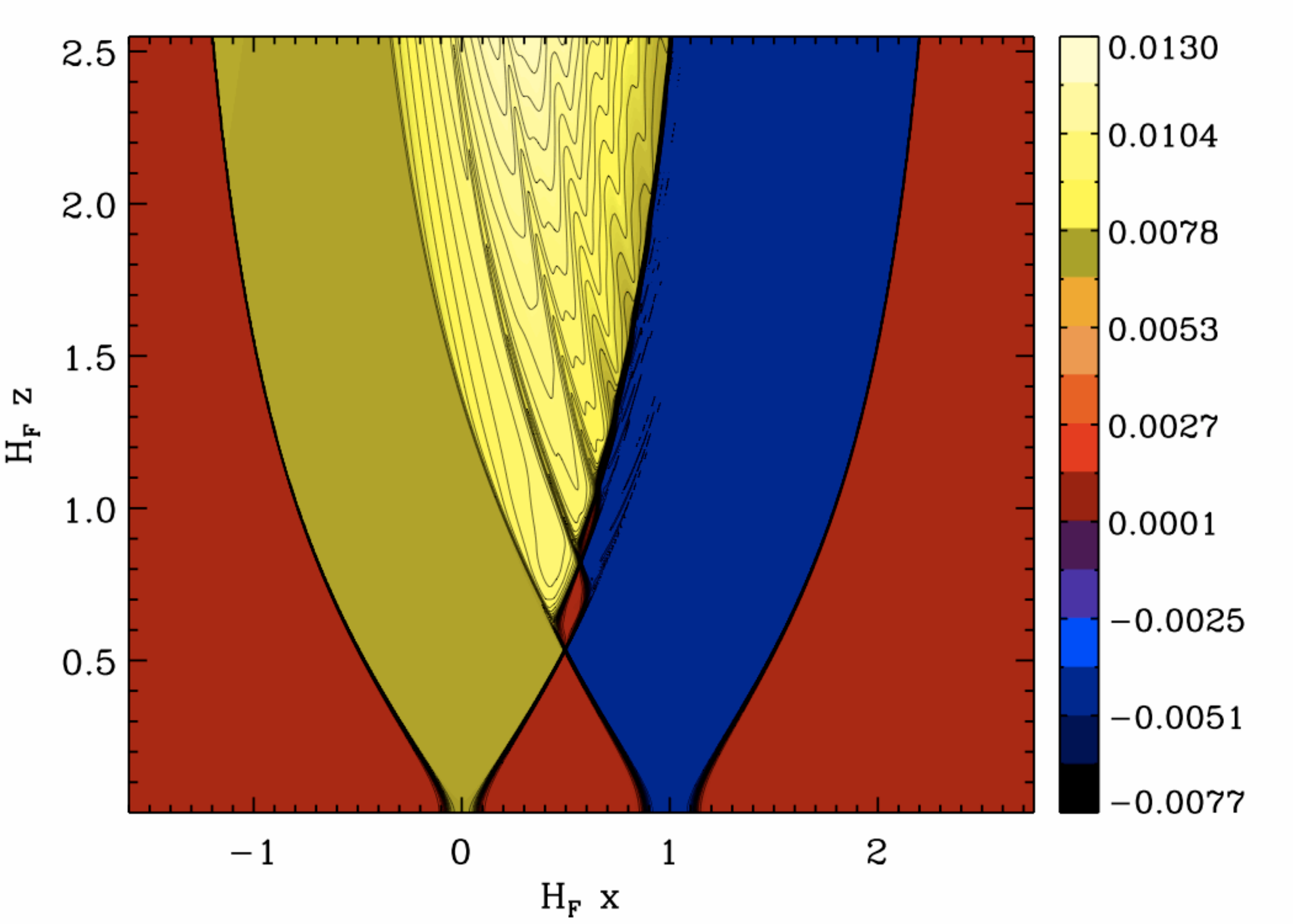} 
\caption{A contour plot of $\varphi$ for the collision of two different bubbles formed from the potential L2. In the immediate future of the collision event, the inflaton inside the Observation bubble is perturbed, and a pocket of the false vacuum is formed. This pocket of false vacuum is a breathing mode of the post-collision domain wall, which once formed, accelerates out of the Observation bubble. The breathing mode further perturbs the inflaton inside the Observation bubble. The amplitude of perturbations to the inflaton is small compared to the total field excursion during inflation in this model.}
\label{fig-L4A_oc}
\end{figure}

In Fig.~\ref{fig-L4A_oc}, we show a simulated collision between the two bubbles formed from the potential L2. Since in this case the energy is lower everywhere inside the Observation bubble, we expect the post-collision domain wall to accelerate out of the Observation bubble at large $z$; this is confirmed in the simulation. Since the potential barriers are nearly symmetric in this case, the free-passage approximation predicts that internal modes of the post-collision domain wall are excited; this is also observed in the simulation. In the future of the collision, the field is pushed down the inflationary plateau, getting an extra kick each time the post-collision domain wall undergoes an oscillation. The amplitude of these perturbations is somewhat smaller than it was for the collision between identical bubbles. This is to be expected, as the energy of the collision is transferred to the creation of the post-collision domain wall and the excitation of breathing modes. Increasing the  center of mass energy of the collision by increasing the initial bubble separation mainly excites larger amplitude breathing modes, as opposed to larger perturbations of the inflaton inside the Observation bubble. This is consistent with the results of Sec.~\ref{sec:vacdiffbubs}, where it was found that for vacuum bubbles very little energy is released in scalar radiation after the collision over a wide range of kinematics. 

\begin{figure}
   \includegraphics[width=8.9 cm]{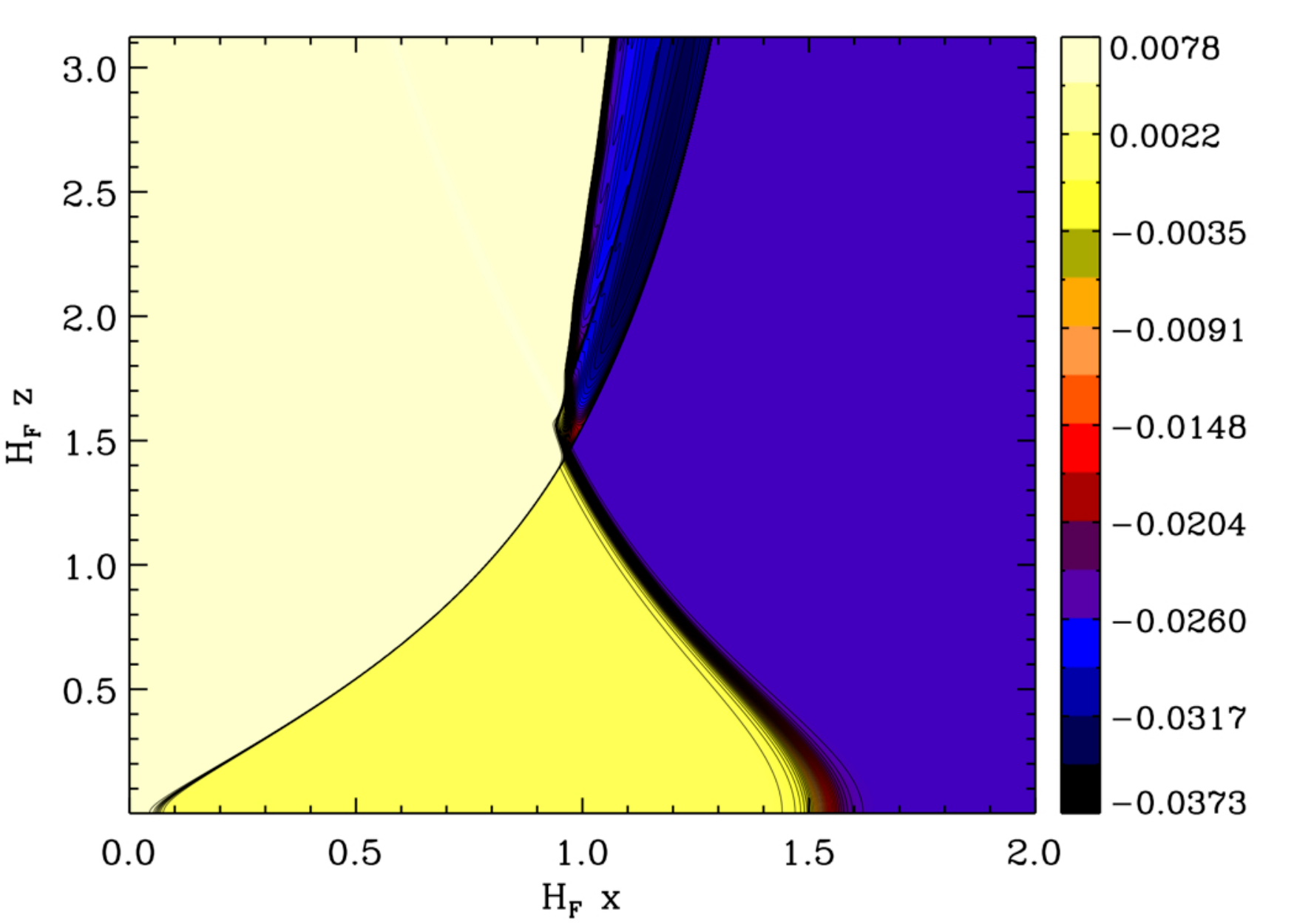}
   \includegraphics[width=8.9 cm]{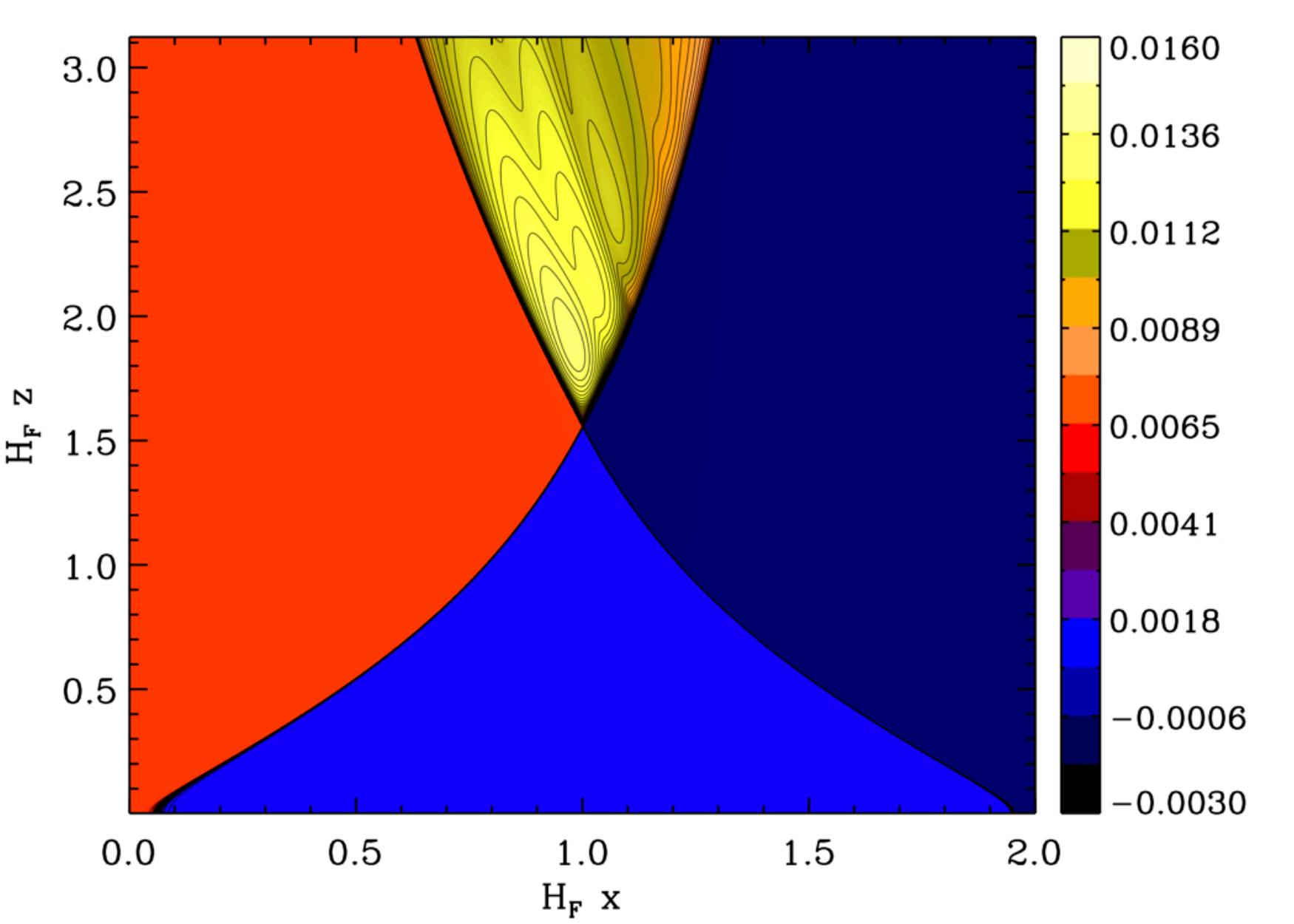}
\caption{On the left, we show a contour plot of $\varphi$ for the collision of bubbles in a potential with a barrier $T1$ twice as large as that of the potential L2. Initially, the post-collision domain wall moves into the Observation bubble, but then turns around. There is very little disturbance to the inflaton inside the Observation bubble. On the right, we show a contour plot of $\varphi$ for the collision of bubbles in a potential with a barrier $T1$ half as large as that of the potential L2. The post-collision domain wall forms accelerating out of the Observation bubble, and no breathing modes are excited. These two examples are consistent with the predictions of the free passage approximation. In both panels, the boundaries of the simulation lie at the center of each bubble.}
\label{fig-thickandthin}
\end{figure}

As in the case of colliding vacuum bubbles, we observe that fewer internal modes of the post-collision domain wall are excited when the potential barriers making up the two different bubbles are asymmetric. Again, we adjust the width of the $T1$ barrier while keeping the other properties of the potential L2 constant: in one case stretching it by a factor of two and in another case compressing it by a factor of two. A contour plot of the field is shown in Fig.~\ref{fig-thickandthin} for the thick barrier $T1$ on the left, and the thin barrier $T1$ on the right. In the case where $T1$ is twice as thick as $T2$, the inflaton inside the Observation bubble is hardly perturbed at all! This corroborates the existence of the ``mild-collisions" first discussed in Ref.~\cite{Aguirre:2007wm}. In the case where $T1$ is half as thick as $T2$, the perturbation is somewhat smaller than for symmetric potential barriers. 

We have studied a number of different initial separations for the bubbles, finding that the field excursion in the aftermath of the collision is bounded by the sum of the field profiles; this confirms the free-passage approximation is valid. What changes with increasing bubble separation is the size of the spacetime region inside which the profiles superpose. Thus, once the bubbles are sufficiently relativistic, increasing the energy in the collision only goes into creating larger regions inside which the field is displaced a constant distance. This suggests that the amplitude of the observable signature of bubble collisions is more sensitive to the structure of the potential than to the kinematics of the collision.  

\begin{figure}
   \includegraphics[width=10 cm]{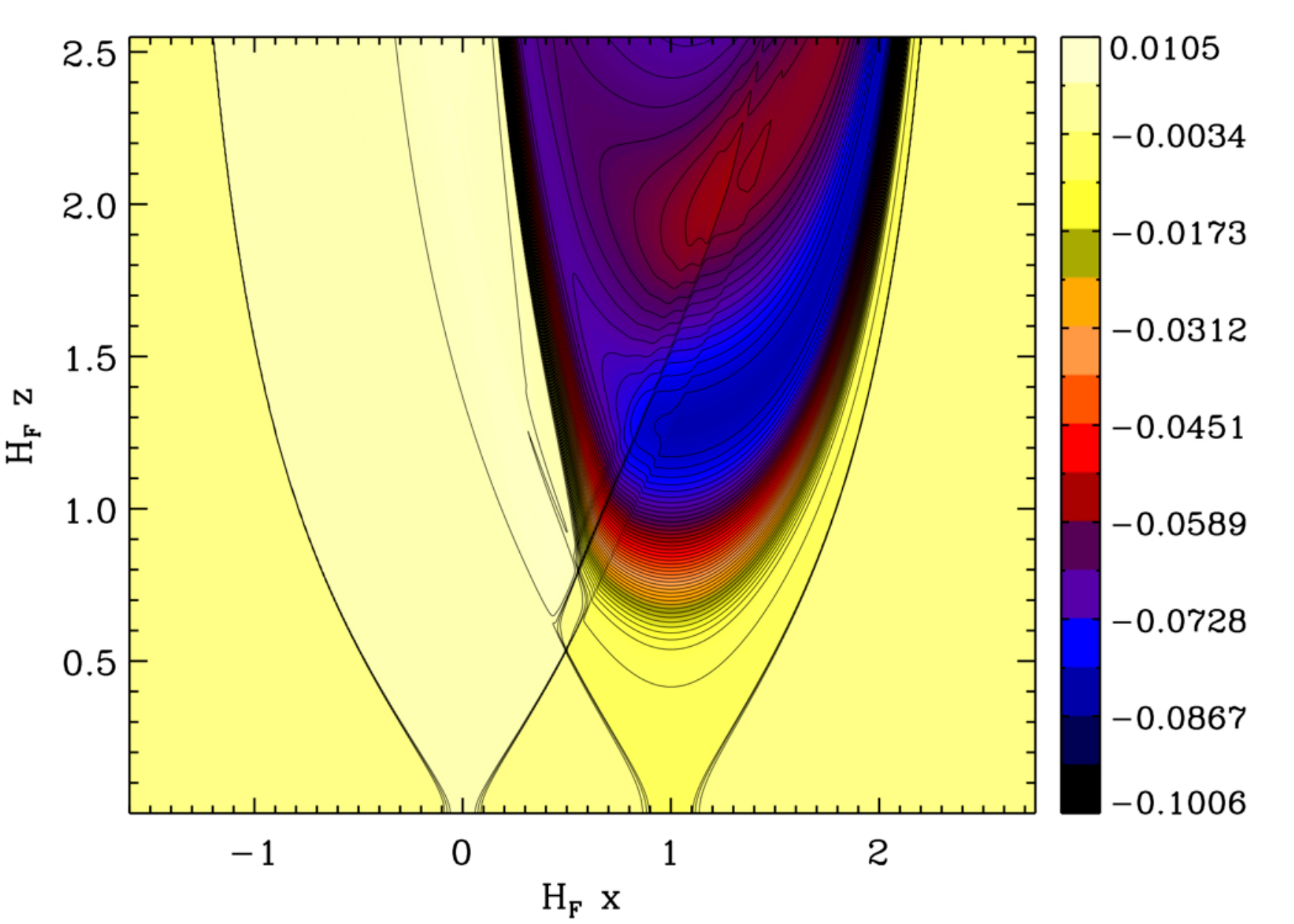} 
   \caption{A contour plot of $\varphi$ for the collision of bubbles in the potential L1. Because the slope $C1$ away from the potential barrier is much steeper inside the Collision bubble than inside the Observation bubble, the post-collision domain wall moves into the Observation bubble during its inflationary phase.}
\label{fig-L2A_oc}
\end{figure}

We now add a cosmology inside the Collision bubble, simulating collisions in the potential L1. This can drastically affect the outcome of a bubble collision, as shown in Fig.~\ref{fig-L2A_oc}. In this example, there is a relatively steep potential slope inside the Collision bubble. Even though the energy inside the Collision bubble is initially higher than inside the Observation bubble, the collision causes the field to quickly roll to a much lower energy than the inflationary plateau inside the Observation bubble. Thus, the post-collision domain wall begins to eat into the Observation bubble. After inflation, since the ultimate vacuum in the Observation bubble is much lower than the vacuum inside the Collision bubble, we expect that the wall will retreat and move into the Collision bubble. Our current implementation cannot reliably track the evolution for such long timescales, and so we cannot confirm this numerically. 

In Ref.~\cite{Aguirre:2008wy}, it was suggested that inflationary potentials of the ``small-field" type would be more susceptible to disruption by collisions than potentials of the large-field type studied above. However, since this work did not include gravitational effects, this was not conclusively established. The free-passage approximation discussed above already provides some idea of which models of inflation are susceptible to disruption by a collision. Focusing for the moment on a collision between identical bubbles, the field in the immediate aftermath of the collision is displaced a distance equal to the width of the barrier making up the bubble. Models of inflation which have an inflationary plateau smaller than or comparable to this distance are susceptible to complete disruption by a collision. This implies that there should be a hierarchy in scales between the width of the barrier $T2$ and the inflationary potential $C2$ for inflation to survive bubble collisions.  

The two types of small-field inflation we use in our numerics are shown in Fig.~\ref{fig:SRpots}. For the potential S1, inflation occurs near an inflection point (see e.g. Refs.~\cite{Baumann:2007np,Linde:2007jn} for a discussion of similar models). For the potential S2, inflation occurs near a potential maximum (see e.g. Ref.~\cite{Boubekeur:2005zm}). The inflaton has a much smaller range in S1 than S2, although in both cases, the field excursion during inflation is sub-Planckian. 

\begin{figure}
   \includegraphics[width=8.9 cm]{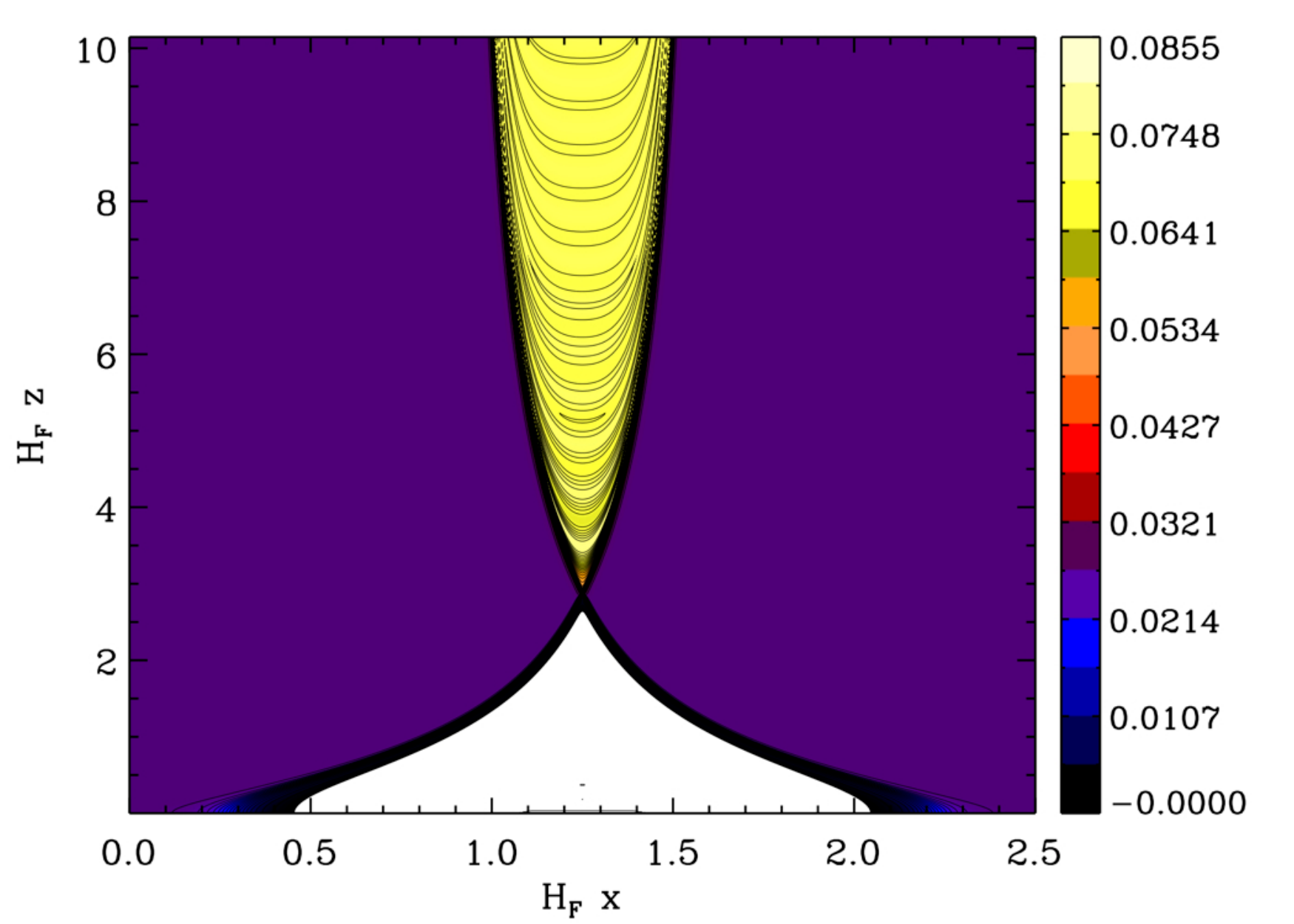}
   \includegraphics[width=8.9 cm]{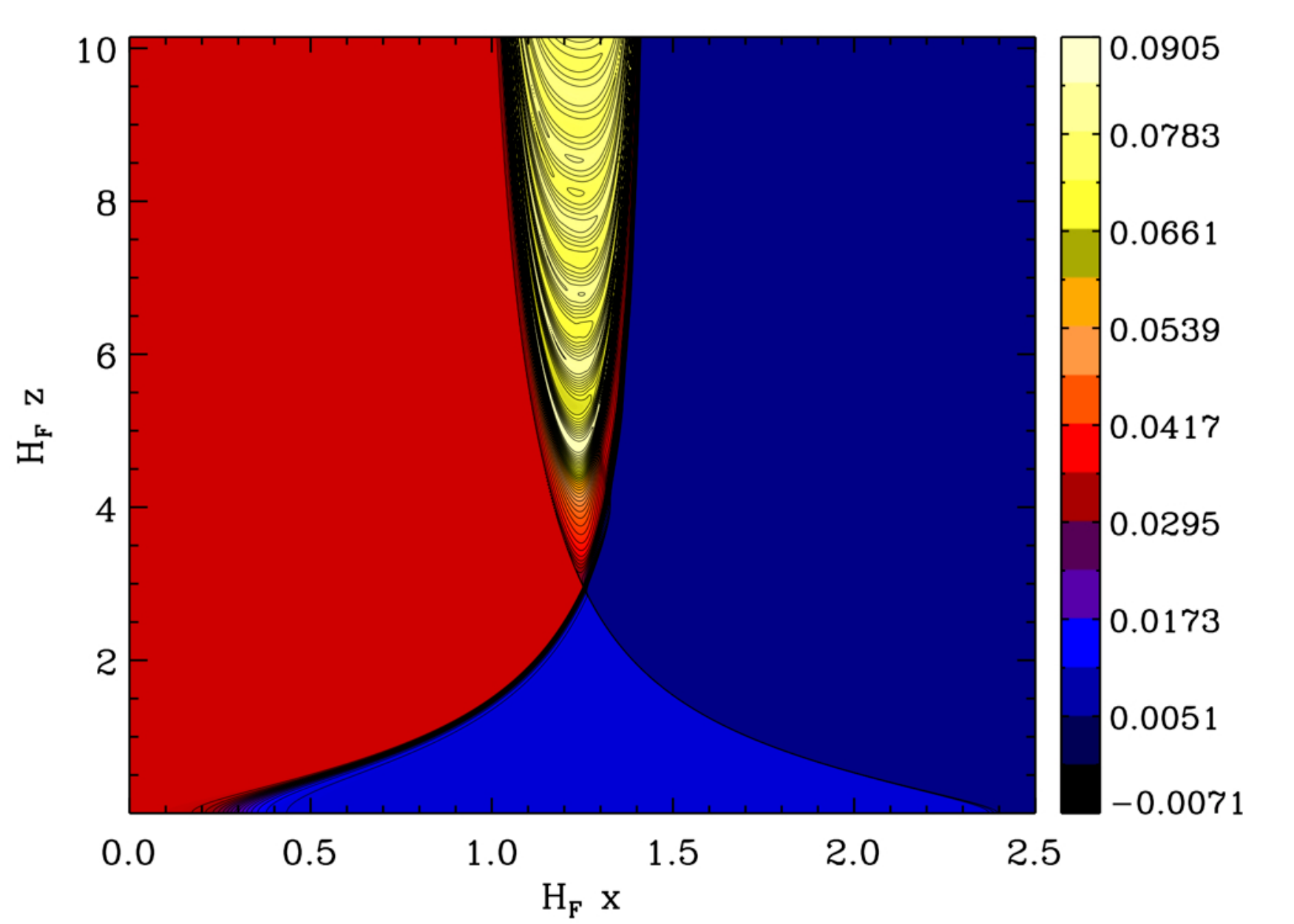}
   \includegraphics[width=8.9 cm]{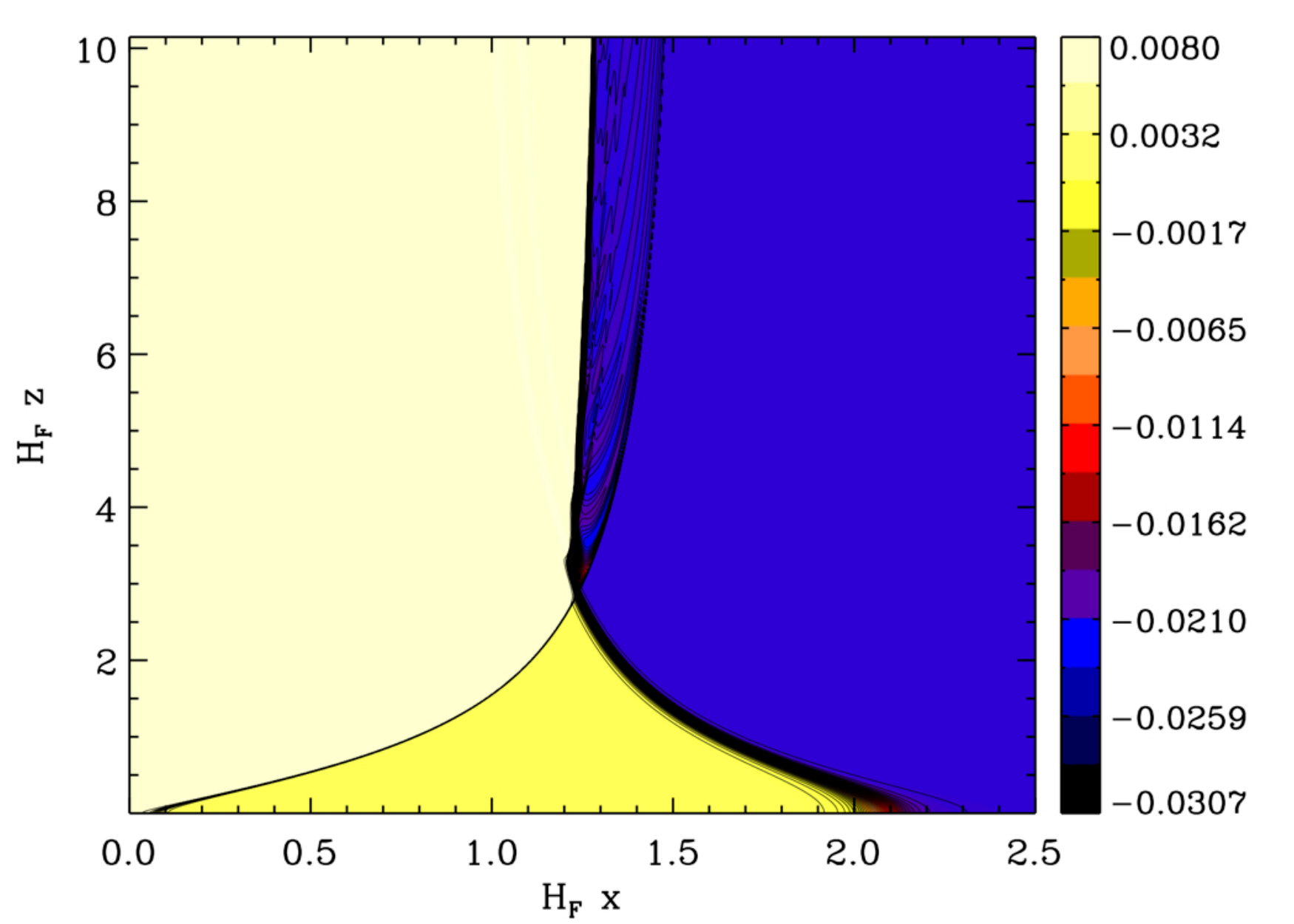}    
\caption{Contour plots of $\varphi$ for the collision of bubbles in the potential S1. On the top left, we simulate the collision between two identical bubbles. In the future light cone of the collision, inflation is completely disrupted (as can be seen by comparing the color bar to the field values on the potential in Fig.~\ref{fig:SRpots}). On the top right, we simulate the collision of two different bubbles in the case where the width of the barrier $T1$ is adjusted to be a quarter of the width of the barrier $T2$. Again, inflation is completely disrupted by the collision. On the bottom, we simulate the collision of two different bubbles in the case where the width of the barrier $T1$ is adjusted to be four times the width of the barrier $T2$. In this case, the inflaton inside the Observation bubble is hardly perturbed, and inflation continues to the future of the collision. In all cases, the boundaries of the simulation lie at the center of each bubble.}
\label{fig-inflection}
\end{figure}

In Fig.~\ref{fig-inflection}, we show contour plots for the collision of bubbles in the potential S1. On the top left is the case where the colliding bubbles are identical. In agreement with the non-gravitational simulations of Ref.~\cite{Aguirre:2008wy}, the collision completely ends inflation inside its future light cone. This is expected since the width of the barrier $T2$ is much wider than the region of the potential $C2$ where the slow-roll parameters are satisfied. In the top right panel, we adjust the barrier width $T1$ to be one fourth the width of the barrier $T2$. Again, inflation is completely disrupted by the collision. However, adjusting the barrier width $T1$ to be four times the width of the barrier $T2$, the inflaton is hardly disturbed, as shown in the bottom panel. We found a similar effect for the large-field models discussed above. Therefore, the sensitivity of inflection point models of inflation to bubble collisions depends quite sensitively on the structure of the potential barriers making up the colliding bubbles. 

We have also simulated collisions in the potential S2. In this case, the barrier width is somewhat smaller than the field excursion during inflation, and we expect that collisions will not completely disrupt inflation inside the Observation bubble. This is indeed what we observe in our simulations. The qualitative picture is nearly identical to the large-field cases discussed above. However, the duration of inflation inside the region affected by the collision will be somewhat less than it was in the large-field examples. 

Let us summarize the results of this section. For the collision between identical bubbles, the inflaton is displaced down the inflationary potential a distance comparable to the barrier width, in agreement with the free passage approximation. Inflation has a shorter duration to the future of the collision than in the undisturbed portions of the bubble. In large-field models, this displacement is rather small compared to the total field excursion. In small-field models, the displacement can be so large that inflation is completely disrupted. Increasing the initial bubble separation appears to have a relatively small effect on the number of $e$-folds in the future of the collision. For non-identical bubbles, the structure of the potential barriers is the dominant factor in determining the outcome of a collision. This is again in agreement with the free passage approximation. When the barrier is symmetric, breathing modes of the wall are excited. When the potential barrier $T1$ is thinner than the barrier $T2$, the post-collision domain wall is formed moving out of the Observation bubble without breathing modes. When the potential barrier $T1$ is thicker than the barrier $T2$, the inflaton inside the Observation bubble is minimally disturbed. Even the inflection point model of inflation (potential S1) can survive a collision in this case. If there is a cosmology in both colliding bubbles, the late time behavior of the post-collision domain wall depends on the details of the field evolution. For example, in the potential L1 the energy density in the Collision bubble quickly falls below that of the Observation bubble (which is undergoing slow-roll), causing the post-collision domain wall to have multiple turning points. For the post-collision domain wall to monotonically progress  out of the Observation bubble, the Collision bubble must be undergoing slow-roll at an energy higher than that of the Observation bubble. Finally, we note that in each of the models we have studied, there are always {\em fewer} $e$-folds of inflation to the future of the collision than in the undisturbed portions of the bubble. 

\section{Conclusions}

This work establishes the groundwork for rigorously determining the connection between an underlying potential landscape giving rise to eternal inflation and the cosmological signal of bubble collisions. Such a goal requires the ability to study the coupled Einstein and scalar field equations without approximations. To this end we have developed a robust code that implements the full system of relativistic field equations, thoroughly tested it on a number of cases and applied it to determine the outcome of collisions between bubbles.

Our results highlight the utility of two complementary analytic tools for predicting the outcome of bubble collisions. For the collision of vacuum bubbles, we have confirmed that the thin-wall solutions obtained from the Israel junction condition formalism approximate the numerical solutions to good accuracy. In particular, the junction condition solutions give a good quantitative prediction for the metric functions to the future of the collision, as well as a good qualitative prediction for the late-time behavior of the post-collision domain wall. The limitation is that the junction condition formalism cannot capture the dynamics of the collision event. This is particularly problematic when considering bubbles with an interior cosmology. The dynamics are, however, captured to a certain extent by the free passage approximation~\cite{Giblin:2010bd} which states that, in the immediate future of the collision, the field profiles making up the two bubbles simply superpose. This illustrates that the structure of the potential can be more important for determining the outcome of a collision than the kinematics. The free passage approximation breaks down shortly after the collision, at which point the full field dynamics must be considered. A full numerical solution is necessary to bridge the gap between the realm of validity of the free passage approximation and the junction condition formalism. 

Some of the main conclusions we have reached based on our numerical solutions are as follows:
\begin{itemize}
\item The energy released in the collision of identical vacuum bubbles goes mostly into the formation localized field configurations such as oscillons and pockets of the false vacuum. 
\item Classical transitions occur in the presence of gravity, and lasting regions of higher energy density can be created through gravitational effects.
\item The structure of the potential is the dominant factor in determining the immediate outcome of a collision. For the collision between identical bubbles, the field is initially displaced a distance equal to the barrier width towards its minimum. For the collision of non-identical bubbles, when the potential barriers are symmetric, the collision excites a breathing mode of the post-collision domain wall. For asymmetric barriers, the post-collision domain wall forms with some initial velocity in the direction of the thicker barrier, with no breathing modes.
\item We have demonstrated conclusively that slow-roll inflation can occur to the future of a collision. Large field models of inflation are more robust to collisions than small-field models, which are completely disrupted unless the potential barriers are highly asymmetric (with the barrier $T1$ much thicker than the barrier $T2$).
\item The direction in which the post-collision wall accelerates is determined not only by the height of the potential at the exit from the barrier, but also by the slope of the potential as the field tends towards the minimum. Therefore, if there is an inflationary potential in the Observation bubble, for the Collision bubble not to intrude, the interior of the latter is required to be either a vacuum (no cosmology) or an inflationary plateau that is either higher or broader than the one in the Observation bubble.
\item The amplitude of the perturbation to the inflaton is set mostly by the barrier widths in the potential, with only a mild dependence on the kinematics. In addition, for the examples we have studied, there are always fewer $e$-folds to the future of the collision than in the undisturbed portions of the bubble. 
\end{itemize}

Based on these observations, we can determine which potentials are most likely to give rise to cosmologies including bubble collisions that are compatible with our current observable universe, yet leave detectable signatures. Potentials resembling L2 in Fig.~\ref{fig:SRpots} are good candidates. Since this potential includes a large-field model of inflation, the collision does not disrupt cosmological evolution inside the bubble. Adjusting the relative width of the two barriers changes the amplitude of the perturbation to the inflaton. The intrinsic amplitude of the perturbation along with the total duration of inflation inside the Observation bubble determines the amplitude of the signal in the CMB or other cosmological observables.

While the existing code is able to accurately obtain the dynamics of the collision, it cannot yet be used for studying the ensuing cosmology: the uniform grid employed and coordinate conditions adopted are unable to track the increasingly narrow structures that develop around walls in a computationally efficient manner. To address this issue, we will extend our code to adopt adaptive mesh refinement (see e.g. Refs.~\cite{1989fnr..book..206C,1993PhRvL..70....9C} for a first example of its application in numerical relativity). We will present the derivation of the observational signatures from these simulations in a forthcoming publication.

\section*{Acknowledgments}
We thank Anthony Aguirre, Jim Cline, Richard Easther, Stephen Feeney, Eugene Lim, and David Neilsen for valuable input at various stages of this project. This work was partially supported by a grant from the Foundational Questions Institute (FQXi) Fund, a donor advised fund of the Silicon Valley Community Foundation on the basis of proposal FQXi-RFP3-1015 to the Foundational Questions Institute. The work was also supported by NSERC through a Discovery Grant. MCJ and HVP acknowledge the hospitality of the Benasque Science Center for Science, where this project was initiated, and UC Santa Cruz, for hospitality. MCJ thanks University College London, where part of this work was completed, for its hospitality. Research at Perimeter Institute is supported by the Government of Canada through Industry Canada and by the Province of Ontario through the Ministry of Research and Innovation. HVP is supported in part by Marie Curie grant MIRG-CT-2007-203314 from the European Commission, and by STFC and the Leverhulme Trust. LL is supported in part by CIFAR. 

\appendix 

\section{General equations of motion}\label{sec:eomappendix}

In this appendix, we derive the general form of the equations of motion 
assuming the line element Eq.~\ref{eq:generalmetric}. In this appendix, we work in units where $M_{\rm pl} = 1$. The results of this appendix are used in Sec.~\ref{sec:equationsofmotion} to derive the gauge-fixed equations of motion used in the simulations.

\subsubsection{Stress energy tensor and constraints}
We begin by finding the energy momentum tensor and field equations for the scalar field. The energy momentum tensor for the scalar field is given by:
\begin{equation}
T^{\mu \nu} = g^{\mu \lambda} g^{\nu \sigma} \partial_{\lambda} \varphi \partial_{\sigma} \varphi - g^{\mu \nu} \left[ \frac{g^{\alpha \beta}}{2} \partial_{\alpha} \varphi \partial_{\beta} \varphi + V(\varphi) \right] \, .
\end{equation}
The non-trivial components of this tensor are,
\begin{eqnarray}
T^{zz}
&=& \frac{\Phi^2 + \Pi^2}{2 a^2 \alpha^2} + \frac{V(\varphi)}{\alpha^2} \, , \\
T_{zx} 
&=& \frac{\alpha}{a} \Pi \Phi + \beta \frac{\Phi^2+\Pi^2}{2} - a^2 \beta V(\varphi)  \, , \\
T_{xx} 
&=& \frac{\Phi^2+\Pi^2}{2} - a^2 V(\varphi)  \, , \\
T_{\chi \chi} 
&=& - z^2 b^2 \left[ \frac{\Phi^2 - \Pi^2}{2 a^2} + V(\varphi) \right] \label{eq:Tchichi} \, .
\end{eqnarray}
where we have introduced the fields $\{\Phi,\Pi\}$ defined as,
\begin{eqnarray}
\Phi &\equiv& \varphi' \, , \\
\Pi &\equiv& \frac{a}{\alpha} \left( \dot{\varphi} - \beta \varphi' \right) \, .
\end{eqnarray}

The density, current and spatial stress associated with $T^{\mu \nu}$ determined by a 
normal observer are given by,
\begin{eqnarray}
\rho &\equiv& n_{\mu} n_{\nu} T^{\mu \nu} = \frac{\Phi^2 + \Pi^2}{2 a^2 } + V(\varphi) \\
J_i &\equiv& - n_{\mu} {T^{\mu}}_i  = -\frac{\Pi \Phi}{a} \delta^x_i \, , \\
{S^x}_x &\equiv& \gamma^{xk} T_{kx} = \frac{\Phi^2 + \Pi^2}{2 a^2} - V \, ,\\
{S^\chi}_\chi &\equiv& \gamma^{\chi k} T_{k \chi} = - \frac{\Phi^2 - \Pi^2}{2 a^2} - V \, .
\end{eqnarray}

The Hamiltonian and momentum constraint equations are given by:
\begin{eqnarray}
R - {K^i}_j {K^j}_i + K^2 = 16 \pi \rho \, , \\
D_j {K_i}^j - D_i K = 8 \pi J_i \, ; 
\end{eqnarray}
which for our problem of interest reduce to
\begin{eqnarray}
16 \pi \left[ \frac{\Phi^2 + \Pi^2}{2 a^2 } + V(\varphi) \right] &=&  -\frac{2}{z^2a^3b^2} \left[ a^3 - 2 z^2 b b' a' + z^2 a \left( b'^2+2bb''  \right) \right] + 4 {{K^x}_x} {K^\chi}_\chi + 2 {{K^\chi}_\chi}^2 \, , \\
\partial_x {K^\chi}_\chi + \frac{(b^2)'}{2 b^2}  \left[ {K^\chi}_\chi - {K^x}_x \right] &=& 4 \pi \frac{\Pi \Phi}{a} \, .
\end{eqnarray}
(only the $x$ component of the momentum constraint is non-trivial).

\subsubsection{Evolution equations}
The evolution equation for the scalar field  is given by
\begin{equation}
\square \varphi =\frac{1}{\sqrt{-g}} \partial_{\mu} \left( \sqrt{-g} g^{\mu \nu} \partial_{\nu} \varphi \right) = \partial_{\varphi} V \, ,
\end{equation}
This yields the evolution equation for $\Pi$,
\begin{equation}
\dot{\Pi} = \left( \frac{2 \beta b'}{b} + \beta' - \frac{2}{z} - \frac{2 \dot{b}}{b} \right) \Pi + \beta \Pi' + \left( \frac{2 b' \alpha}{a b} + \frac{\alpha'}{a} - \frac{a' \alpha}{a^2} \right) \Phi + \frac{\alpha}{a} \Phi' - \alpha a \partial_\varphi V \, .
\end{equation}
The evolution equation for $\Phi$ is obtained by taking the spatial derivative of $\dot{\varphi}$:
\begin{equation}
\dot{\Phi} = \partial_x \left( \beta \Phi + \frac{\alpha}{a} \Pi \right) \, , 
\end{equation}
where, in terms of our variables, 
\begin{eqnarray}
\partial_\varphi V &=&  \frac{1}{a \alpha b^2 z^2} \partial_z \left[  a \alpha b^2 z^2  \left( g^{zz} \partial_z \varphi + g^{zx} \partial_x \varphi \right)  \right] +  \frac{1}{a \alpha b^2} \partial_x \left[  a \alpha b^2 \left( g^{xx} \partial_x \varphi + g^{xz} \partial_z \varphi \right)  \right] \\
&=& - \frac{1}{a \alpha b^2 z^2} \partial_z \left[  b^2 z^2  \Pi \right] +  \frac{1}{a \alpha b^2} \partial_x \left[  \alpha b^2 \left( \frac{\Phi}{a} + \frac{\beta}{\alpha} \Pi  \right)  \right] \\
&=& - \frac{1}{a \alpha} \dot{\Pi} -  \frac{2}{z a \alpha} \Pi - \frac{2 \dot{b}}{a \alpha b} \Pi  + \frac{2 \beta b'}{a \alpha b} \Pi + \frac{\beta'}{a \alpha} \Pi + \frac{\beta}{a \alpha} \Pi' - \frac{a'}{a^3} \Phi + \frac{2 b'}{a^2 b} \Phi + \frac{a'}{a^2 \alpha} \Phi + \frac{1}{a^2} \Phi' \, .
\end{eqnarray}
This, together with the trivial equation
\begin{equation}
\dot{\varphi} = \frac{\alpha}{a} \Pi + \beta \Phi \, ,
\end{equation}
completely determine the equations of motion for the scalar field $\varphi$.

The evolution equations for the three-metric are obtained from the definition of the extrinsic curvature. The $z-z$ and $\chi-\chi$ components yield:
\begin{eqnarray}
\dot{a} &=& - \alpha a {K^{x}}_x + \beta a' + a \beta' \, , \\
\dot{b} &=& - \alpha b {K^{\chi}}_{\chi} + \beta b' - \frac{b}{z} \, .
\end{eqnarray}

Einstein equations determine the evolution equations for the extrinsic curvature,
\begin{equation}
{{\dot{K}^i}}_j = \beta^k \partial_k {K^{i}}_j - \partial_k \beta^i {K^k}_j + \partial_j \beta^k {K^i}_k - D^i D_j \alpha + \alpha \left( {R^i}_j + K {K^i}_j + 4 \pi (S-\rho) {\delta^i}_j - 8 \pi {S^i}_j \right) \, .
\end{equation}
which reduce in our case to,
\begin{eqnarray}
\dot{{K^x}_x} &=& \beta {{K^x}_x}' - \frac{1}{a} \left( \frac{\alpha'}{a} \right)' + \alpha \left( - \frac{2}{ab} \left[ \frac{b'}{a} \right]' + K {K^x}_x - 8 \pi \frac{\Phi^2}{a^2} - 8 \pi V(\varphi)  \right) \, , \\
\dot{{K^\chi}_\chi} &=& \beta {{K^\chi}_\chi}' - \frac{\alpha'b'}{a^2 b} + \alpha \left[ -\frac{1}{z^2 b^2} + \frac{-a b'^2 + b (a'b'-ab'')}{a^3 b^2} + K K^\chi_\chi  - 8 \pi V \right] \, .
\end{eqnarray}

\section{Recovering hyperbolic dS}\label{sec:HdSsection}

The system of equations Eq.~\ref{eq:adot},~\ref{eq:alphadot},~\ref{eq:pi} and~\ref{eq:phi} should recover the hyperbolic foliation of dS space in the case where $\Phi = \Pi = 0$ and $V = {\rm const.}$ In this appendix, we work in units where $M_{\rm pl} = 1$. The metric in hyperbolic dS is given by:
\begin{equation}
ds^2 = - (1+H^2 z^2)^{-1} dz^2 + (1+H^2 z^2) dx^2 + z^2 dH_2^2 \, .
\end{equation}
The evolution equation for $b$, Eq. (\ref{basicbdot}), defines
\begin{equation}
{K^\chi}_\chi = - \frac{1}{\alpha z} \, . \label{eq:appKchichi}
\end{equation}
This expression can be replaced in the Hamiltonian constraint:
\begin{equation}
16 \pi V(\varphi)  =  -\frac{2}{z^2}+ 4 {{K^x}_x} {K^\chi}_\chi + 2 {{K^\chi}_\chi}^2 \, ,
\end{equation}
to determine an equation for ${K^x}_x$:
\begin{equation}
{K^x}_x = \frac{1- \alpha^2}{2 \alpha z} - \alpha z 4 \pi V (\varphi) \, .
\end{equation}
This can now be exploited by noting the momentum constraint,
\begin{equation}
\partial_x {K^x}_x = 0\, ,
\end{equation}
implies that $\alpha = \alpha (z)$. One can then find $\alpha(z)$ by solving 
the evolution equation for ${K^{\chi}}_{\chi}$: 
\begin{equation}
\dot{{K^\chi}_\chi} = \frac{3 - (1+8 \pi V(\varphi) z^2)}{2 z^2 \alpha} \, .
\end{equation}
Exploiting again Eq. (\ref{eq:appKchichi}), one obtains an equation for $\alpha$:
\begin{equation}
\dot{\alpha} = \frac{\alpha - (1+8 \pi V(\varphi) z^2) \alpha^3}{2 z} \, .
\end{equation}
The general solution of this equation is:
\begin{equation}
\alpha = \left( 1 + \frac{8 \pi V}{3} z^2 - \frac{2 M}{z} \right)^{-1/2}\, ,
\end{equation}
where we have set an integration constant to $M$. For the case $M=0$, 
and identifying $H^2 = 8 \pi V(\varphi) / 3$, this is the metric function we expected. 

The evolution equation for $a$ yields:
\begin{equation}
\dot{a} = - \alpha a {K^{x}}_x = \frac{a (\alpha^2-1)}{2 z} +a \alpha^2 z 4 \pi V (\varphi) \, .
\end{equation}
Substituting now with the expression obtained for $\alpha$,
\begin{equation}
\dot{a} = \frac{8 \pi z V(\varphi)}{3 + 8 \pi z^2 V(\varphi)} a \, ,
\end{equation}
whose solution is given by:
\begin{equation}
a = \left( 1 + \frac{8 \pi V}{3} z^2 - \frac{2 M}{z} \right)^{1/2} \, ,
\end{equation}
which is again the expected solution.

\section{Field equations and characteristic structure in hyperbolic de Sitter}\label{sec:fieldeqnsHdS}

The equation of motions for a scalar field configuration possessing hyperbolic symmetry 
in a background HdS spacetime is given, in terms of the first order variables $\{\varphi,\Pi,\Phi\}$ by:
\begin{eqnarray}\label{eq:Hbgrndeom}
\dot{\varphi} &=& (1+H^2 z^2)^{-1} \Pi \\
\dot{\Pi} &=& (1+H^2 z^2)^{-1} \Phi' - \frac{2}{z} \Pi - \partial_{\varphi} V(\varphi) \\ 
\dot{\Phi} & = & (1+H^2 z^2)^{-1} \Pi' \, .
\end{eqnarray}
having identified $H^2 = 8 \pi V/3$. The eigenvalues of the principal part of these
equations determine the characteristic speeds of the system, which are given by
\begin{equation}
c = \pm (1+H^2 z^2)^{-1} \, ; \, 0 \, .
\end{equation}
Thus, the physical (non-zero) characteristic speeds range from unity at small $z$ to 
approaching zero as $z$ becomes large compared to $H$.

\section{Numerical algorithm for finding CDL instantons}\label{sec:numericalCDL}

In this appendix, we outline a numerical method for determining the initial data for single bubbles. The initial data for a single bubble spacetime is given by the analytic continuation of the Coleman-de Luccia (CDL) instanton~\cite{Coleman:1980aw}. The CDL instanton is a solution to the Euclidean Einstein and field equations that describes the formation of a bubble. First, we outline a convenient set of dimensionless variables and the associated equations of motion. Then, we describe our numerical algorithm for finding the CDL instanton. Finally, as an aside, we apply our numerics to test the validity of the thin-wall approximation. 

\subsection{Variables and equations of motion}

Consider the class of potentials defined by
\begin{equation}
V(\varphi ) = \mu^4 v\left( \varphi /M \right)\, ,
\end{equation}
We define the following dimensionless variables:
\begin{equation}\label{eq:CDLvariables}
x = \frac{\varphi}{M}\, , \,  r \equiv \frac{\mu^2 \rho}{M} \, , \,
s \equiv \frac{\mu^2 t}{M} \, , \,
\epsilon^2 \equiv \frac{8 \pi M^2}{3 M_{\rm Pl}^{2}} \, .
\end{equation}
Substituting these into the Euclidean Einstein (Eq.~\ref{eq:instanton}) and field (Eq.~\ref{eq:eucfield}) equations, we obtain:
\begin{equation}\label{ddotx}
\ddot{x} + \frac{3 \dot{r}}{r}\dot{x} - v' = 0\, ,
\end{equation}
\begin{equation}\label{dotr2}
\ddot{r} = - \epsilon^2 r \left( \dot{x}^2 +v \right) \, ,
\end{equation}
where the primes and dots, respectively, refer to $x-$ and $s-$derivatives. To find the CDL instanton, we must solve a double-boundary value problem with initial and final conditions given by:
\begin{equation}
x(s=0) \simeq x_T, \ \ \dot{x}(s=0) = 0, \ \ r(s=0) = 0, \ \ x(s=s_{\rm max}) \simeq x_F, \ \ \dot{x}(s=s_{\rm max}) = 0, \ \ r(s=s_{\rm max}) = 0\, .
\end{equation}
Such solutions interpolate between the basin of attraction of the true and false vacuum as the scale factor $r$ goes between its two zeroes at $s=0$ and $s=s_{\rm max}$. An example is shown in Fig.~\ref{fig-inst_examples}.

\begin{figure*}
   \includegraphics[width=8 cm]{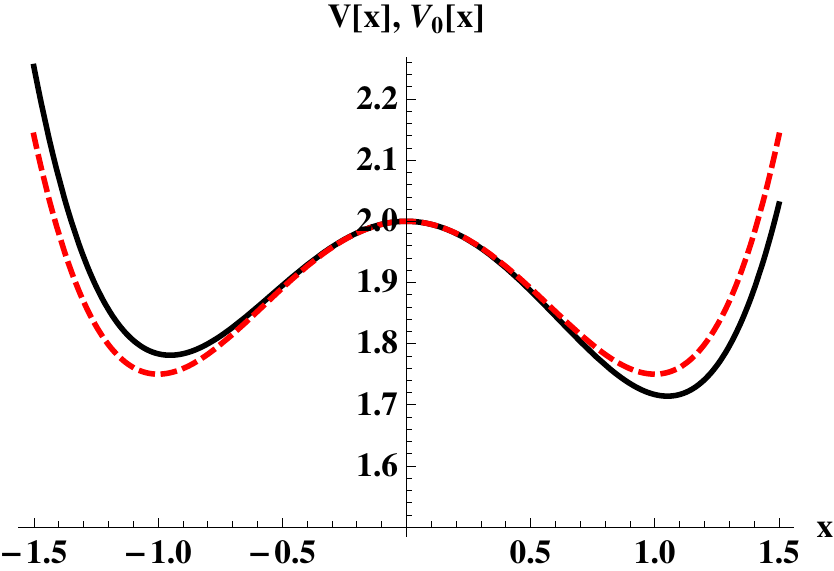}
\caption{$v$ (black) and $v_0$ (red dashed) with $a=0.1$ and $C=2$}
\label{fig-VV0}
\end{figure*}

In this appendix, we consider potentials of the form
\begin{equation}\label{eq:V}
v = -\frac{x^2}{2} + \frac{x^4}{4} + C  - \frac{a x^3}{3} \, ,
\end{equation}
where $C$ and $a$ are constants. An example is shown in Fig.~\ref{fig-VV0}.

Since the instanton solution interpolates between the false and true vacua in the classically forbidden region of the potential, we can use this solution to obtain the probability per unit four-volume that a bubble forms from the false vacuum in the WKB approximation. This is given by~\cite{Coleman:1980aw}:
\begin{equation}
\Gamma = A e^{-B},
\end{equation}
where
\begin{equation}\label{eq:B}
B = S_{I} - S_{BG}\, .
\end{equation}
The first term is the Euclidean action, evaluated for the instanton solution. Substituting the equations of motion into the Euclidean action for a canonically normalized scalar field minimally coupled to gravity, the instanton action is given by
\begin{equation}\label{eq:action}
S_{I} = -4 \pi^2 \left( \frac{M^4}{\mu^4} \right) \int_{s=0}^{s=s_{\rm max}} ds \left( - r^3 v + \frac{r}{\epsilon^2}\right)\, .
\end{equation}
where $r(s)$ is the instanton solution. The second term in Eq.~\ref{eq:B} is known as the background subtraction, and is simply the Euclidean action of the false vacuum de Sitter space: 
\begin{equation}\label{SBG}
S_{BG} = - \frac{8 \pi^2}{3 \epsilon^4 v_F}\, .
\end{equation}

\subsection{Numerical strategy}

To solve for the CDL instanton, we utilize the over/under-shoot argument of Coleman. We have formulated a numerical implementation of the algorithm described in Ref.~\cite{Aguirre:2006ap}, which involves the following steps: 
\begin{enumerate}
\item Start the field at $x_0 = x_T - 10^{-12}$. We start the numerical evolution at $s=ds$ to avoid the singular nature of the friction term at $r=0$. The initial conditions for the evolution at $s=ds$ can be solved for by Taylor expanding the solution in $s$ near $s=0$. The first non-trivial dependence on the potential arises at 2nd and 3rd order respectively for $x$ and $r$. Expanding to 3rd order, and solving for the coefficients in the Taylor expansion
(the first and third-order terms in $x$ are zero; the 2nd order term in $r$ is zero), we obtain:
\begin{equation}
x(ds) = x_0 + \frac{\partial_x v(x_0)  ds^2}{8}, \ \ \dot{x}(ds) = \frac{\partial_x v(x_0)  ds}{4}, \ \ r(ds) = ds - \frac{\epsilon^2 v(x_0) ds^3}{6}.
\end{equation}

We then evolve the field and test for an overshoot. 
\item If an overshoot is found on the original test:
\begin{enumerate}
\item Define $x_L = x_{\rm max}$ and $x_R = x_{T}$, then bisect and set $x_0$ equal to the midpoint: $x_0 = x_L + (x_R - x_L)/2$. If there is an overshoot, $x_0$ is too close to $x_R$, so redefine $x_R = x_0$ and test again. If there is an undershoot, $x_0$ is too close to $x_L$, so redefine $x_L = x_0$ and test again. Repeat these tests until $(x_R - x_L)/2$ is at a pre-specified tolerance (we use $10^{-14}$, which is of order machine precision). In doing so, we iterate to the solution of the instanton end-point in the vicinity of the true vacuum.
\item The evolution is terminated either when $x=x_F$ or $\dot{x}$ changes sign. Either way, the scale factor has not yet reached its second zero. Therefore, we match onto a solution where:
\begin{equation}
x = x(s_{\rm end}), \ \ \ r(s) = \frac{1}{\epsilon \sqrt{v(x_F)}} \sin \left[ \epsilon \sqrt{v(x_F)} (s - s^*)  \right]
\end{equation}
with
\begin{eqnarray}
s^* &=& s_{\rm end} - \arcsin \left[ \epsilon \sqrt{v(x_F)} r(s_{\rm end}) \right] /(\epsilon \sqrt{v(x_F)}), \ \ \dot{r} > 0 \\
s^* &=& s_{\rm end} - \pi + \arcsin \left[ \epsilon \sqrt{v(x_F)} r(s_{\rm end}) \right] /(\epsilon \sqrt{v(x_F)}), \ \ \dot{r} < 0 
\end{eqnarray}
where we have been careful to take the correct root of the $\arcsin$ function. This solution is used until $r=0$.
\end{enumerate}
\item If an overshoot is not found on the original test, the field must be too close to the true vacuum for us to resolve numerically.
\begin{enumerate}
\item We use the solution
\begin{equation}
x = x_T, \ \ \ r(s) =   \frac{1}{\epsilon \sqrt{v(x_T)}} \sin \left[ \epsilon \sqrt{v(x_T)} s \right]
\end{equation}
for a time $\Delta s_1$, substitute in the new initial conditions for $x$ and $r$ at that time, and test for overshoot. If no overshoot is found, use the solution for another period $\Delta s_1$, and re-test for overshoot. Repeat until overshoot is found.
\item Now, find the instanton end-point by bisection as above.
\item As above, find the portion of the solution near $x_F$.
\end{enumerate}
\item Calculate the instanton action and export the initial data.
\end{enumerate}

The output of the instanton calculation is the post-tunnelling field configuration:
\begin{equation}
\frac{\varphi (t = \frac{M}{\mu^2} s) }{M} \, .
\end{equation}
We then convert the solution into units appropriate to the simulation:
\begin{equation}
\tilde{\varphi} = \frac{M}{M_\mathrm{Pl}} \frac{\varphi}{M} = \left( \frac{3 \epsilon^2}{8 \pi} \right)^{1/2} \frac{\varphi}{M} \, .
\end{equation}
Additionally, we have 
\begin{eqnarray}
\tilde{t} = \epsilon \sqrt{v(\varphi_F)} \ s \, \\
\tilde{V} = \frac{3}{8 \pi} \frac{v(\varphi)}{v(\varphi_F)} \, .
\end{eqnarray}

\subsection{Checking the thin-wall approximation}

Potentials where the field loiters in a close neighborhood of the true vacuum for a time much longer than the Euclidean transition time from the true to the false vacuum are known to satisfy the ``thin-wall" approximation~\cite{Coleman:1980aw}. In this approximation, the bubble wall is assumed to be a rigid membrane with tension $\sigma$. The instanton can be constructed by pasting a portion of a true vacuum dS 4-sphere to a portion of a false vacuum dS 4-sphere. The tension accounts for the discontinuity in the derivative of the metric across this junction. We can check our numerical solutions against the thin-wall approximation. In this limit, equations for the critical radius of the bubble and tunnelling exponent were first derived in Ref.~\cite{Parke:1982pm}, and are given by
\begin{equation}\label{eq:thinwallcriticalradius}
r_{\rm crit}^2 = \frac{9 \sigma^2}{ \left(v(x_F) - v(x_T)\right)^2 } \times \left[ 1 + 2 \left( \frac{3 \epsilon \sigma}{2} \frac{( v(x_F) -v(x_T) )^{1/2}}{(v(x_F) + v(x_T) )} \right)^2 + \left( \frac{3 \epsilon \sigma}{2 ( v(x_F) - v(x_T) )^{1/2} }  \right)^4  \right]^{-1}
\end{equation}
and
\begin{equation}\label{eq:thinwallexponent}
B = \frac{M^4}{\mu^4} \frac{27 \pi^2 \sigma^4}{2 (v(x_F) - v(x_T))^3 } \times r \left[ \left( \frac{3 \epsilon \sigma}{2 ( v(x_F) - v(x_T) )^{1/2} }  \right)^2 , \frac{ v(x_F) + v(x_T) }{ v(x_F) - v(x_T) } \right] \, ,
\end{equation}
where
\begin{equation}
r(x,y) = \frac{2 \left[ \left( 1 + xy \right) - \sqrt{ 1+ 2xy +x^2 }  \right]}{ x^2 (y^2 - 1) \sqrt{ 1+ 2xy +x^2 }  } \, .
\end{equation}

The vacuum energy is determined by the potential, and therefore to evaluate the thin-wall formulas above we need only calculate the tension, defined by:
\begin{equation}\label{eq:tension}
\sigma = 2 \int ds (v_0 [\phi(s)] - v_0 [\phi_F]    )  \, ,
\end{equation}
where $v_0$ is a potential without the symmetry-breaking terms. For potentials of the form Eq.~\ref{eq:V}, $v_0$ is just
\begin{equation}
v_0 = -\frac{x^2}{2} + \frac{x^4}{4} + C \, .
\end{equation}
This is shown in Fig.~\ref{fig-VV0}. We obtain slightly different estimates for the tension by changing the limits of integration (the numerical values we obtain are $\sigma_{\rm low} = 0.94$ and $\sigma_{\rm high} = 1.02$). The numerical value of the tunnelling exponent is quite sensitive to any errors, since it depends on the fourth power of the tension. 

\begin{figure*}
   \includegraphics[width=8 cm]{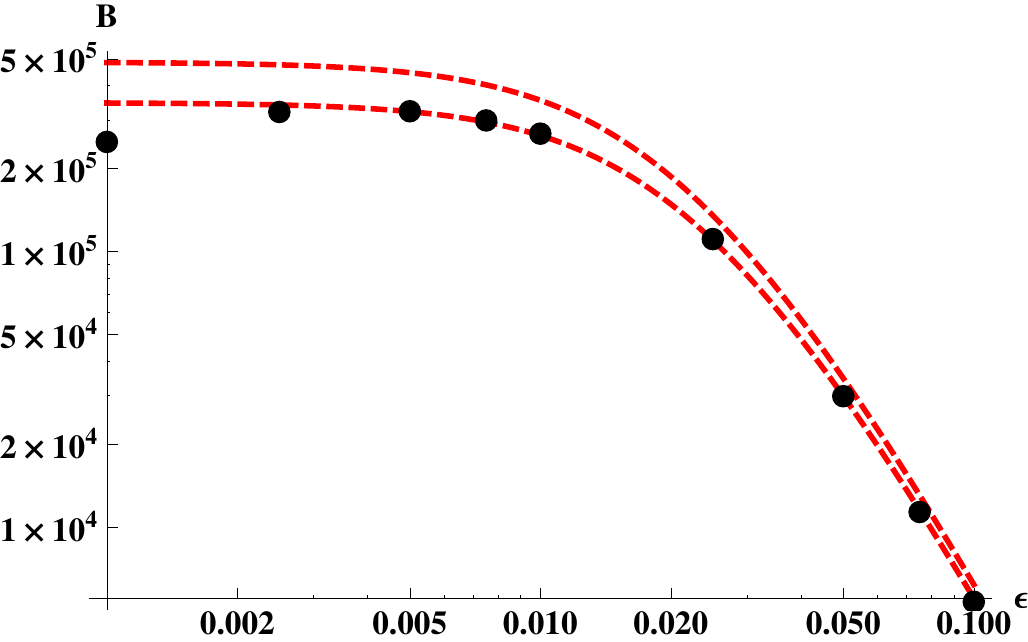}
\caption{The tunnelling exponent as a function of $\epsilon$ computed for a few examples numerically (dots) and from Eq.~\ref{eq:thinwallexponent} using the upper and lower estimates of the tension (red-dashed lines).}
\label{fig-actionvsepsilon}
\end{figure*}

\begin{figure*}
   \includegraphics[width=8 cm]{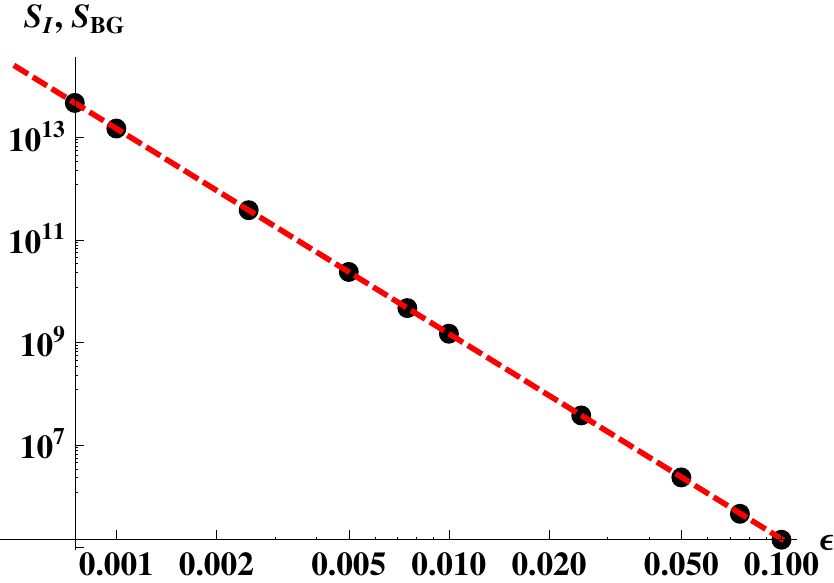}
\caption{The numerically computed instanton action (dots) plotted against the background subtraction (red-dashed line). Both are growing like $\epsilon^{-4}$ as $\epsilon \rightarrow 0$. }
\label{fig-sibgepsilon}
\end{figure*}

In Fig.~\ref{fig-actionvsepsilon}, we compare the tunnelling exponent obtained from Eq.~\ref{eq:thinwallexponent} to that obtained by  integrating the numerically determined instanton solution (using Eq~\ref{eq:action} and Eq.~\ref{SBG}) as a function of $\epsilon$. The data points follow the curve well for large values of $\epsilon$, but then fall off for smaller $\epsilon$ due to truncation error. In Fig.~\ref{fig-sibgepsilon}, the numerically computed instanton action (Eq~\ref{eq:action}) and the background subtraction are plotted separately. Both  grow like $\epsilon^{-4}$ as $\epsilon \rightarrow 0$ as expected, and must cancel increasingly more precisely as $\epsilon$ goes to zero (where the instanton action should approach a constant).

Our numerical solutions corroborate the validity of the thin-wall equations for the critical radius and tunnelling exponent. As far as the authors are aware, this is the first explicit numerical test of the validity of the thin-wall approximation. This result contradicts recent claims~\cite{Copsey:2011wy,Copsey:2011zj} in the literature that Eq.~\ref{eq:thinwallexponent} does not correctly reproduce the instanton action Eq.~\ref{eq:action} computed using full solutions to the equations of motion. For all of the examples we have studied, the thin-wall formulas reproduce the numerically computed tunnelling exponents to high accuracy. We conclude that the thin-wall approximation is indeed valid and highly accurate.

\section{Potential Parameters}\label{sec:potentialpars}
We present here the parameters used to generate the potentials defined in Eq.~\ref{eq:pot1} and shown in Figs.~\ref{fig:vacpots} and~\ref{fig:SRpots}. These are specified in
 Table~\ref{tab:potparams}.

\begin{table*}[h]
\begin{tabular}[t]{c c c c c c c c c c c c}
\hline
\hline
 type \ & \ $a_1$ \ & \  $a_2$ \ & \ $M$ \ & \ $V_{01}$ \ & \ $\varphi_{m1}$ \ & \ $\varphi_{m2}$ \ & \ $\varphi_{j1}$ \ & \ $\varphi_{j2}$ \ & \ $C_1$ \ & \ $\nu$  \ & \ $\lambda$\\
\hline
$V1$ & $0.5$ & $0.75$ & $3.45 \times 10^{-3}$ & $5.0 \times 10^{-11}$ & N/A & N/A & N/A & N/A & $2.0$ & N/A & N/A \\
$V2$ & $0.749$ & $0.75$ & $3.45 \times 10^{-3}$ & $5.0 \times 10^{-11}$ & N/A & N/A & N/A & N/A & $2.0$ & N/A & N/A \\
$V3$ & $-0.5$ & $0.50$ & $3.45 \times 10^{-3}$ & $1.0 \times 10^{-12}$ & N/A & N/A & N/A & N/A & $1.0$ & N/A & N/A \\
$V4$ & $-0.5$ & $1.0 \times 10^{-6}$ & $3.45 \times 10^{-3}$ & $1.0 \times 10^{-12}$ & N/A & N/A & N/A & N/A & $0.51$ & N/A & N/A \\
$L1$ & $0.5$ & $0.75$ & $3.45 \times 10^{-3}$ & $5.0 \times 10^{-11}$ & $-0.08$ & $2.75$ & $-0.01$ & $0.015$ & $2.0$ & N/A & N/A \\
$L2$ & $0.5$ & $0.75$ & $3.45 \times 10^{-3}$ & $5.0 \times 10^{-11}$ & N/A & $2.75$ & N/A & $0.015$ & $2.0$ & N/A & N/A \\
$S1$ & $0.5$ & $0.75$ & $3.45 \times 10^{-3}$ & $5.0 \times 10^{-11}$ & $-0.08$ & $2.75$ & $-0.01$ & $0.015$ & $2.0$ & $5.0 \times 10^{-3}$ & $8.7 \times 10^{-5}$ \\
$S2$ & $0.5$ & $0.75$ & $3.45 \times 10^{-3}$ & $5.0 \times 10^{-11}$ & N/A & $0.6$ & N/A & $0.5$ & $2.0$ & N/A & N/A \\
\hline
\hline
 \end{tabular} 
  \begin{center}
 \caption{The parameters used to generate the potentials determined by Eq.~\ref{eq:pot1} and shown in Figs.~\ref{fig:vacpots} and~\ref{fig:SRpots}.
   \label{tab:potparams}}
 \end{center}
\end{table*}

\section{Size of the simulation and resolution requirements}\label{sec:box_size}

Here, we discuss the maximum size of the simulation box necessary to contain two colliding bubbles. Curves of constant acceleration in a background HdS space take the following form when centered about $x=0$:
\begin{equation}
\tilde{x} = \arccos \left[ \frac{\cos \tilde{x_0}}{\sqrt{1+\tilde{z}^2}} \right]\, .
\end{equation}
At large $z$, this asymptotes to $\tilde{x} = \pm \pi/2$ independent of the value of $x_0$. Now, consider the wall of the Observation bubble at $\Delta \tilde{x}_o$ and the wall of a colliding bubble of initial radius $\Delta \tilde{x}_c$ centered about $\tilde{x} = \tilde{x}_c$. The two curves are parameterized by:
\begin{equation}
\tilde{x} = \arccos \left[ \frac{\cos \Delta \tilde{x}_o}{\sqrt{1+\tilde{z}^2}} \right], \ \ \ \tilde{x} = \tilde{x}_c +  \arccos \left[ - \frac{\cos \Delta \tilde{x}_c}{\sqrt{1+\tilde{z}^2}} \right]\, .
\end{equation}
The maximum separation between the centers that allows for a collision occurs for $x_c = \pi$. We can also conclude that the size of the box that entirely contains both bubbles is given by
\begin{equation}
-\frac{\pi}{2} \leq \tilde{x} \leq \tilde{x}_c + \frac{\pi}{2} \, .
\end{equation}

Since the bubbles are embedded in an expanding spacetime, the resolution in terms of physical distance is decreasing with $z$. The physical distance between two points at constant $z$ is given from the metric by:
\begin{equation}
H_F \Delta s =  \sqrt{1 + \tilde{z}^2} \Delta \tilde{z} \, .
\end{equation}
At large-$z$, the resolution in terms of physical coordinates is decreasing roughly linearly. 

To accurately simulate bubble collisions, we must have enough resolution to describe the bubble walls at the time of the collision. We can estimate the required resolution as follows. If we define the wall by an inner and outer surface of constant field, we can make the approximation that they follow lines of constant acceleration in the background false vacuum de Sitter space. If these constant field surfaces are at the positions $\tilde{x}_{\rm in}$ and $\tilde{x}_{\rm out}$ on the $z=0$ surface, they are separated by a distance
\begin{equation}
\Delta \tilde{x} (z) = \arccos \left[  \frac{\cos \tilde{x}_{\rm out}}{\sqrt{1+\tilde{z}^2}} \right] -  \arccos \left[  \frac{\cos \tilde{x}_{\rm in}}{\sqrt{1+\tilde{z}^2}} \right] \simeq \frac{1}{z} \left( \cos \tilde{x}_{\rm in} - \cos \tilde{x}_{\rm out}  \right) \, ,
\end{equation}
 where the last equality holds at large $\tilde{z}$. We can therefore estimate that if $N_0$ grid points are necessary to resolve the wall at small-$z$, we need $z_{c} N_0$ grid points to resolve the wall at the time of the collision, and  $z_{\rm max} N_0$ grid points to resolve the walls everywhere in the simulation. For our standard resolution of $32 \times 10^{3}$ spatial grid points, there are typically on the order of $100$ spatial grid points describing each bubble wall on the initial timeslice. Therefore, to retain tens of grid points in the wall at late times, runs are limited to a total duration of order $z \sim 10$.

\bibliography{NRbubbles}

\begin{thebibliography}{63}
\expandafter\ifx\csname natexlab\endcsname\relax\def\natexlab#1{#1}\fi
\expandafter\ifx\csname bibnamefont\endcsname\relax
  \def\bibnamefont#1{#1}\fi
\expandafter\ifx\csname bibfnamefont\endcsname\relax
  \def\bibfnamefont#1{#1}\fi
\expandafter\ifx\csname citenamefont\endcsname\relax
  \def\citenamefont#1{#1}\fi
\expandafter\ifx\csname url\endcsname\relax
  \def\url#1{\texttt{#1}}\fi
\expandafter\ifx\csname urlprefix\endcsname\relax\def\urlprefix{URL }\fi
\providecommand{\bibinfo}[2]{#2}
\providecommand{\eprint}[2][]{\url{#2}}

\bibitem[{\citenamefont{Coleman}(1977)}]{Coleman:1977py}
\bibinfo{author}{\bibfnamefont{S.~R.} \bibnamefont{Coleman}},
  \bibinfo{journal}{Phys. Rev.} \textbf{\bibinfo{volume}{D15}},
  \bibinfo{pages}{2929} (\bibinfo{year}{1977}).

\bibitem[{\citenamefont{Coleman and De~Luccia}(1980)}]{Coleman:1980aw}
\bibinfo{author}{\bibfnamefont{S.~R.} \bibnamefont{Coleman}} \bibnamefont{and}
  \bibinfo{author}{\bibfnamefont{F.}~\bibnamefont{De~Luccia}},
  \bibinfo{journal}{Phys. Rev.} \textbf{\bibinfo{volume}{D21}},
  \bibinfo{pages}{3305} (\bibinfo{year}{1980}).

\bibitem[{\citenamefont{Aguirre}(2008)}]{Aguirre:2007gy}
\bibinfo{author}{\bibfnamefont{A.}~\bibnamefont{Aguirre}}, in
  \emph{\bibinfo{booktitle}{Beyond the Big Bang}}
  (\bibinfo{publisher}{Springer}, \bibinfo{year}{2008}).

\bibitem[{\citenamefont{Aguirre et~al.}(2007)\citenamefont{Aguirre, Johnson,
  and Shomer}}]{Aguirre:2007an}
\bibinfo{author}{\bibfnamefont{A.}~\bibnamefont{Aguirre}},
  \bibinfo{author}{\bibfnamefont{M.~C.} \bibnamefont{Johnson}},
  \bibnamefont{and} \bibinfo{author}{\bibfnamefont{A.}~\bibnamefont{Shomer}},
  \bibinfo{journal}{Phys. Rev.} \textbf{\bibinfo{volume}{D76}},
  \bibinfo{pages}{063509} (\bibinfo{year}{2007}), \eprint{arXiv:0704.3473
  [hep-th]}.

\bibitem[{\citenamefont{Gott and Statler}(1984)}]{Gott:1984ps}
\bibinfo{author}{\bibfnamefont{J.~R.} \bibnamefont{Gott}} \bibnamefont{and}
  \bibinfo{author}{\bibfnamefont{T.~S.} \bibnamefont{Statler}},
  \bibinfo{journal}{Phys. Lett.} \textbf{\bibinfo{volume}{B136}},
  \bibinfo{pages}{157} (\bibinfo{year}{1984}).

\bibitem[{\citenamefont{Garriga et~al.}(2007)\citenamefont{Garriga, Guth, and
  Vilenkin}}]{Garriga:2006hw}
\bibinfo{author}{\bibfnamefont{J.}~\bibnamefont{Garriga}},
  \bibinfo{author}{\bibfnamefont{A.~H.} \bibnamefont{Guth}}, \bibnamefont{and}
  \bibinfo{author}{\bibfnamefont{A.}~\bibnamefont{Vilenkin}},
  \bibinfo{journal}{Phys. Rev.} \textbf{\bibinfo{volume}{D76}},
  \bibinfo{pages}{123512} (\bibinfo{year}{2007}), \eprint{hep-th/0612242}.

\bibitem[{\citenamefont{Hawking et~al.}(1982)\citenamefont{Hawking, Moss, and
  Stewart}}]{Hawking:1982ga}
\bibinfo{author}{\bibfnamefont{S.~W.} \bibnamefont{Hawking}},
  \bibinfo{author}{\bibfnamefont{I.~G.} \bibnamefont{Moss}}, \bibnamefont{and}
  \bibinfo{author}{\bibfnamefont{J.~M.} \bibnamefont{Stewart}},
  \bibinfo{journal}{Phys. Rev.} \textbf{\bibinfo{volume}{D26}},
  \bibinfo{pages}{2681} (\bibinfo{year}{1982}).

\bibitem[{\citenamefont{Wu}(1983)}]{Wu:1984eda}
\bibinfo{author}{\bibfnamefont{Z.-C.} \bibnamefont{Wu}},
  \bibinfo{journal}{Phys. Rev.} \textbf{\bibinfo{volume}{D28}},
  \bibinfo{pages}{1898} (\bibinfo{year}{1983}).

\bibitem[{\citenamefont{Moss}(1994)}]{Moss:1994pi}
\bibinfo{author}{\bibfnamefont{I.~G.} \bibnamefont{Moss}}
  (\bibinfo{year}{1994}), \eprint{gr-qc/9405045}.

\bibitem[{\citenamefont{Freivogel et~al.}(2007)\citenamefont{Freivogel,
  Horowitz, and Shenker}}]{Freivogel:2007fx}
\bibinfo{author}{\bibfnamefont{B.}~\bibnamefont{Freivogel}},
  \bibinfo{author}{\bibfnamefont{G.~T.} \bibnamefont{Horowitz}},
  \bibnamefont{and} \bibinfo{author}{\bibfnamefont{S.}~\bibnamefont{Shenker}},
  \bibinfo{journal}{JHEP} \textbf{\bibinfo{volume}{0705}}, \bibinfo{pages}{090}
  (\bibinfo{year}{2007}), \eprint{hep-th/0703146}.

\bibitem[{\citenamefont{Aguirre and Johnson}(2008)}]{Aguirre:2007wm}
\bibinfo{author}{\bibfnamefont{A.}~\bibnamefont{Aguirre}} \bibnamefont{and}
  \bibinfo{author}{\bibfnamefont{M.~C.} \bibnamefont{Johnson}},
  \bibinfo{journal}{Phys. Rev.} \textbf{\bibinfo{volume}{D77}},
  \bibinfo{pages}{123536} (\bibinfo{year}{2008}), \eprint{0712.3038}.

\bibitem[{\citenamefont{Aguirre et~al.}(2009)\citenamefont{Aguirre, Johnson,
  and Tysanner}}]{Aguirre:2008wy}
\bibinfo{author}{\bibfnamefont{A.}~\bibnamefont{Aguirre}},
  \bibinfo{author}{\bibfnamefont{M.~C.} \bibnamefont{Johnson}},
  \bibnamefont{and} \bibinfo{author}{\bibfnamefont{M.}~\bibnamefont{Tysanner}},
  \bibinfo{journal}{Phys. Rev.} \textbf{\bibinfo{volume}{D79}},
  \bibinfo{pages}{123514} (\bibinfo{year}{2009}).

\bibitem[{\citenamefont{Chang et~al.}(2008)\citenamefont{Chang, Kleban, and
  Levi}}]{Chang:2007eq}
\bibinfo{author}{\bibfnamefont{S.}~\bibnamefont{Chang}},
  \bibinfo{author}{\bibfnamefont{M.}~\bibnamefont{Kleban}}, \bibnamefont{and}
  \bibinfo{author}{\bibfnamefont{T.~S.} \bibnamefont{Levi}},
  \bibinfo{journal}{JCAP} \textbf{\bibinfo{volume}{0804}}, \bibinfo{pages}{034}
  (\bibinfo{year}{2008}).

\bibitem[{\citenamefont{Chang et~al.}(2009)\citenamefont{Chang, Kleban, and
  Levi}}]{Chang:2008gj}
\bibinfo{author}{\bibfnamefont{S.}~\bibnamefont{Chang}},
  \bibinfo{author}{\bibfnamefont{M.}~\bibnamefont{Kleban}}, \bibnamefont{and}
  \bibinfo{author}{\bibfnamefont{T.~S.} \bibnamefont{Levi}},
  \bibinfo{journal}{JCAP} \textbf{\bibinfo{volume}{0904}}, \bibinfo{pages}{025}
  (\bibinfo{year}{2009}), \eprint{0810.5128}.

\bibitem[{\citenamefont{Czech et~al.}(2010)\citenamefont{Czech, Kleban, Larjo,
  Levi, and Sigurdson}}]{Czech:2010rg}
\bibinfo{author}{\bibfnamefont{B.}~\bibnamefont{Czech}},
  \bibinfo{author}{\bibfnamefont{M.}~\bibnamefont{Kleban}},
  \bibinfo{author}{\bibfnamefont{K.}~\bibnamefont{Larjo}},
  \bibinfo{author}{\bibfnamefont{T.~S.} \bibnamefont{Levi}}, \bibnamefont{and}
  \bibinfo{author}{\bibfnamefont{K.}~\bibnamefont{Sigurdson}},
  \bibinfo{journal}{JCAP} \textbf{\bibinfo{volume}{1012}}, \bibinfo{pages}{023}
  (\bibinfo{year}{2010}), \eprint{1006.0832}.

\bibitem[{\citenamefont{Freivogel et~al.}(2009)\citenamefont{Freivogel, Kleban,
  Nicolis, and Sigurdson}}]{Freivogel:2009it}
\bibinfo{author}{\bibfnamefont{B.}~\bibnamefont{Freivogel}},
  \bibinfo{author}{\bibfnamefont{M.}~\bibnamefont{Kleban}},
  \bibinfo{author}{\bibfnamefont{A.}~\bibnamefont{Nicolis}}, \bibnamefont{and}
  \bibinfo{author}{\bibfnamefont{K.}~\bibnamefont{Sigurdson}},
  \bibinfo{journal}{JCAP} \textbf{\bibinfo{volume}{0908}}, \bibinfo{pages}{036}
  (\bibinfo{year}{2009}).

\bibitem[{\citenamefont{Easther et~al.}(2009)\citenamefont{Easther, Giblin,
  Hui, and Lim}}]{Easther:2009ft}
\bibinfo{author}{\bibfnamefont{R.}~\bibnamefont{Easther}},
  \bibinfo{author}{\bibfnamefont{J.~T.} \bibnamefont{Giblin},
  \bibfnamefont{Jr}}, \bibinfo{author}{\bibfnamefont{L.}~\bibnamefont{Hui}},
  \bibnamefont{and} \bibinfo{author}{\bibfnamefont{E.~A.} \bibnamefont{Lim}},
  \bibinfo{journal}{Phys. Rev.} \textbf{\bibinfo{volume}{D80}},
  \bibinfo{pages}{123519} (\bibinfo{year}{2009}), \eprint{0907.3234}.

\bibitem[{\citenamefont{Giblin et~al.}(2010)\citenamefont{Giblin, Hui, Lim, and
  Yang}}]{Giblin:2010bd}
\bibinfo{author}{\bibfnamefont{J.~T.} \bibnamefont{Giblin}, \bibfnamefont{Jr}},
  \bibinfo{author}{\bibfnamefont{L.}~\bibnamefont{Hui}},
  \bibinfo{author}{\bibfnamefont{E.~A.} \bibnamefont{Lim}}, \bibnamefont{and}
  \bibinfo{author}{\bibfnamefont{I.-S.} \bibnamefont{Yang}},
  \bibinfo{journal}{Phys. Rev.} \textbf{\bibinfo{volume}{D82}},
  \bibinfo{pages}{045019} (\bibinfo{year}{2010}), \eprint{1005.3493}.

\bibitem[{\citenamefont{Kleban et~al.}(2011)\citenamefont{Kleban, Levi, and
  Sigurdson}}]{Kleban:2011yc}
\bibinfo{author}{\bibfnamefont{M.}~\bibnamefont{Kleban}},
  \bibinfo{author}{\bibfnamefont{T.~S.} \bibnamefont{Levi}}, \bibnamefont{and}
  \bibinfo{author}{\bibfnamefont{K.}~\bibnamefont{Sigurdson}}
  (\bibinfo{year}{2011}), \eprint{1109.3473}.

\bibitem[{\citenamefont{Lim and Simon}(2011)}]{Lim:2011kd}
\bibinfo{author}{\bibfnamefont{E.~A.} \bibnamefont{Lim}} \bibnamefont{and}
  \bibinfo{author}{\bibfnamefont{D.}~\bibnamefont{Simon}}
  (\bibinfo{year}{2011}), \eprint{1103.4300}.

\bibitem[{\citenamefont{Dahlen}(2010)}]{Dahlen:2008rd}
\bibinfo{author}{\bibfnamefont{A.}~\bibnamefont{Dahlen}},
  \bibinfo{journal}{Phys.Rev.} \textbf{\bibinfo{volume}{D81}},
  \bibinfo{pages}{063501} (\bibinfo{year}{2010}), \eprint{0812.0414}.

\bibitem[{\citenamefont{Johnson and Yang}(2010)}]{Johnson:2010bn}
\bibinfo{author}{\bibfnamefont{M.~C.} \bibnamefont{Johnson}} \bibnamefont{and}
  \bibinfo{author}{\bibfnamefont{I.-S.} \bibnamefont{Yang}},
  \bibinfo{journal}{Phys. Rev.} \textbf{\bibinfo{volume}{D82}},
  \bibinfo{pages}{065023} (\bibinfo{year}{2010}), \eprint{1005.3506}.

\bibitem[{\citenamefont{Aguirre and Johnson}(2011)}]{Aguirre:2009ug}
\bibinfo{author}{\bibfnamefont{A.}~\bibnamefont{Aguirre}} \bibnamefont{and}
  \bibinfo{author}{\bibfnamefont{M.~C.} \bibnamefont{Johnson}},
  \bibinfo{journal}{Rept.Prog.Phys.} \textbf{\bibinfo{volume}{74}},
  \bibinfo{pages}{074901} (\bibinfo{year}{2011}), \eprint{0908.4105}.

\bibitem[{\citenamefont{Bennett et~al.}(2003)}]{Bennett:2003ba}
\bibinfo{author}{\bibfnamefont{C.~L.} \bibnamefont{Bennett}}
  \bibnamefont{et~al.} (\bibinfo{collaboration}{WMAP}),
  \bibinfo{journal}{Astrophys. J.} \textbf{\bibinfo{volume}{583}},
  \bibinfo{pages}{1} (\bibinfo{year}{2003}), \eprint{astro-ph/0301158}.

\bibitem[{\citenamefont{Komatsu et~al.}(2011)}]{Komatsu:2010fb}
\bibinfo{author}{\bibfnamefont{E.}~\bibnamefont{Komatsu}} \bibnamefont{et~al.}
  (\bibinfo{collaboration}{WMAP Collaboration}),
  \bibinfo{journal}{Astrophys.J.Suppl.} \textbf{\bibinfo{volume}{192}},
  \bibinfo{pages}{18} (\bibinfo{year}{2011}), \eprint{1001.4538}.

\bibitem[{\citenamefont{Feeney et~al.}(2011{\natexlab{a}})\citenamefont{Feeney,
  Johnson, Mortlock, and Peiris}}]{Feeney:2010dd}
\bibinfo{author}{\bibfnamefont{S.~M.} \bibnamefont{Feeney}},
  \bibinfo{author}{\bibfnamefont{M.~C.} \bibnamefont{Johnson}},
  \bibinfo{author}{\bibfnamefont{D.~J.} \bibnamefont{Mortlock}},
  \bibnamefont{and} \bibinfo{author}{\bibfnamefont{H.~V.}
  \bibnamefont{Peiris}}, \bibinfo{journal}{Phys.Rev.}
  \textbf{\bibinfo{volume}{D84}}, \bibinfo{pages}{043507}
  (\bibinfo{year}{2011}{\natexlab{a}}), \eprint{1012.3667}.

\bibitem[{\citenamefont{Feeney et~al.}(2011{\natexlab{b}})\citenamefont{Feeney,
  Johnson, Mortlock, and Peiris}}]{Feeney:2010jj}
\bibinfo{author}{\bibfnamefont{S.~M.} \bibnamefont{Feeney}},
  \bibinfo{author}{\bibfnamefont{M.~C.} \bibnamefont{Johnson}},
  \bibinfo{author}{\bibfnamefont{D.~J.} \bibnamefont{Mortlock}},
  \bibnamefont{and} \bibinfo{author}{\bibfnamefont{H.~V.}
  \bibnamefont{Peiris}}, \bibinfo{journal}{Phys.Rev.Lett.}
  \textbf{\bibinfo{volume}{107}}, \bibinfo{pages}{071301}
  (\bibinfo{year}{2011}{\natexlab{b}}), \eprint{1012.1995}.

\bibitem[{\citenamefont{{Lehner}}(2001)}]{2001CQGra..18R..25L}
\bibinfo{author}{\bibfnamefont{L.}~\bibnamefont{{Lehner}}},
  \bibinfo{journal}{Classical and Quantum Gravity}
  \textbf{\bibinfo{volume}{18}}, \bibinfo{pages}{25} (\bibinfo{year}{2001}),
  \eprint{arXiv:gr-qc/0106072}.

\bibitem[{\citenamefont{Adams et~al.}(1990)\citenamefont{Adams, Freese, and
  Widrow}}]{Adams:1989su}
\bibinfo{author}{\bibfnamefont{F.~C.} \bibnamefont{Adams}},
  \bibinfo{author}{\bibfnamefont{K.}~\bibnamefont{Freese}}, \bibnamefont{and}
  \bibinfo{author}{\bibfnamefont{L.~M.} \bibnamefont{Widrow}},
  \bibinfo{journal}{Phys. Rev.} \textbf{\bibinfo{volume}{D41}},
  \bibinfo{pages}{347} (\bibinfo{year}{1990}).

\bibitem[{\citenamefont{Garriga and Vilenkin}(1992)}]{Garriga:1991tb}
\bibinfo{author}{\bibfnamefont{J.}~\bibnamefont{Garriga}} \bibnamefont{and}
  \bibinfo{author}{\bibfnamefont{A.}~\bibnamefont{Vilenkin}},
  \bibinfo{journal}{Phys. Rev.} \textbf{\bibinfo{volume}{D45}},
  \bibinfo{pages}{3469} (\bibinfo{year}{1992}).

\bibitem[{\citenamefont{Garriga and Vilenkin}(1991)}]{Garriga:1991ts}
\bibinfo{author}{\bibfnamefont{J.}~\bibnamefont{Garriga}} \bibnamefont{and}
  \bibinfo{author}{\bibfnamefont{A.}~\bibnamefont{Vilenkin}},
  \bibinfo{journal}{Phys. Rev.} \textbf{\bibinfo{volume}{D44}},
  \bibinfo{pages}{1007} (\bibinfo{year}{1991}).

\bibitem[{\citenamefont{Aguirre and Johnson}(2005)}]{Aguirre:2005sv}
\bibinfo{author}{\bibfnamefont{A.}~\bibnamefont{Aguirre}} \bibnamefont{and}
  \bibinfo{author}{\bibfnamefont{M.~C.} \bibnamefont{Johnson}},
  \bibinfo{journal}{Phys. Rev.} \textbf{\bibinfo{volume}{D72}},
  \bibinfo{pages}{103525} (\bibinfo{year}{2005}), \eprint{gr-qc/0508093}.

\bibitem[{\citenamefont{Blanco-Pillado
  et~al.}(2004)\citenamefont{Blanco-Pillado, Bucher, Ghassemi, and
  Glanois}}]{Blanco-Pillado:2003hq}
\bibinfo{author}{\bibfnamefont{J.~J.} \bibnamefont{Blanco-Pillado}},
  \bibinfo{author}{\bibfnamefont{M.}~\bibnamefont{Bucher}},
  \bibinfo{author}{\bibfnamefont{S.}~\bibnamefont{Ghassemi}}, \bibnamefont{and}
  \bibinfo{author}{\bibfnamefont{F.}~\bibnamefont{Glanois}},
  \bibinfo{journal}{Phys. Rev.} \textbf{\bibinfo{volume}{D69}},
  \bibinfo{pages}{103515} (\bibinfo{year}{2004}), \eprint{hep-th/0306151}.

\bibitem[{\citenamefont{Carone and Guth}(1990)}]{Carone:1989nj}
\bibinfo{author}{\bibfnamefont{C.~D.} \bibnamefont{Carone}} \bibnamefont{and}
  \bibinfo{author}{\bibfnamefont{A.~H.} \bibnamefont{Guth}},
  \bibinfo{journal}{Phys. Rev.} \textbf{\bibinfo{volume}{D42}},
  \bibinfo{pages}{2446} (\bibinfo{year}{1990}).

\bibitem[{\citenamefont{Israel}(1966)}]{Israel:1966rt}
\bibinfo{author}{\bibfnamefont{W.}~\bibnamefont{Israel}},
  \bibinfo{journal}{Nuovo Cim.} \textbf{\bibinfo{volume}{B44S10}},
  \bibinfo{pages}{1} (\bibinfo{year}{1966}).

\bibitem[{\citenamefont{Vilenkin}(1983)}]{Vilenkin:1984hy}
\bibinfo{author}{\bibfnamefont{A.}~\bibnamefont{Vilenkin}},
  \bibinfo{journal}{Phys. Lett.} \textbf{\bibinfo{volume}{B133}},
  \bibinfo{pages}{177} (\bibinfo{year}{1983}).

\bibitem[{\citenamefont{Ipser and Sikivie}(1984)}]{Ipser:1983db}
\bibinfo{author}{\bibfnamefont{J.}~\bibnamefont{Ipser}} \bibnamefont{and}
  \bibinfo{author}{\bibfnamefont{P.}~\bibnamefont{Sikivie}},
  \bibinfo{journal}{Phys. Rev.} \textbf{\bibinfo{volume}{D30}},
  \bibinfo{pages}{712} (\bibinfo{year}{1984}).

\bibitem[{\citenamefont{Langlois et~al.}(2002)\citenamefont{Langlois, Maeda,
  and Wands}}]{Langlois:2001uq}
\bibinfo{author}{\bibfnamefont{D.}~\bibnamefont{Langlois}},
  \bibinfo{author}{\bibfnamefont{K.-i.} \bibnamefont{Maeda}}, \bibnamefont{and}
  \bibinfo{author}{\bibfnamefont{D.}~\bibnamefont{Wands}},
  \bibinfo{journal}{Phys. Rev. Lett.} \textbf{\bibinfo{volume}{88}},
  \bibinfo{pages}{181301} (\bibinfo{year}{2002}), \eprint{gr-qc/0111013}.

\bibitem[{\citenamefont{Lyth}(1997)}]{Lyth:1996im}
\bibinfo{author}{\bibfnamefont{D.~H.} \bibnamefont{Lyth}},
  \bibinfo{journal}{Phys. Rev. Lett.} \textbf{\bibinfo{volume}{78}},
  \bibinfo{pages}{1861} (\bibinfo{year}{1997}), \eprint{hep-ph/9606387}.

\bibitem[{\citenamefont{{Choptuik}}(1993)}]{1993PhRvL..70....9C}
\bibinfo{author}{\bibfnamefont{M.~W.} \bibnamefont{{Choptuik}}},
  \bibinfo{journal}{Physical Review Letters} \textbf{\bibinfo{volume}{70}},
  \bibinfo{pages}{9} (\bibinfo{year}{1993}).

\bibitem[{\citenamefont{Aguirre et~al.}(2006)\citenamefont{Aguirre, Banks, and
  Johnson}}]{Aguirre:2006ap}
\bibinfo{author}{\bibfnamefont{A.}~\bibnamefont{Aguirre}},
  \bibinfo{author}{\bibfnamefont{T.}~\bibnamefont{Banks}}, \bibnamefont{and}
  \bibinfo{author}{\bibfnamefont{M.}~\bibnamefont{Johnson}},
  \bibinfo{journal}{JHEP} \textbf{\bibinfo{volume}{08}}, \bibinfo{pages}{065}
  (\bibinfo{year}{2006}), \eprint{hep-th/0603107}.

\bibitem[{\citenamefont{Banks and Johnson}(2005)}]{Banks:2005ru}
\bibinfo{author}{\bibfnamefont{T.}~\bibnamefont{Banks}} \bibnamefont{and}
  \bibinfo{author}{\bibfnamefont{M.}~\bibnamefont{Johnson}}
  (\bibinfo{year}{2005}), \eprint{hep-th/0512141}.

\bibitem[{\citenamefont{Banks}(2002)}]{Banks:2002nm}
\bibinfo{author}{\bibfnamefont{T.}~\bibnamefont{Banks}} (\bibinfo{year}{2002}),
  \eprint{hep-th/0211160}.

\bibitem[{\citenamefont{Dong and Harlow}(2011)}]{Dong:2011gx}
\bibinfo{author}{\bibfnamefont{X.}~\bibnamefont{Dong}} \bibnamefont{and}
  \bibinfo{author}{\bibfnamefont{D.}~\bibnamefont{Harlow}},
  \bibinfo{journal}{JCAP} \textbf{\bibinfo{volume}{1111}}, \bibinfo{pages}{044}
  (\bibinfo{year}{2011}), \eprint{1109.0011}.

\bibitem[{\citenamefont{Calabrese et~al.}(2004)\citenamefont{Calabrese, Lehner,
  Reula, Sarbach, and Tiglio}}]{Calabrese:2003vx}
\bibinfo{author}{\bibfnamefont{G.}~\bibnamefont{Calabrese}},
  \bibinfo{author}{\bibfnamefont{L.}~\bibnamefont{Lehner}},
  \bibinfo{author}{\bibfnamefont{O.}~\bibnamefont{Reula}},
  \bibinfo{author}{\bibfnamefont{O.}~\bibnamefont{Sarbach}}, \bibnamefont{and}
  \bibinfo{author}{\bibfnamefont{M.}~\bibnamefont{Tiglio}},
  \bibinfo{journal}{Class.Quant.Grav.} \textbf{\bibinfo{volume}{21}},
  \bibinfo{pages}{5735} (\bibinfo{year}{2004}), \eprint{gr-qc/0308007}.

\bibitem[{\citenamefont{Lehner et~al.}(2004)\citenamefont{Lehner, Neilsen,
  Reula, and Tiglio}}]{Lehner:2004cf}
\bibinfo{author}{\bibfnamefont{L.}~\bibnamefont{Lehner}},
  \bibinfo{author}{\bibfnamefont{D.}~\bibnamefont{Neilsen}},
  \bibinfo{author}{\bibfnamefont{O.}~\bibnamefont{Reula}}, \bibnamefont{and}
  \bibinfo{author}{\bibfnamefont{M.}~\bibnamefont{Tiglio}},
  \bibinfo{journal}{Class.Quant.Grav.} \textbf{\bibinfo{volume}{21}},
  \bibinfo{pages}{5819} (\bibinfo{year}{2004}), \eprint{gr-qc/0406116}.

\bibitem[{\citenamefont{Richardson}(1911)}]{richardson}
\bibinfo{author}{\bibfnamefont{L.~F.} \bibnamefont{Richardson}},
  \bibinfo{journal}{Phil. Trans. Roy. Soc. A} \textbf{\bibinfo{volume}{210}}
  (\bibinfo{year}{1911}).

\bibitem[{\citenamefont{Copeland et~al.}(1995)\citenamefont{Copeland, Gleiser,
  and Muller}}]{Copeland:1995fq}
\bibinfo{author}{\bibfnamefont{E.~J.} \bibnamefont{Copeland}},
  \bibinfo{author}{\bibfnamefont{M.}~\bibnamefont{Gleiser}}, \bibnamefont{and}
  \bibinfo{author}{\bibfnamefont{H.-R.} \bibnamefont{Muller}},
  \bibinfo{journal}{Phys.Rev.} \textbf{\bibinfo{volume}{D52}},
  \bibinfo{pages}{1920} (\bibinfo{year}{1995}), \eprint{hep-ph/9503217}.

\bibitem[{\citenamefont{Fodor et~al.}(2010)\citenamefont{Fodor, Forgacs, and
  Mezei}}]{Fodor:2009kg}
\bibinfo{author}{\bibfnamefont{G.}~\bibnamefont{Fodor}},
  \bibinfo{author}{\bibfnamefont{P.}~\bibnamefont{Forgacs}}, \bibnamefont{and}
  \bibinfo{author}{\bibfnamefont{M.}~\bibnamefont{Mezei}},
  \bibinfo{journal}{Phys.Rev.} \textbf{\bibinfo{volume}{D81}},
  \bibinfo{pages}{064029} (\bibinfo{year}{2010}), \eprint{0912.5351}.

\bibitem[{\citenamefont{Amin et~al.}(2011)\citenamefont{Amin, Easther, Finkel,
  Flauger, and Hertzberg}}]{Amin:2011hj}
\bibinfo{author}{\bibfnamefont{M.~A.} \bibnamefont{Amin}},
  \bibinfo{author}{\bibfnamefont{R.}~\bibnamefont{Easther}},
  \bibinfo{author}{\bibfnamefont{H.}~\bibnamefont{Finkel}},
  \bibinfo{author}{\bibfnamefont{R.}~\bibnamefont{Flauger}}, \bibnamefont{and}
  \bibinfo{author}{\bibfnamefont{M.~P.} \bibnamefont{Hertzberg}}
  (\bibinfo{year}{2011}), \eprint{1106.3335}.

\bibitem[{\citenamefont{Amin et~al.}(2010)\citenamefont{Amin, Easther, and
  Finkel}}]{Amin:2010dc}
\bibinfo{author}{\bibfnamefont{M.~A.} \bibnamefont{Amin}},
  \bibinfo{author}{\bibfnamefont{R.}~\bibnamefont{Easther}}, \bibnamefont{and}
  \bibinfo{author}{\bibfnamefont{H.}~\bibnamefont{Finkel}},
  \bibinfo{journal}{JCAP} \textbf{\bibinfo{volume}{1012}}, \bibinfo{pages}{001}
  (\bibinfo{year}{2010}), \eprint{1009.2505}.

\bibitem[{\citenamefont{Amin}(2010)}]{Amin:2010xe}
\bibinfo{author}{\bibfnamefont{M.~A.} \bibnamefont{Amin}}
  (\bibinfo{year}{2010}), \eprint{1006.3075}.

\bibitem[{\citenamefont{Amin and Shirokoff}(2010)}]{Amin:2010jq}
\bibinfo{author}{\bibfnamefont{M.~A.} \bibnamefont{Amin}} \bibnamefont{and}
  \bibinfo{author}{\bibfnamefont{D.}~\bibnamefont{Shirokoff}},
  \bibinfo{journal}{Phys.Rev.} \textbf{\bibinfo{volume}{D81}},
  \bibinfo{pages}{085045} (\bibinfo{year}{2010}), \eprint{1002.3380}.

\bibitem[{\citenamefont{Dymnikova et~al.}(2000)\citenamefont{Dymnikova, Koziel,
  Khlopov, and Rubin}}]{Dymnikova:2000dy}
\bibinfo{author}{\bibfnamefont{I.}~\bibnamefont{Dymnikova}},
  \bibinfo{author}{\bibfnamefont{L.}~\bibnamefont{Koziel}},
  \bibinfo{author}{\bibfnamefont{M.}~\bibnamefont{Khlopov}}, \bibnamefont{and}
  \bibinfo{author}{\bibfnamefont{S.}~\bibnamefont{Rubin}},
  \bibinfo{journal}{Grav. Cosmol.} \textbf{\bibinfo{volume}{6}},
  \bibinfo{pages}{311} (\bibinfo{year}{2000}), \eprint{hep-th/0010120}.

\bibitem[{\citenamefont{Hindmarsh and Salmi}(2008)}]{Hindmarsh:2007jb}
\bibinfo{author}{\bibfnamefont{M.}~\bibnamefont{Hindmarsh}} \bibnamefont{and}
  \bibinfo{author}{\bibfnamefont{P.}~\bibnamefont{Salmi}},
  \bibinfo{journal}{Phys.Rev.} \textbf{\bibinfo{volume}{D77}},
  \bibinfo{pages}{105025} (\bibinfo{year}{2008}), \eprint{0712.0614}.

\bibitem[{\citenamefont{Khlopov et~al.}(1998)\citenamefont{Khlopov, Konoplich,
  Rubin, and Sakharov}}]{Khlopov:1998nm}
\bibinfo{author}{\bibfnamefont{M.}~\bibnamefont{Khlopov}},
  \bibinfo{author}{\bibfnamefont{R.}~\bibnamefont{Konoplich}},
  \bibinfo{author}{\bibfnamefont{S.}~\bibnamefont{Rubin}}, \bibnamefont{and}
  \bibinfo{author}{\bibfnamefont{A.}~\bibnamefont{Sakharov}}
  (\bibinfo{year}{1998}), \eprint{hep-ph/9807343}.

\bibitem[{\citenamefont{Baumann et~al.}(2007)\citenamefont{Baumann, Dymarsky,
  Klebanov, McAllister, and Steinhardt}}]{Baumann:2007np}
\bibinfo{author}{\bibfnamefont{D.}~\bibnamefont{Baumann}},
  \bibinfo{author}{\bibfnamefont{A.}~\bibnamefont{Dymarsky}},
  \bibinfo{author}{\bibfnamefont{I.~R.} \bibnamefont{Klebanov}},
  \bibinfo{author}{\bibfnamefont{L.}~\bibnamefont{McAllister}},
  \bibnamefont{and} \bibinfo{author}{\bibfnamefont{P.~J.}
  \bibnamefont{Steinhardt}}, \bibinfo{journal}{Phys. Rev. Lett.}
  \textbf{\bibinfo{volume}{99}}, \bibinfo{pages}{141601}
  (\bibinfo{year}{2007}), \eprint{0705.3837}.

\bibitem[{\citenamefont{Linde and Westphal}(2008)}]{Linde:2007jn}
\bibinfo{author}{\bibfnamefont{A.}~\bibnamefont{Linde}} \bibnamefont{and}
  \bibinfo{author}{\bibfnamefont{A.}~\bibnamefont{Westphal}},
  \bibinfo{journal}{JCAP} \textbf{\bibinfo{volume}{0803}}, \bibinfo{pages}{005}
  (\bibinfo{year}{2008}), \eprint{0712.1610}.

\bibitem[{\citenamefont{Boubekeur and Lyth}(2005)}]{Boubekeur:2005zm}
\bibinfo{author}{\bibfnamefont{L.}~\bibnamefont{Boubekeur}} \bibnamefont{and}
  \bibinfo{author}{\bibfnamefont{D.~H.} \bibnamefont{Lyth}},
  \bibinfo{journal}{JCAP} \textbf{\bibinfo{volume}{0507}}, \bibinfo{pages}{010}
  (\bibinfo{year}{2005}), \eprint{hep-ph/0502047}.

\bibitem[{\citenamefont{{Choptuik}}(1989)}]{1989fnr..book..206C}
\bibinfo{author}{\bibfnamefont{M.~W.} \bibnamefont{{Choptuik}}},
  \emph{\bibinfo{title}{{Experiences with an adaptive mesh refinement algorithm
  in numerical relativity.}}} (\bibinfo{year}{1989}), pp.
  \bibinfo{pages}{206--221}.

\bibitem[{\citenamefont{Parke}(1983)}]{Parke:1982pm}
\bibinfo{author}{\bibfnamefont{S.~J.} \bibnamefont{Parke}},
  \bibinfo{journal}{Phys. Lett.} \textbf{\bibinfo{volume}{B121}},
  \bibinfo{pages}{313} (\bibinfo{year}{1983}).

\bibitem[{\citenamefont{Copsey}(2011{\natexlab{a}})}]{Copsey:2011wy}
\bibinfo{author}{\bibfnamefont{K.}~\bibnamefont{Copsey}}
  (\bibinfo{year}{2011}{\natexlab{a}}), \eprint{1108.2255}.

\bibitem[{\citenamefont{Copsey}(2011{\natexlab{b}})}]{Copsey:2011zj}
\bibinfo{author}{\bibfnamefont{K.}~\bibnamefont{Copsey}}
  (\bibinfo{year}{2011}{\natexlab{b}}), \eprint{1109.4931}.

\end{thebibliography}

\end{document}